\documentclass[10pt,fleqn]{article}

\usepackage{latexsym}
\usepackage{amsbsy}
\usepackage{amsfonts}
\usepackage{amssymb}
\usepackage{amscd}
\usepackage{mathabx}  
\usepackage{t1enc}
\usepackage{epsfig}
\usepackage{color}
\usepackage{cite}

\usepackage{capt-of}
\usepackage{caption}

\def\appendix{\par
  \setcounter{section}{0}%
  \def\@chapapp{\appendixname}%
\def\thesection{Appendix \Alph{section}}}

\renewcommand{\appendixname}{Appendix}

\mathchardef\Gamma="0100
\mathchardef\Delta="0101
\mathchardef\Theta="0102
\mathchardef\Lambda="0103
\mathchardef\Xi="0104
\mathchardef\Pi="0105
\mathchardef\Sigma="0106
\mathchardef\Upsilon="0107
\mathchardef\Phi="0108
\mathchardef\Psi="0109
\mathchardef\Omega="010A

\newcommand{\sq}{\hbox{\rlap{$\sqcap$}$\sqcup$}}
\newcommand{\qed}{\ifmmode\sq\else{\unskip\nobreak\hfil
  \penalty50\hskip1em\null\nobreak\hfil\sq
  \parfillskip=0pt\finalhyphendemerits=0\endgraf}\fi{}}

\def\I{{\rm i}}

\def\D{{\rm d}}
\def\E{{\rm e}}

\def\vec{\boldsymbol}

\def\cal{\mathcal}

\def\rap{\!\!\!\! } 

\def\its{\it}

\def\qed{\hfill $\Box$}

\def\primespe{\kern .08em '}

\def\boxlim#1#2{\buildrel\hbox{\scriptsize $#1$}\over {\hbox{\scriptsize $#2$}}}
\def\doublelimite#1#2{\lim_{\boxlim{#1}{#2}}}

\newtheorem{theorem}{Theorem}{\bf}{\it}
\newtheorem{proposition}{Proposition}
{\bf}{\it}

\newtheorem{remark}{Remark}[section]
\newtheorem{definition}{Definition}

\input{diagxy}

\begin{document}

\title{Fractional-order Fourier transformations as pseudo-differential operators. Applications to diffraction and imaging by optical systems}

\date{}

\maketitle

\begin{center}

\vskip -1.2cm

{
\renewcommand{\thefootnote}{}
 {\bf   Pierre Pellat-Finet\footnote{\hskip -.53cm Laboratoire de Math\'ematiques de Bretagne Atlantique (LMBA) UMR CNRS 6205,

\noindent Universit\'e de Bretagne Sud, CS 60573, 56017 Vannes, France.

\noindent pierre.pellat-finet@univ-ubs.fr     
 }}
}
\setcounter{footnote}{0}

\medskip
    {\sl \small Universit\'e  Bretagne Sud,    UMR CNRS 6205, LMBA, F-56000 Vannes, France}

    \smallskip

\end{center}

\vskip .45cm 

\begin{center}
\begin{minipage}{12cm}
\hrulefill

\smallskip
{\small
  {\bf Abstract.}  Circular and hyperbolic fractional-order Fourier transformations are actually Weyl pseudo-differential operators.  Their associated kernels and symbols are  written explicitly. Products of  fractional-order Fourier transformations are obtained by composing their kernels or symbols, in accordance with the Weyl calculus.  On the other hand, optical field transfers by diffraction, from spherical emitters to  receivers, are mathematically expressed using fractional-order Fourier transforms and are classified into three categories, for which we provide geometrical characterizations. Respecting the Huygens--Fresnel principle requires that the associated transformations be adequately composed, which is achieved through the composition rules of pseudo-differential operators. By applying the Weyl calculus, we prove that, in general, imaging through a refracting spherical cap can be described as the composition of two fractional-order Fourier transformations, but only if those transformations are of the same kind. Basic laws of coherent geometrical imaging are thus deduced from the composition of two circular fractional-order Fourier transformations. We also prove that the arrangement of a sequence of points along the axis of a centered optical system is preserved by imaging, up to a circular permutation.
This recovers a classical result from geometrical optics, but here it is derived  within the framework of  diffraction theory.


}
\end{minipage}
\end{center}

\begin{center}
\begin{minipage}{12cm}
  
\smallskip

{\small \noindent {\bf Contents}

\smallskip

\noindent Introduction \dotfill \pageref{intro}

\noindent Part I. Fractional Fourier transformations and pseudo-differential operators \dotfill \pageref{part1}

\hskip .3cm \ref{sect1}. Weyl calculus\dotfill \pageref{sect1}

\hskip .7cm \ref{sect11}. Two-dimensional Fourier transformation\dotfill \pageref{sect11}

\hskip .7cm \ref{sect12}. Pseudo-differential operators\dotfill \pageref{sect12}

\hskip .7cm \ref{sect13}. Weyl pseudo-differential operators\dotfill \pageref{sect13}

\hskip .7cm \ref{sect14}. Symbols in ${\cal S}'({\mathbb R}^2\times {\mathbb R}^2)$
\dotfill \pageref{sect14}

\hskip .7cm \ref{sect15}. The parity operator\dotfill \pageref{sect15}

\hskip .3cm \ref{sect2}. Fractional-order Fourier transformations as Weyl pseudo-differential  

\hskip .7cm operators\dotfill \pageref{sect2}

\hskip .7cm \ref{sect21}. Three classes of fractional-order Fourier transformations \dotfill \pageref{sect21}

\hskip .7cm \ref{sect22}.  Circular fractional-order Fourier transformations \dotfill \pageref{sect22}

\hskip .7cm \ref{sect23}.  Hyperbolic fractional-order Fourier transformations of the first kind\dotfill\pageref{sect23}

\hskip .7cm \ref{sect24}.  Hyperbolic fractional-order Fourier transformations of the second kind\dotfill\pageref{sect24}

\hskip .7cm \ref{sect25}.  Concluding remark\dotfill\pageref{sect25}

\null
\vfill

}\end{minipage}
\end{center}

\begin{center}
  \begin{minipage}{12cm}

    {\small 

            \hskip .3cm \ref{sect3}. Compositions of fractional-order Fourier transformations of different kinds\dotfill \pageref{sect3}

      \hskip .7cm \ref{sect31}.  Circular transformations and hyperbolic  transformations of the first

      \hskip 1.35cm  kind \dotfill \pageref{sect31}

       \hskip .7cm \ref{sect32}.  Circular transformations and hyperbolic  transformations of the second

       \hskip 1.35cm kind \dotfill \pageref{sect32}

       \hskip .7cm \ref{sect33}.  Hyperbolic  transformations of the first and the second kind\dotfill \pageref{sect33}

       \hskip .7cm \ref{sect34}.  Fractional Fourier  transformation algebra\dotfill \pageref{sect34}

       \hskip .7cm \ref{sect35}. Identity and parity operators generally cannot be the product of two 

       \hskip 1.35cm fractional-order Fourier transformations of different kinds \dotfill \pageref{sect35}

\noindent Part II. Applications to diffraction and optical imaging \dotfill \pageref{part2}

\hskip .3cm \ref{sect4}. Elements of fractional Fourier optics \dotfill \pageref{sect4}

\hskip .7cm \ref{sect41}. Field transfer by diffraction\dotfill\pageref{sect41}

\hskip .7cm \ref{sect42}. Field-transfer representation by a fractional-order Fourier

\hskip 1.35cm transformation\dotfill\pageref{sect42}

\hskip .7cm \ref{sect43}. Geometrical characterization of a field transfer by diffraction\dotfill\pageref{sect43}

\hskip .7cm \ref{sect44}. Huygens--Fresnel principle\dotfill\pageref{sect44}

\hskip .3cm \ref{sect5}. Application to geometrical imaging by a centered system\dotfill\pageref{sect5}

\hskip .7cm \ref{sect51}. Coherent geometrical imaging\dotfill\pageref{sect51}

\hskip .7cm \ref{sect52}. Basic laws of coherent geometrical imaging\dotfill\pageref{sect52}

\hskip .7cm \ref{sect53}. Imaging a diffraction phenomenon\dotfill\pageref{sect53}

\hskip .7cm \ref{sect54}. Transforming a sequence of points on the optical axis\dotfill\pageref{sect54}

\hskip .3cm \ref{sect6}. Imaging as result of the composition of two  fractional-order

\hskip .7cm transformations\dotfill \pageref{sect6}

\hskip .7cm \ref{sect61}. Condition for a refracting spherical cap to form an image\dotfill \pageref{sect61}

\hskip .7cm \ref{sect62}. Imaging by a refracting spherical cap by composition of two circular

\hskip 1.35cm   fractional-order Fourier transformations\dotfill\pageref{sect62}

\hskip .7cm \ref{sect63}. Imaging by a centered optical system\dotfill\pageref{sect63}

\noindent Conclusion \dotfill \pageref{conc}

\ref{appenA}. Notation on distributions\dotfill\pageref{appenA}

\ref{appenB}. Composition of field-transfer operators: proof of Equation (\ref{eq2.23})\dotfill\pageref{appenB}

\noindent References \dotfill \pageref{refe2} 

}
\hrulefill
\end{minipage}
\end{center}

{\Large \bigskip \centerline{\sc Introduction}}\label{intro}  \bigskip\smallskip

\noindent Fractional-order Fourier transformations were systematically formalized
in 1980 by Namias \cite{Nam}, following precursor works by Wiener \cite{Wie}, Condon \cite{Con}, Kober \cite{Kob} and Patterson \cite{Pat}. Namias's work was complemented in 1987 by McBride and Kerr \cite{Mcb}.  The applications of Namias's work involved sol\-ving partial differential equations arising in quantum mechanics.   The first uses of fractional Fourier transformations in optics were by Khare in 1974 \cite{Kha} and by Ludwig in 1988 \cite{Lud} (independently of Namias's article). However, a decisive impulse for applying them to optics was given in 1993 by Mendlovic and Ozaktas \cite{Man,Oza1} and by Lohmann \cite{Loh}. Fractional Fourier transformations have been associated with Fresnel diffraction   since 1993-94 \cite{PPFbog,PPF1,PPF2,PPF3}, as well as with 
light propagation through optical systems 
\cite{Ali}, leading to the realm of fractional Fourier optics \cite{Oza2,PPF4}. Short after, the operational calculus associated with those transformations has also been applied to signal processing
\cite{Alm,Oza3}.

Hyperbolic fractional-order Fourier transformations were recently introduced in the expression of Fresnel diffraction, when the geometry of the underlying configuration was not adapted to usual transformations (from now on called ``circular transformations'') \cite{PPF7,PPF5}.

Fractional-order Fourier transformations do not constitute an isolated field of mathematics; they may be connected to various areas. For example, the circular transformation or order $\alpha$ can be shown to be the operator $\exp (\I\alpha{\cal H})$, where ${\cal H}$ denotes the harmonic oscillator operator \cite{Nam,PPF4}.
In this article we show that fractional-order Fourier transformations---circular as well as hyper\-bolic---are, in fact, Weyl pseudo-differential operators. Usual method of pseudo-differential calculus can then be used to establish properties of fractional Fourier transformations. We will consider circular transformations and hyperbolic transformations of the first and second kind and will give the corresponding kernels and symbols in the sense of pseudo-differential operators.

Using fractional-order Fourier transformations in optics leads to classify diffraction phenomena into three classes \cite{PPF5}. We will provide a geometrical characterization for each class, deduce some basic properties of optical imaging and prove  a general theorem stating that imaging preserves the class of a given diffraction phenomenon.

An important feature of diffraction is related to the Huygens--Fresnel principle, according to which the field transfer from an emitter ${\cal A}$ to a receiver ${\cal B}$ can be seen as the composition 
of the transfer from ${\cal A}$ to ${\cal C}$ and the transfer from ${\cal C}$ to ${\cal B}$, where ${\cal C}$ is an intermediate surface located between ${\cal A}$ and ${\cal B}$. If field transfers by diffraction are represented by fractional-order Fourier transformations, then the Huygens--Fresnel principle is expressed in terms of compositions of these transformations. For this reason, various types of compositions are emphasized in the present article; they are treated mathematically  within the framework of pseudo-differential calculus, specifically through the compositions of the corresponding symbols or kernels.

Imaging through an optical system is usually described as a ``double diffraction'' process, from the object to the entrance pupil of the objective lens and from the exit pupil to the image \cite{Mar}. We will examine the simpler and basic case of imaging through a refracting spherical cap. The field transfer from an emitter to its image is described as the composition of two transfers: from the object to the refracting cap followed by the transfer from the refracting cap to the image. On the basis of pseudo-differential calculus, we will show that if both transfers are described by fractional-order Fourier transformations, apart from trivial cases,  these transformations are necessarily of the same kind. This finding limits the number of cases to be studied in applications to image formation.

\bigskip
\bigskip
{\Large 
\centerline{\sc Part I -- Fractional Fourier transformations}
\centerline{\sc and pseudo-differential operators}\label{part1}
}

\bigskip
\bigskip
\noindent The aim of this part is to include fractional-order Fourier transformations in the framework of Weyl pseudo-differential calculus and to prove Theorem \ref{th1}, which will be applied to diffraction and imaging by a lens in the second part.

In this part, we use the conventional notation and definitions found in some basic texts on pseudo-differential operators \cite{Ego,Hor3,Unt,Nou}. One exception concerns the Fourier transformation, for which we adopt the definition  we use in Fourier optics \cite{PPF4}.

\section{Weyl calculus \cite{Ego,Hor3,Unt,Nou}}\label{sect1}

\subsection{Two-dimensional Fourier transformation}\label{sect11}

Since the mathematical concepts developped in this section will be apply to diffraction theory and to optics, we will consider complex-valued functions of a two-dimensional real variable.

If $f$ is a function belonging to the space of rapid descent functions ${\cal S}({\mathbb R}^2$), the Fourier transform of $f$, denoted $\widehat f$, is defined by
\begin{equation}
  \widehat f(\vec \xi)=\int_{{\mathbb R}^2}f(\vec x)\,\E^{2\I\pi \vec x\vec\cdot\vec \xi}\,\D\vec x\,,\label{eq1}\end{equation}
where $\vec x\vec \cdot\vec\xi$ denotes the Euclidean scalar product of two-dimensional real vectors $\vec x$ and $\vec \xi$. In con\-nec\-tion with fractional-order Fourier transformations, we also write  ${\cal F}[f]=\widehat f$, or ${\cal F}[f](\vec \xi )=\widehat f(\vec \xi )$, and ${\cal F}$ is called the standard Fourier transformation. (We will see that ${\cal F}\equiv{\cal F}_{\pi /2}$.)

By inverse Fourier transformation, we obtain
\begin{equation}
  f(\vec x)=\int_{{\mathbb R}^2}\widehat f(\vec \xi)\,\E^{-2\I\pi \vec x\vec\cdot\vec \xi}\,\D\vec \xi\,. \label{eq2}\end{equation}

The Fourier transformation is extended to tempered distributions, elements of the vector space ${\cal S}'({\mathbb R}^2)$, the topological dual of ${\cal S}({\mathbb R}^2)$: if $T\in {\cal S}'({\mathbb R}^2)$, then $\langle\widehat T,f\rangle=\langle T,\widehat f\,\rangle$.

The Dirac distribution $\delta$ is defined by $\langle \delta,\varphi\rangle =\varphi (0)$, for every test function $\varphi$. For every $f\in{\cal S}({\mathbb R}^2)$ we have
\begin{equation}
  \langle \widehat \delta ,f\rangle =  \langle \delta ,  \widehat f\,\rangle  = \widehat f (0)=\int_{{\mathbb R}^2}f(\vec x)\,\D\vec x =\langle 1, f\rangle\,,
\end{equation}
so that $\widehat \delta =1$.

We say that $f$ and $\widehat f$ form a Fourier pair, and we write $f\rightleftharpoons \widehat f$, or $f (\vec x)\rightleftharpoons \widehat f(\vec \xi )$. (Distributions $T$ and $\widehat T$ also form a Fourier pair.)

For two-dimensional variables, we shall use the following Fourier pair\footnote{For actual use, Eq.\ (\ref{eq3}) can be written:
$\exp\left(\displaystyle{\I\pi x^2\over A}\right)\;\rightleftharpoons\;\I A\,\exp (-\I\pi A\xi^2)$.}
\begin{equation}
  -{\I\over A}\exp\left({\I\pi x^2\over A}\right)\;\rightleftharpoons\;\exp (-\I\pi A\xi^2)\,,\label{eq3}
  \end{equation}
with $A\in {\mathbb R}$ and where  $x=\| \vec x\|$ is the Euclidean norm of vector $\vec x$ (and $\xi =\|\vec \xi \|$).

Since $\exp (-\I\pi A\xi^2)$ tends to 1, when $A$ tends to 0, and since $\delta \rightleftharpoons 1$, in the limit we obtain
\begin{equation}\lim_{A\rightarrow \,0}{-\I\over A}\exp\left({\I\pi x^2\over A}\right)=\delta\,.\label{eq4}\end{equation}
(In Eq.\ (\ref{eq4}), $x$ is the norm of the  two-dimensional variable $\vec x$ and the equation is valid for the two-dimensional Dirac distribution only. By comparison, $\widehat \delta =1$ holds, regardless of the dimension.)

Finally, we point out (see \ref{appenA} for notation)
\begin{equation}
  \exp (-2\I\pi \vec x\vec\cdot\vec \xi_0)\;\rightleftharpoons \;\delta_{\vec \xi_0}=\delta (\vec \xi-\vec \xi_0)\,,
  \label{eq3b} \end{equation}
and more generally
\begin{equation}
  f(\vec x)\exp (-2\I\pi \vec x\vec\cdot\vec \xi_0)\;\rightleftharpoons \;\widehat f(\vec \xi-\vec \xi_0)\,.
  \label{eq3t} \end{equation}
(The support of $\delta_{\vec \xi_0}=\delta (\vec\xi -\vec \xi_0)$ is $\{\vec \xi_0\}$. The support of $\delta$ is $\{\vec 0\}=\{ (0,0)\}$.)

\subsection{Pseudo-differential operators}\label{sect12}

\subsubsection{Definition}

Let $a$ be a function belonging to  ${\cal S}({\mathbb R}^2\!\!\times\! {\mathbb R}^2)$. We associate  an operator with $a$, denoted ${\rm Op}(a)$, defined by
\begin{equation}
  {\rm Op}(a)[f](\vec x)=\int_{{\mathbb R}^2}a(\vec x,\vec\xi )\widehat f(\vec\xi)\,\E^{-2\I\pi \vec x\vec\cdot\vec\xi}\,\D\vec\xi\,, \hskip 1cm f\in{\cal S}({\mathbb R}^2)\,.\label{eq5}\end{equation}

The operator ${\rm Op}(a)$ is called a pseudo-differential operator. The reason is that the previous definition includes partial differential operators as special cases, once it has been extended to functions $a$ belonging to ${\cal S}'({\mathbb R}^2\!\!\times\! {\mathbb R}^2)$, as will be done later (see Section \ref{sect14}). For example, if $\vec \xi =(\xi_1,\xi_2)$,  the operator associated with $a(\vec x,\vec \xi)=(-2\I\pi\xi_1)^p$ ($p\in{\mathbb N}$) is
\begin{equation}
  {\rm Op}(a)[f](\vec x)=\int_{{\mathbb R}^2}(-2\I\pi \xi_1)^p\widehat f(\vec\xi)\,\E^{-2\I\pi \vec x\vec\cdot\vec\xi}\,\D\vec\xi
  ={\partial^p\over \partial x_1^p}f(\vec x)\,,\label{eq6}\end{equation}
as deduced  from Eq.\ (\ref{eq2}).

If we substitute Eq.\ (\ref{eq1}) into Eq.\ (\ref{eq5}), we obtain
\begin{equation}
  {\rm Op}(a)[f](\vec x)=\int_{{\mathbb R}^2\times {\mathbb R}^2}\!\!\!\!\! \!\!a(\vec x,\vec\xi ) f(\vec y)\,\E^{-2\I\pi (\vec x-\vec y)\vec\cdot\vec\xi}\,\D\vec\xi\,\D\vec y\,.\label{eq7}\end{equation}

Finally, if we make the function $a$ to also depend on $\vec y$, we obtain a more general form for a pseudo-differential operator, that is
\begin{equation}
  {\rm Op}(a)[f](\vec x)=\int_{{\mathbb R}^2\times {\mathbb R}^2}\!\!\!\!\!\!\! a(\vec x,\vec y,\vec\xi ) f(\vec y)\,\E^{-2\I\pi (\vec x-\vec y)\vec\cdot\vec\xi}\,\D\vec\xi\,\D\vec y\,.\label{eq8}\end{equation}
The function $a(\vec x,\vec y,\vec \xi)$ is called the symbol of the pseudo-differential operator ${\rm Op} (a)$.
(We abusively use the same letter for $a(\vec x,\vec \xi )$ and $a(\vec x,\vec y,\vec \xi )$.)

Equation (\ref{eq8}) can also be written
\begin{equation}
  {\rm Op}(a)[f](\vec x)=\int_{{\mathbb R}^2}K(\vec x,\vec y)f(\vec y)\,\D\vec y\,,\label{eq9}\end{equation}
where
\begin{equation}
  K(\vec x,\vec y)=\int_{{\mathbb R}^2} \!\!\!
  a(\vec x,\vec y,\vec\xi )\,\E^{-2\I\pi (\vec x-\vec y)\vec\cdot\vec\xi}\,\D\vec\xi\label{eq12a}\end{equation}
is called the kernel of the pseudo-differential operator $ {\rm Op}(a)$.

\subsubsection{Composition of kernels}
Let ${\rm Op} (a)$ and ${\rm Op} (b)$ be two pseudo-differential operators with respective kernels $K_a$ and $K_b$.  Then
\begin{eqnarray}
  {\rm Op} (a)\circ {\rm Op} (b)[f](\vec x)\rap &=& \rap \int_{{\mathbb R}^2}\!\!K_a(\vec x,\vec z)\,{\rm Op} (b)[f](\vec z)\,\D\vec z \nonumber \\ \rap &=& \rap
  \int_{{\mathbb R}^2}\!\!\!K_a(\vec x,\vec z)\left(\int_{{\mathbb R}^2}\!\!\!K_b(\vec z,\vec y)\,f(\vec y)\,\D\vec y \right)\,\D\vec z\nonumber \\
  \rap & =&\rap \int_{{\mathbb R}^2}\left(\int_{{\mathbb R}^2}\!\!K_a(\vec x,\vec z)K_b(\vec z,\vec y)\,\D\vec z\right)f(\vec y)\,\D\vec y \nonumber \\ \rap & =&\rap
  \int_{{\mathbb R}^2}\!\!K(\vec x,\vec y)f(\vec y)\,\D\vec y \,,
\end{eqnarray}
where
\begin{equation}
  K(\vec x,\vec y)=\int_{{\mathbb R}^2}K_a(\vec x,\vec z)K_b(\vec z,\vec y)\,\D\vec z\label{eq22a}\end{equation}
is the kernel of the composed operator ${\rm Op} (a)\circ {\rm Op} (b)$.

\subsection{Weyl pseudo-differential operators}\label{sect13}

\subsubsection{Definition}

Weyl pseudo-differential operators constitute a subclass of pseudo-differential operators for which the symbols depend on $\vec x$ and $\vec y$ through $(\vec x+\vec y)/2$, that is
\begin{equation}
  a(\vec x,\vec y,\vec \xi )=a\left({\vec x+\vec y\over 2},\vec \xi \right)\,.\end{equation}
(Once more, we abusively use the same letter $a$ for two different functions.)

A Weyl pseudo-differential operator with symbol $a$ is denoted $a^{\rm w}$ and takes the form
\begin{eqnarray}
   a^{\rm w}[f](\vec x)={\rm Op}(a)[f](\vec x)\rap &=&\rap \int_{{\mathbb R}^2\times {\mathbb R}^2}\!\!\! a\left({\vec x+\vec y\over 2},\vec\xi \right) f(\vec y)\,\E^{-2\I\pi (\vec x-\vec y)\vec\cdot\vec\xi}\,\D\vec\xi\,\D\vec y\nonumber \\
\rap &=&\rap \int_{{\mathbb R}^2}f(\vec y) \int_{{\mathbb R}^2}  a\left({\vec x+\vec y\over 2},\vec\xi \right) \E^{-2\I\pi (\vec x-\vec y)\vec\cdot\vec\xi} \,\D\vec\xi\, \D\vec y\,,
\label{eq13}\end{eqnarray}
so that its kernel is
\begin{equation}
  K(\vec x,\vec y)=\int_{{\mathbb R}^2}  a\left({\vec x+\vec y\over 2},\vec\xi \right) \E^{-2\I\pi (\vec x-\vec y)\vec\cdot\vec\xi} \,\D\vec\xi\,,\label{eq14}
  \end{equation}
and appears to be the inverse partial-Fourier transform of $a$, with respect to  variable $\vec \xi$, and taken at point $\vec x-\vec y$. We write
\begin{equation}
  K(\vec x,\vec y)=\,\widecheck{\hskip -0.15ex a} \left({\vec x+\vec y\over 2},\vec x-\vec y\right)\,.\label{eq15}\end{equation}

For actual derivations of the corresponding symbol, the kernel $K$ can be written as follows. Let
 $\vec u=(\vec x+\vec y)/2$ and $\vec t=\vec x-\vec y$,
so that
\begin{equation}
  \vec x=\vec u+{\vec t\over 2}\,,\hskip .5cm \vec y=\vec u -{\vec t\over 2}\,.
\end{equation}
Equation (\ref{eq14}) gives
\begin{equation}
  K\left(\vec u +{\vec t\over 2},\vec u -{\vec t\over 2}\right)=\int_{{\mathbb R}^2}a(\vec u,\vec \xi )\,\E^{-2\I\pi \vec t\vec \cdot\vec\xi}\,\D\vec \xi\,,\label{eq17}\end{equation}
 which is the inverse Fourier transform of $a(\vec u,\vec \xi )$ with respect to $\vec \xi$, so that, by direct Fourier transformation, we obtain
\begin{equation}
  a (\vec u,\vec \xi )=\int_{{\mathbb R}^2} K\left(\vec u +{\vec t\over 2},\vec u -{\vec t\over 2}\right)\,\E^{2\I\pi \vec t\vec \cdot\vec\xi}\,\D\vec t 
  \,.\label{eq18}
\end{equation}
In general, we will use $\vec x$ as a variable in place of $\vec u$, and we will write Eq.\ (\ref{eq18}) as
\begin{equation}
  a (\vec x,\vec \xi )=\int_{{\mathbb R}^2} K\left(\vec x +{\vec t\over 2},\vec x -{\vec t\over 2}\right)\,\E^{2\I\pi \vec t\vec \cdot\vec\xi}\,\D\vec t
  \,.\label{eq19}\end{equation}

We observe that Eqs. (\ref{eq13}), (\ref{eq14}), (\ref{eq17}), and (\ref{eq18}) are equivalent, so that Eq.\ (\ref{eq19}) is characteristic of a Weyl pseudo-differential operator, that is, if $a(\vec x,\vec \xi)$ can be written as in Eq.\ (\ref{eq19}), then $a$ is the symbol of a Weyl pseudo-differential operator $a^{\rm w}$.

\subsubsection{Composition of symbols}
Let us consider the composition $a^{\rm w}\!\circ b^{\rm w}$ of operators $a^{\rm w}$ and $b^{\rm w}$, whose symbols are $a$ and $b$. The corresponding kernels are $K_a$ and $K_b$.
Then
\begin{eqnarray}
  a^{\rm w}\!\circ b^{\rm w}[f](\vec x)\rap &=&\rap \int_{{\mathbb R}^2}K_a(\vec x,\vec z)\left(\int_{{\mathbb R}^2}K_b(\vec z,\vec y) f(\vec y) \,\D\vec y \right)\D\vec z \nonumber \\
  \rap &=&\rap  \int_{{\mathbb R}^2\times {\mathbb R}^2}a\left({\vec x+\vec z\over 2},\vec\eta\right)\E^{-2\I\pi (\vec x-\vec z)\vec\cdot\vec\eta}\nonumber \\
   & &\hskip 1cm \times \;\;
  \left(\int_{{\mathbb R}^2\times {\mathbb R}^2}b\left({\vec z+\vec y\over 2},\vec\tau\right)\E^{-2\I\pi (\vec z-\vec y)\vec\cdot\vec\tau} f(\vec y) \,\D\vec y\,\D\vec \tau\right)\D\vec z\,\D\vec \eta \nonumber\\
   \rap &=&\rap \int_{{\mathbb R}^2} K(\vec x,\vec y) f(\vec y)\,\D\vec y\,,
\end{eqnarray}
where
\begin{equation}
   K(\vec x,\vec y)=\int_{{\mathbb R}^6} a\left({\vec x+\vec z\over 2},\vec\eta\right) b\left({\vec z+\vec y\over 2},\vec\tau\right)\E^{-2\I\pi [(\vec x-\vec z)\vec\cdot\vec\eta]+(\vec z-\vec y)\vec\cdot\vec\tau)}\D\vec z\,\D\vec\tau \,\D\vec\eta\,.
\end{equation}
The corresponding symbol, donoted $a\# b$, is given by Eq.\ (\ref{eq19}), that is
\begin{eqnarray}
  a\#b(\vec x,\vec \xi)\rap &=&\rap\int_{{\mathbb R}^2} K\left(\vec x +{\vec t\over 2},\vec x -{\vec t\over 2}\right)\,\E^{2\I\pi \vec t\vec \cdot\vec\xi}\,\D\vec t \nonumber\\
  \rap &=&\rap \int_{{\mathbb R}^8}\E^{2\I\pi \vec {t\cdot\xi}}\;a\!\left({\vec x+\vec z\over 2}+{\vec t\over 4},\vec\eta\right)\;b\!\left({\vec x+\vec z\over 2}-{\vec t\over 4},\vec\tau\right)\nonumber \\
  & & \hskip 1cm \times \;\; \exp \left[-2\I\pi\left(\vec x-\vec z+{\vec t\over 2}\right)\vec\cdot\vec\eta -2\I\pi\left(\vec z-\vec x+{\vec t\over 2}\right)\vec\cdot\vec\tau\right]\, \D\vec z\,\D\vec\tau \,\D\vec\eta\,\D\vec t
  \nonumber\\
  \rap &=&\rap \int_{{\mathbb R}^8}\E^{2\I\pi\phi}\, a\!\left({\vec x+\vec z\over 2}+{\vec t\over 4},\vec\eta\right)\;b\!\left({\vec x+\vec z\over 2}-{\vec t\over 4},\vec\tau\right)\, \D\vec z\,\D\vec\tau \,\D\vec\eta\,\D\vec t
   \,,\label{eq22}
\end{eqnarray}
where $\phi$ depends on $\vec x$, $\vec z$, $\vec t$, $\vec \xi$, $\vec \tau$  and $\vec \eta$. 
Let us set
\begin{equation} \vec z'={\vec x+\vec z\over 2}+{\vec t\over 4}\,,\hskip 1cm \vec t'={\vec x+\vec z\over 2}-{\vec t\over 4}\,,\label{eq20}\end{equation}
so that
\begin{equation}\vec t=2(\vec z'-\vec t')\,,\hskip 1cm
  \vec x-\vec z'={\vec x-\vec z\over 2}-{\vec t\over 4}\,,\hskip 1cm \vec x-\vec t'={\vec x-\vec z\over 2}+{\vec t\over 4}=\vec x-\vec z'+{\vec t\over 2}\,.
\end{equation}
We obtain
\begin{eqnarray}2\phi \rap &=&\rap 
  \vec t\vec \cdot\vec\xi -\Bigl(\vec x-\vec z+{\vec t\over 2}\Bigr)\vec\cdot\vec\eta
  -\Bigl(\vec z-\vec x+{\vec t\over 2}\Bigr)\vec\cdot\vec\tau \nonumber \\ \rap &=&\rap
  2(\vec z'-\vec t')\vec\cdot\vec\xi -2(\vec x-\vec t')\vec\cdot\vec\eta +2(\vec x-\vec z')\vec\cdot\vec\tau \nonumber \\
  \rap &=&\rap 2(\vec x-\vec z')\vec\cdot(\vec \tau-\vec\xi ) -2(\vec x-\vec t')\vec\cdot (\vec\eta -\vec\xi )\,.
\end{eqnarray}
In the integral in Eq.\ (\ref{eq22}), we make the change of variables $(\vec z,\vec t)\longmapsto (\vec z',\vec t')$, given by Eq.\ (\ref{eq20}). Since $\vec z$ and $\vec t$ are two dimensional vectors, the Jacobian is
\begin{equation}
  |J|=\left|\begin{matrix}{1/2 & 0 & 1/4 & 0\cr 0 &1/2& 0 &1/4\cr 1/2 & 0 & -1/4 & 0\cr 0 & 1/2 & 0 & -1/4}\end{matrix}\right|={1\over 2^4}\,,\end{equation}
and  $\D\vec z '\,\D \vec t'=|J|\D\vec z\,\D\vec t=2^{-4}\D\vec z\,\D\vec t$.

We eventually obtain
\begin{equation}
  a\# b(\vec x,\vec \xi)=2^4\int_{{\mathbb R}^8} \E^{4\I\pi \left[(\vec x-\vec z')\vec\cdot(\vec \tau-\vec\xi ) -(\vec x-\vec t')\vec\cdot (\vec\eta -\vec\xi )\right]}\,a( \vec z',\vec \eta )\,b(\vec t',\vec \tau )
  \,\D\vec \eta\,\D\vec \tau\,\D\vec z'\,\D\vec t'\,.\label{eq28a}
    \end{equation}

The first line of Eq.\ (\ref{eq22}) shows that $a\#b $ is the symbol of a Weyl pseudo-differential  operator, so that we may write  $a^{\rm w}\!\circ b^{\rm w} =(a\#b)^{\rm w}$.

\subsection{Symbols in  ${\cal S'}({\mathbb R}^2\times {\mathbb R}^2)$ \cite{Hor3,Nou}}\label{sect14}

 Let us consider the operator $a^{\rm w}$ whose symbol is $a$. So far, the symbol $a$ belongs to ${\cal S}({\mathbb R}^2\times {\mathbb R}^2)$ and $K$ also belongs to ${\cal S}({\mathbb R}^2\times {\mathbb R}^2)$. For every $f\in {\cal S}({\mathbb R}^2)$ and $g\in {\cal S}({\mathbb R}^2)$,  we have
 \begin{equation}
   \int_{{\mathbb R}^2}a^{\rm w}[f](\vec x)\,g(\vec x)\,\D\vec x
   =\int_{{\mathbb R}^2\times {\mathbb R}^2}\!\!\!\!\! K(\vec x,\vec y)\,f(\vec y)\,g(\vec x)\,\D \vec x\,\D\vec y\,.
   \end{equation}
 The kernel $K$ can be seen like a tempered distribution operating on ${\cal S}({\mathbb R}^2\times {\mathbb R}^2)$, that is
 \begin{equation}
   \langle K(\vec x,\vec y),f(\vec y)\, g(\vec x)\rangle =\int_{{\mathbb R}^2\times {\mathbb R}^2}K(\vec x,\vec y)\,f(\vec y)\,g(\vec x)\,\D \vec x\,\D\vec y\,.\label{eq32a}
 \end{equation}
 (For notation on distributions, see \ref{appenA}.)

 Equation (\ref{eq32a}) allows us to introduce symbols that are tempered distributions, namely elements of ${\cal S'}({\mathbb R}^2\times {\mathbb R}^2)$.  We proceed as follows.
 Let $a$ be  a tempered distribution  on ${\cal S}({\mathbb R}^2\times{\mathbb R}^2)$. We write $a(\vec x,\vec \xi)$ to indicate the variables on which $a$ operates. The inverse Fourier transform of $a(\vec x,\vec\xi )$ with respect to variable $\xi$ is  $\,\widecheck{\hskip -0.15ex a}(\vec x,\vec z)$, such that
 \begin{equation}
   \langle \,\widecheck{\hskip -0.15ex a}(\vec x,\vec z),\varphi (\vec z)\rangle =
    \langle a(\vec x,\vec \xi), \,\widecheck{\hskip -0.15ex \varphi}(\vec \xi)\rangle\,,
 \end{equation}
 where $\varphi\in{\cal S'}({\mathbb R}^2$).
We then define $K$ as a tempered distribution on ${\cal S}({\mathbb R}^2\times{\mathbb R}^2)$ by
 \begin{equation}
   K(\vec x,\vec y)= \,\widecheck{\hskip -0.15ex a} \left({\vec x+\vec y\over 2},\vec x-\vec y\right)\,.\label{eq15a}
   \end{equation}
 Equation (\ref{eq15a}) is Eq.\ (\ref{eq15}) extended to tempered distributions.

 For every $f\in{\cal S}({\mathbb R}^2)$, we define then a tempered distribution $T_f$, such that for every $g\in{\cal S}({\mathbb R}^2)$
 \begin{equation}
   \langle T_f,g\rangle =  \langle K(\vec x,\vec y),f(\vec y)\, g(\vec x)\rangle\,.\label{eq35}
 \end{equation}
 The mapping $f\longmapsto T_f$ is an operator ${\cal S}({\mathbb R}^2)\longrightarrow  {\cal S}'({\mathbb R}^2)$, denoted  $a^{\rm w}$; we write $T_f=a^{\rm w}[f]$. Equation (\ref{eq35}) becomes
 \begin{equation}
   \langle a^{\rm w}[f],g\rangle =  \langle K(\vec x,\vec y),f(\vec y) g(\vec x)\rangle\,.
 \end{equation}
 The operator $a^{\rm w}$ is the Weyl pseudo-differential operator associated with the symbol $a$ (which is a tempered distribution).

 In particular $a$ may be a polynomial.

 \begin{remark}\label{rem11}{\rm If $a \in {\cal S}'({\mathbb R}^2\times{\mathbb R}^2)$, the method used in  this section to define the Weyl operator $a^{\rm w}$
     can be applied to define a (more general)  pseudo-differential operator ${\rm Op}(a)$.
   }
   \end{remark}

\subsection{The parity operator}\label{sect15}
\subsubsection{Definition}
Let $\widetilde f$ denote  the function $f$ symmetrized, defined by $\widetilde f(\vec x)= f(-\vec x)$. The parity operator ${\cal P}$ is defined by ${\cal P}\; :\; f\longmapsto \widetilde f$, that is, ${\cal P}[f](\vec x)=\widetilde f(\vec x)=f(-\vec x)$.

\subsubsection{Kernel and symbol}\label{sect152}
We proceed  by comparison with the identity operator ${\cal I}$, whose symbole is $a_{\cal I}(\vec x,\vec\xi)= 1$. We check
\begin{equation}
  {a_{\cal I}}^{\rm w}[f](\vec x)=\int_{{\mathbb R}^2}f(\vec y)\int_{{\mathbb R}^2}\E^{-2\I\pi (\vec x-\vec y)\vec\cdot\vec \xi}\,\D\vec\xi\,\D\vec y =\int_{{\mathbb R}^2}f(\vec y)\,\delta (\vec x-\vec y)\,\D\vec y =f(\vec x)\,,\label{eq37}
\end{equation}
so that the kernel of ${\cal I}$ is ${\cal K}_{\cal I}(\vec x,\vec y)=\delta (\vec x-\vec y)$. (See \ref{appenA} for notation.)

From Eq.\ (\ref{eq37}) we deduce
\begin{equation}
  f(-\vec x)
  =\int_{{\mathbb R}^2}f(\vec y)\,\delta (\vec x+\vec y)\,\D\vec y\,,\end{equation}
so that the kernel of the operator ${\cal P}$ is $K_{\cal P}(\vec x,\vec y)= \delta (\vec x+\vec y)$. Then
\begin{equation}
K_{\cal P}\left(\vec u+{\vec t\over 2}, \vec u-{\vec t\over 2}\right)=\delta (2\vec u)\,,\end{equation}
and from Eq.\ (\ref{eq18})
we deduce that the symbol of the parity operator is
\begin{equation}
  {a_{\cal P}}(\vec u,\vec \xi)= \delta (2\vec u)\int_{{\mathbb R}^2}\E^{2\I\pi \vec t\vec\cdot\vec \xi}\,\D\vec t = {1\over 4}\delta (\vec u)\,\delta (\vec \xi) =
  {1\over 4}\delta (\vec u)\otimes\delta(\vec \xi)
  \label{eq40}\,,
\end{equation}
that is, ${a_{\cal P}}(\vec x,\vec \xi )= (1/4)\,\delta (\vec x)\otimes\delta (\vec \xi)$.

We check
\begin{eqnarray}
  {a_{\cal P}}^{\rm w}[f](\vec x)\rap & = &\rap {1\over 4}\int_{{\mathbb R}^2}f(\vec y)\,\delta\left({\vec x+\vec y\over 2}\right)\left(\int_{{\mathbb R}^2}\delta (\vec \xi)
  \,\E^{-2\I\pi (\vec x-\vec y)\vec\cdot\vec \xi}\,\D\vec\xi\right)\,\D\vec y \nonumber \\
  \rap & = &\rap \int_{{\mathbb R}^2}f(\vec y)\,\delta (\vec x+\vec y)
  \,\D\vec y
  = f(-\vec x)={\cal P}[f](\vec x)\,.\label{eq41}
\end{eqnarray}

In conclusion, the parity operator ${\cal P}$ is a Weyl pseudo-differential operator whose  kernel is $K(\vec x,\vec y)=\delta (\vec x+\vec y)$ and  symbol is $a_{\cal P}(\vec x,\vec \xi )=(1/4)\,\delta (\vec x)\,\delta (\vec \xi)$.

\section{Fractional-order Fourier transformations as Weyl pseudo-differential operators}\label{sect2}

\subsection{Three classes of fractional-order Fourier transformations}\label{sect21}

Three classes of fractional-order Fourier transformations have been introduced to represent diffraction phenomena\cite{PPF5}:
\begin{enumerate}
\item Circular transformations, denoted ${\cal F}_\alpha$ and  defined as follows, for $f\in {\cal S}({\mathbb R}^2)$ (adapted from Namias \cite{Nam})
  \begin{equation}
    {\cal F}_\alpha[f](\vec x)={\I\E^{-\I\alpha}\over \sin\alpha}\,\E^{-\I\pi x^2\cot\alpha}\int_{{\mathbb R}^2}\E^{-\I\pi y^2\cot\alpha}\,\exp\left({2\I\pi\over\sin\alpha}\vec x\vec\cdot\vec y\right)\,f(\vec y )\,\D\vec y\,,
 \label{eq26} \end{equation}
  where the order $\alpha$ is a real number ($-\pi <\alpha <\pi$). (The definition can been extended to every real $\alpha$ (see Sect.\ \ref{sect225}), and also to complex orders.)
\item Hyperbolic transformations of the first kind, denoted ${\cal H}_\beta$ ($\beta \in {\mathbb R}$) and  defined as follows, for $f\in {\cal S}({\mathbb R}^2)$
   \begin{equation}
    {\cal H}_\beta[f](\vec x)={\I\E^{\beta}\over \sinh\beta}\,\E^{-\I\pi x^2\coth\beta}\int_{{\mathbb R}^2}\E^{-\I\pi y^2\coth\beta}\,\exp\left({2\I\pi\over\sinh\beta}\vec x\vec\cdot\vec y\right)\,f(\vec y)\,\D\vec y\,.
   \label{eq27}\end{equation}
   \item Hyperbolic transformations of the second kind, denoted ${\cal K}_\beta$ ($\beta \in {\mathbb R}$) and  defined as follows, for $f\in {\cal S}({\mathbb R}^2)$
   \begin{equation}
    {\cal K}_\beta[f](\vec x)={\I\E^{\beta}\over \cosh\beta}\,\E^{\I\pi x^2\tanh\beta}\int_{{\mathbb R}^2}\E^{-\I\pi y^2\tanh\beta}\,\exp\left({2\I\pi\over\cosh\beta}\vec x\vec\cdot\vec y\right)\,f(\vec y )\,\D\vec y\,.
  \label{eq28}\end{equation}
\end{enumerate}

We will show that each transformation above is a Weyl pseudo-differential operator, and we will provide its kernel and its symbol.

\subsection{Circular fractional-order Fourier transformations}\label{sect22}

\subsubsection{Kernel}
According to Eq.\ (\ref{eq26}), for $\alpha \in ]-\pi, \pi [$, the kernel of ${\cal F}_\alpha$ is
\begin{equation}
  K(\vec x,\vec y )=C_\alpha\,\E^{-\I\pi x^2\cot\alpha}\E^{-\I\pi y^2\cot\alpha}\exp\left({2\I\pi\over \sin\alpha}\vec x\vec\cdot\vec y\right)\,,\label{eq34}
\end{equation}
where
\begin{equation}
  C_\alpha={\I\E^{-\I\alpha}\over \sin\alpha}\,.\label{eq46a}\end{equation}

\subsubsection{Symbol}
\begin{proposition}\label{prop1} {When $\alpha\in ]-\pi ,\pi [$, the circular fractional-order Fourier transformation ${\cal F}_\alpha$ is a Weyl pseudo-differential operator whose symbol is
\begin{equation}
  a(\vec x,\vec \xi )= \E^{-\I\alpha}\left(1+\tan^2{\alpha\over 2}\right)\exp \left(2\I\pi x^2\tan{\alpha \over 2}\right)\exp \left(2\I\pi \xi^2\tan{\alpha \over 2}\right)\,.\label{eq36}
\end{equation}
}
\end{proposition}

\noindent{\its Proof.} We first write the kernel in the form 
\begin{eqnarray}
  K\left(\vec x+{\vec t\over 2},\vec x-{\vec t\over 2}\right)\rap &=&\rap C_\alpha
  \exp\left(-\I\pi \Bigl\|\vec x+{\vec t\over 2}\Bigr\|^2\cot\alpha\right) \exp\left(-\I\pi \Bigl\|\vec x-{\vec t\over 2}\Bigr\|^2\cot\alpha\right)\nonumber \\
  & & \hskip 3cm \times \;\;\exp\left[{2\I\pi\over \sin\alpha}\left(\vec x+{\vec t\over 2}\right)\vec\cdot \left(\vec x-{\vec t\over 2}\right)\right]\nonumber \\
  \rap &=&\rap C_\alpha \exp (-2\I\pi x^2\cot\alpha )\exp \left(-\I\pi {t^2\over 2}\cot\alpha\right)
  \exp\left({2\I\pi x^2\over \sin\alpha}\right)\exp\left({-\I\pi t^2\over 2\sin\alpha}\right) \nonumber \\
  \rap &=&\rap C_\alpha\exp\left[-2\I\pi x^2\left(\cot\alpha-{1\over \sin\alpha}\right)\right]
  \exp\left[-\I\pi {t^2\over 2}\left(\cot\alpha+{1\over \sin\alpha}\right)\right]\,.
  \end{eqnarray}
We use
\begin{equation}
  \cot\alpha -{1\over \sin\alpha}={\cos\alpha -1\over \sin\alpha}=-{2\tan^2(\alpha /2)\over 2\tan (\alpha /2)}=-\tan{\alpha \over 2}\,,\end{equation}
and
\begin{equation}
  \cot\alpha +{1\over \sin\alpha}={\cos\alpha +1\over \sin\alpha}={1\over \tan(\alpha /2)}=\cot{\alpha \over 2}\,,\end{equation}
and we obtain
\begin{equation}
  K\left(\vec x+{\vec t\over 2},\vec x-{\vec t\over 2}\right)=C_\alpha\exp \left(2\I\pi x^2\tan{\alpha \over 2}\right)\exp \left(-{\I\pi t^2\over 2}\cot{\alpha \over 2}\right)\,.
\end{equation}
According to Eq.\ (\ref{eq19}), the symbol is the Fourier transform of $K$, considered as a function of the two-dimensional variable $\vec t$. We apply Eq.\ (\ref{eq3}) with $A=-2/\cot (\alpha /2)$ and we obtain
\begin{equation}
  \exp \left(-{\I\pi t^2\over 2}\cot{\alpha \over 2}\right)\;\rightleftharpoons\; {-2\I  \tan{\alpha \over  2}}\exp\left(2\I\pi \xi^2\tan{\alpha\over 2}\right)\,.
\end{equation}
Since
\begin{equation}
  -2\I C_\alpha\tan {\alpha \over 2}={2\, \E^{-\I\alpha}\over \sin\alpha}\tan{\alpha \over  2}=\E^{-\I\alpha}\left(1+\tan^2{\alpha\over 2}\right)\,,
\end{equation}
we obtain
\begin{equation}
  a(\vec x,\vec \xi )= \E^{-\I\alpha}\left(1+\tan^2{\alpha\over 2}\right)\exp \left(2\I\pi x^2\tan{\alpha \over 2}\right)\exp \left(2\I\pi \xi^2\tan{\alpha \over 2}\right)\,,
\end{equation}
which is Eq.\ (\ref{eq36}).
We conclude with ${\cal F}_\alpha = a^{\rm w}$. \qed

\subsubsection{Calculation of ${\cal F}_0$}\label{sect223}
When $\alpha =0$, we obtain $a(\vec x,\vec \xi )=1$, so that, according to Eq.\ (\ref{eq13}), for every $f\in{\cal S}({\mathbb R}^2)$
\begin{eqnarray}
  {\cal F}_0[f](\vec x)=a^{\rm w}[f](\vec x)\rap &=&\rap \int_{{\mathbb R}^2\times{\mathbb R}^2}\!\!\!\!\!\E^{-2\I\pi (\vec x-\vec y)\vec\cdot\vec\xi}\,f(\vec y)\,\D\vec y\,\D\vec \xi\nonumber\\
  \rap &=&\rap \int_{{\mathbb R}^2}\E^{-2\I\pi \vec x\vec\cdot\vec\xi}\left(\int_{{\mathbb R}^2}\E^{2\I\pi\vec y\vec\cdot\vec\xi}\,f(\vec y)\,\D\vec y\right)\D\vec\xi\nonumber \\
  \rap &=&\rap  \int_{{\mathbb R}^2}\E^{-2\I\pi \vec x\vec\cdot\vec\xi} \widehat f(\vec \xi )\,\D\vec \xi =f(\vec x)\,.
\end{eqnarray}
(Compared to the proof in Sect.\ \ref{sect152}, the derivation above avoids improper integrals, but fails to make the kernel explicit.)

We conclude that ${\cal F}_0={\cal I}$ (identity operator) on ${\cal S}({\mathbb R}^2)$. If $T$ is a tempered distribution, then 
$\langle{\cal F}_0[T],f\rangle  =\langle T,{\cal F}_0[f]\rangle =  \langle T,f\rangle$, and ${\cal F_0}={\cal I}$ (identity operator on  ${\cal S}'({\mathbb R}^2)$).

\subsubsection{Calculation of ${\cal F}_{\pi /2}$}\label{sect224}
For $\alpha =\pi /2$, Eq.\ (\ref{eq26}) gives ${\cal F}_{\pi /2}={\cal F}$ (standard Fourier transformation). For an illustration, let us confirm the result by using the corresponding symbol.

For $\alpha =\pi /2$, we obtain from Eq.\ (\ref{eq36})
\begin{equation}
  a(\vec x,\vec \xi )=-2\I\,\E^{2\I\pi (x^2+\xi^2)}\,,\end{equation}
and then, by Eq.\ (\ref{eq13})
\begin{eqnarray}
  a^{\rm w}[f](\vec x)\rap &=& \rap 
  -2\I\int_{{\mathbb R}^2}\!\!f(\vec y) \left(\int_{{\mathbb R}^2} \!\! a\left({\vec x+\vec y\over 2},\vec\xi \right) \E^{-2\I\pi (\vec x-\vec y)\vec\cdot\vec\xi} \,\D\vec\xi\right) \D\vec y\nonumber \\
  \rap &=&\rap -2\I\int_{{\mathbb R}^2}\!\!f(\vec y)\exp\Bigl({\I\pi\over 2}\left\| {\vec x+\vec y}\right\|^2\Bigr)
   \left(\int_{{\mathbb R}^2}\!\!\! \E^{-2\I\pi (\vec x-\vec y)\vec\cdot\vec\xi}\E^{2\I\pi \xi^2}\,\D\vec \xi\right)\,\D\vec y\,.
   \label{eq59}\end{eqnarray}
From $\exp (2\I\pi\xi^2)\rightleftharpoons \displaystyle{\I\over 2}\,\exp (-\I\pi \eta^2/2)$, we deduce
\begin{equation}
  \int_{{\mathbb R}^2} \E^{-2\I\pi (\vec x-\vec y)\vec\cdot\vec\xi}\,\E^{2\I\pi \xi^2}\,\D\vec \xi ={\I\over 2}\exp \Bigl(-{\I\pi\over 2}\| \vec x-\vec y\|^2 \Bigr)\,,
  \end{equation}
and
\begin{eqnarray}
  {\cal F}_{\pi /2}[f](\vec x)=a^{\rm w}[f](\vec x)\rap &=&\rap \int_{{\mathbb R}^2}
  f(\vec y)\exp \Bigl({\I\pi\over 2}\| \vec x+\vec y\|^2\Bigr) \exp \Bigl(-{\I\pi\over 2}\| \vec x-\vec y\|^2 \Bigr)\,\D\vec y \nonumber \\
  \rap &=&\rap  \int_{{\mathbb R}^2} f(\vec y)\,\E^{2\I\pi \vec x\vec\cdot\vec y}\,\D\vec y \nonumber \\
  \rap &=&\rap
  \widehat f(\vec x)
       ={\cal F}[f](\vec x)\,.
\end{eqnarray}
The operator ${\cal F}_{\pi /2}$ is the standard Fourier transformation.

\subsubsection{Calculation of ${\cal F}_{\pi}$. Extension of the domain of $\alpha$}\label{sect225}

We obtain the symbol $a_{\pi}$ of  ${\cal F}_\pi$ by  making $\alpha$ ($\alpha >0$) tend to $\pi$ in Eq.\ (\ref{eq36}). According to
Eq.\ (\ref{eq4}), we obtain
\begin{equation}
  \lim_{\alpha\rightarrow\pi_-}\left[{-\I \sqrt{1+\tan^2{\alpha \over 2}}}\,\exp \left(2\I\pi\| \vec x\|^2\tan{\alpha\over 2}\right)\right]={1\over 2}\,\delta (\vec x)\,,
\end{equation}
and the same result is valid if we replace $\vec x$ with $\vec \xi$. In the limit, we obtain
\begin{equation}
  a_{\pi}(\vec x,\vec \xi )={1\over 4}\,\delta (\vec x)\,\delta (\vec \xi )={a_{\cal P}}(\vec x, \vec \xi)\,,\end{equation}
where $a_{\cal P}$ is the symbol of the parity operator (see Sect.\ \ref{sect152}). We  conclude that ${\cal F}_\pi={\cal P}$. 

By Eq.\ (\ref{eq26}), we obtain ${\cal F}_{-\pi} ={\cal F}_\pi$, so that ${\cal F}_\alpha$ is now defined for $\alpha \in [-\pi, \pi]$. Eventually we remark that for every $\alpha \in [-\pi ,\pi]$  and for every $n\in {\mathbb Z}$, Eq.\ (\ref{eq26}) gives ${\cal F}_{\alpha +2n\pi }={\cal F}_\alpha$, so that ${\cal F}_\alpha$ can be defined for every real $\alpha$.

\subsubsection{Composition of two circular fractional-order Fourier transformations}\label{sect324}

\begin{proposition}\label{prop2} For every $\alpha_1$ and $\alpha_2$ in $]-\pi , \pi [$, the product  ${\cal F}_{\alpha_2}\circ{\cal F}_{\alpha_1}$ is commutative and $ {\cal F}_{\alpha_2}\circ{\cal F}_{\alpha_1}={\cal F}_{\alpha_1+\alpha_2}$.
\end{proposition}

\goodbreak

\noindent{\its Proof.}

\smallskip

\noindent
    {\its i.} We assume first $\alpha_1+\alpha_2\ne \pm\pi$. 
  We use the symbol composition law, given by Eq.\ (\ref{eq28a}). For $j=1,2$, let $a_j$ be the symbol of ${\cal F}_{\alpha_j}$,  that is,
${\cal F}_{\alpha_j}=a^{\rm w}_j$, with 
\begin{equation}
  a_j(\vec x,\vec \xi)=\E^{-\I\alpha_j} (1+\kappa_j^2)\,\exp (2\I\pi \kappa_jx^2)\,\exp (2\I\pi \kappa_j\xi^2)\,,
  \end{equation}
where $\kappa_j=\tan (\alpha_j/2)$. Then ${\cal F}_{\alpha_2}\circ{\cal F}_{\alpha_1}=(a_2\#a_1)^{\rm w}$.

According to Eq.\ (\ref{eq28a}) we have
\begin{eqnarray}
  (a_2\# a_1)(\vec x,\vec \xi)\rap &=&\rap 2^4\int_{{\mathbb R}^8}\!\!\!\exp[4\I\pi (\vec x-\vec z')\vec\cdot (\vec \tau -\vec\xi )]\,\exp
  [-4\I\pi (\vec x-\vec t')\vec\cdot (\vec \eta -\vec\xi)]\nonumber \\
  & & \hskip 5cm \times \;\;a_2(\vec z',\vec\eta )\,a_1 (\vec t',\vec \tau)\,\D\vec \eta\,\D\vec\tau\,\D\vec z'\,\D \vec t'\nonumber \\
  \rap &=&\rap 2^4\int_{{\mathbb R}^4} \!\!\!
  \exp (4\I\pi \vec z'\vec\cdot\vec \xi )\,\exp (-4\I\pi\vec x\vec\cdot\vec\eta )\,a_2(\vec z',\vec\eta )
  \biggl(\int_{{\mathbb R}^4}\exp \bigl[4\I\pi (\vec x-\vec z')\vec\cdot\vec\tau \bigr]
  \nonumber \\
  & & \hskip 2cm \times \;\;\exp \bigl[-4\I\pi \vec t'\vec\cdot (\vec \eta -\vec \xi )\bigr]\,a_1 (\vec t',\vec \tau)\, \D\vec \tau\, \D \vec t'\biggr)\,\D\vec \eta\,\D \vec z'\,.\label{eq43}
\end{eqnarray}
Let us denote
\begin{eqnarray}
  I_1(\vec z',\vec \eta )\rap &=&\rap\int_{{\mathbb R}^4}\exp \bigl[4\I\pi (\vec x-\vec z')\vec\cdot\vec\tau \bigr] \,\exp \bigl[-4\I\pi \vec t'\vec\cdot (\vec \eta -\vec \xi )\bigr]\,a_1 (\vec t',\vec \tau)\, \D\vec \tau\, \D \vec t'\nonumber \\ \label{eq44}
  \rap &=&\rap\E^{-\I\alpha_1}(1+\kappa_1^2) \int_{{\mathbb R}^2}\exp \bigl[4\I\pi (\vec x-\vec z')\vec\cdot\vec\tau \bigr]\,\exp (2\I\pi \kappa_1\,\tau^2 )\,\D\vec \tau\nonumber \\
 \rap & &\rap \hskip 3cm \times \;\; \int_{{\mathbb R}^2}\exp \bigl[-4\I\pi \vec t'\vec\cdot (\vec \eta -\vec \xi )\bigr]\,\exp (2\I\pi \kappa_1\,t'^2 )\,\D\vec t'\,.\label{eq45}
\end{eqnarray}
We apply Eq.\ (\ref{eq3}) and obtain
\begin{equation}
  I_1(\vec z',\vec \eta )= \E^{-\I\alpha_1}(1+\kappa_1^2) \left({\I\over 2\kappa_1}\right)^2\,\exp\left(-{2\I\pi\over \kappa_1}\| \vec x-\vec z'\|^2\right)\,\exp\left(-{2\I\pi\over \kappa_1}\| \vec \xi-\vec \eta\|^2\right)\,.\label{eq46}
\end{equation}
Then
\begin{eqnarray}
  (a_2\# a_1)(\vec x,\vec \xi)\rap &=&\rap  2^4\int_{{\mathbb R}^4} \!\!\!
  \exp (4\I\pi \vec z'\vec\cdot\vec \xi )\,\exp (-4\I\pi\vec x\vec\cdot\vec\eta )\,a_2(\vec z',\vec\eta )\, I_1 (\vec z',\vec \eta )\,\D\vec z'\,\D \vec\eta \nonumber \\ 
  \rap &=&\rap  -{2^4\over 4\kappa_1^2}{\E^{-\I (\alpha_1+\alpha_2) }(1+\kappa_1^2)(1+\kappa_2^2)}\nonumber \\
  & & \hskip .5cm \times \;\;\int_{{\mathbb R}^2} \!\!\!
  \exp (4\I\pi \vec z'\vec\cdot\vec \xi ) \,\exp (2\I\pi \kappa_2z'^2)\,\exp\left(-{2\I\pi\over \kappa_1}\| \vec x-\vec z'\|^2\right)\,\D \vec z'\nonumber \\
  & & \hskip .5cm \times \;\;\int_{{\mathbb R}^2} \!\!\!\exp (-4\I\pi\vec x\vec\cdot\vec\eta ) \,\exp (2\I\pi \kappa_2\eta^2)\exp\left(-{2\I\pi\over \kappa_1}\| \vec \xi-\vec \eta\|^2\right)\,\D\vec \eta \nonumber \\
  \rap &=&\rap  -{2^4\over 4\kappa_1^2}{\E^{-\I (\alpha_1+\alpha_2) }(1+\kappa_1^2)(1+\kappa_2^2)}\nonumber \\
   & & \hskip .5cm \times \;\;\int_{{\mathbb R}^2} \!\!\!
  \exp \left[2\I\pi \left(2\vec z'\vec\cdot\vec \xi +{2\over \kappa_1}\vec z'\vec\cdot \vec x\right) \right] \,\exp \left(-{2\I\pi x^2\over \kappa_1}\right)\nonumber \\
  & &\hskip 6cm \times \;\;\exp\left[2\I\pi z'^2\left(\kappa_2-{1\over \kappa_1}\right)\right]\,\D\vec z'\nonumber \\
  & & \hskip .5cm \times \;\;\int_{{\mathbb R}^2} \!\!\!\exp \left[2\I\pi\left(-2\vec x\vec\cdot\vec\eta +{2\over \kappa_1}\vec\eta\vec\cdot\vec\xi \right)\right] \,\exp \left(-{2\I\pi \xi^2\over \kappa_1}\right)\nonumber \\
  & & \hskip 6cm \times \;\; \exp\left[2\I\pi \eta^2\left(\kappa_2-{1\over \kappa_1}\right)\right]\,\D\vec \eta\nonumber \\
  \rap &=&\rap  -{2^4\over 4\kappa_1^2}{\E^{-\I (\alpha_1+\alpha_2) }(1+\kappa_1^2)(1+\kappa_2^2)}\nonumber \\
  & &\hskip .5cm \times \;\; {\I \kappa_1\over 2(\kappa_1\kappa_2-1)}\,\exp\biggl({-2\I\pi \kappa_1\over \kappa_1\kappa_2-1}\Bigl\|{\vec x\over \kappa_1}+\vec \xi\Bigr\|^2\biggr)\,\exp\left(-{2\I\pi x^2\over \kappa_1}\right)\nonumber \\
  & &\hskip .5cm \times \;\; {\I \kappa_1\over 2(\kappa_1\kappa_2-1)}\,\exp\biggl({-2\I\pi \kappa_1\over \kappa_1\kappa_2-1}\Bigl\| -\vec x+{\vec\xi\over \kappa_1}\Big\|^2\bigg)\,\exp\left(-{2\I\pi \xi^2\over \kappa_1}\right)\nonumber \\ 
  \rap &=&\rap \E^{-\I (\alpha_1+\alpha_2)}{(1+\kappa_1^2)(1+\kappa_2^2)\over (\kappa_1\kappa_2-1)^2}
  \,\exp\left[-{2\I\pi \kappa_1\over \kappa_1\kappa_2-1}\left(1+{1\over \kappa_1^2 }\right) (x^2+\xi^2)\right]\nonumber \\
  & & \hskip 6cm \times \;\;\exp\left[{-2\I\pi\over \kappa_1} (x^2+\xi^2)\right]\nonumber \\
   \rap &=&\rap \E^{-\I (\alpha_1+\alpha_2)}{(1+\kappa_1^2)(1+\kappa_2^2)\over (\kappa_1\kappa_2-1)^2}\,\exp\left[-2\I\pi (x^2+\xi^2){\kappa_1+\kappa_2\over \kappa_1\kappa_2-1}\right]\,.\label{eq48}
\end{eqnarray}
We have
\begin{equation}
  {\kappa_1+\kappa_2\over \kappa_1\kappa_2-1}={\tan \displaystyle{\alpha_1\over 2}+\tan{\alpha_2\over 2}\over \tan \displaystyle{\alpha_1\over 2}\tan{\alpha_2\over 2}-1}=-\tan {\alpha_1+\alpha_2\over 2}\,,\end{equation}
and
\begin{equation}
  {(1+\kappa_1^2)(1+\kappa_2^2)\over (\kappa_1\kappa_2-1)^2}
   =1+{(\kappa_1+\kappa_2)^2\over (\kappa_1\kappa_2-1)^2}
  =1+\tan^2{\alpha_1+\alpha_2\over 2}\,.
\end{equation}

Eventually, we obtain
\begin{equation}
  (a_2\# a_1)(\vec x,\vec \xi)=\E^{-\I(\alpha_1+\alpha_2)}\left(1+\tan^2{\alpha_1+\alpha_2\over 2}\right)
  \,\exp \left[2\I\pi (x^2+\xi^2)\tan{\alpha_1+\alpha_2\over 2}\right]\,,\label{eq81}
\end{equation}
that is
\begin{equation}
{\cal F}_{\alpha_2}\circ{\cal F}_{\alpha_1}=(a_2\# a_1)^{\rm w}={\cal F}_{\alpha_1+\alpha_2}\,.\end{equation}
We have $(a_1\# a_2)=(a_2\# a_1)$, so that ${\cal F}_{\alpha_1}\circ{\cal F}_{\alpha_2}={\cal F}_{\alpha_2}\circ{\cal F}_{\alpha_1}$ (commutativity).

\medskip
\noindent{\its ii.} We assume $\alpha_1+\alpha _2=\pi$. According to Eq.\ (\ref{eq26}) we have
\begin{eqnarray}
  {\cal F}_{\alpha_2}[f](\vec x)\rap & = &\rap {\cal F}_{\pi - \alpha_1}[f](x)\nonumber \\
  \rap &= &\rap {-\I\E^{\I\alpha_1}\over \sin\alpha_1}\,\E^{\I\pi x^2\cot\alpha_1}\int_{{\mathbb R}^2}\E^{\I\pi y^2\cot\alpha_1}\,\exp\left({2\I\pi\over\sin\alpha_1}\vec x\vec\cdot\vec y\right)\,f(\vec y )\,\D\vec y \nonumber \\
  \rap &= &\rap {\cal P}\circ {\cal F}_{-\alpha_1}[f](\vec x)\,.
  \end{eqnarray}
Then ${\cal F}_{\alpha_2}\circ{\cal F}_{\alpha_1}={\cal P}\circ {\cal F}_{-\alpha_1}\circ{\cal F}_{\alpha_1}$. Since $-\alpha_1+\alpha_1=0\ne \pm\pi$, and since ${\cal F}_0={\cal I}$, the first part of the proof gives  ${\cal F}_{\alpha_2}\circ{\cal F}_{\alpha_1}={\cal P}\circ{\cal F}_0={\cal P}={\cal F}_\pi={\cal F}_{\alpha_2+\alpha_1}$.

\medskip
\noindent{\its iii.} The same methods gives ${\cal F}_{\alpha_2}\circ{\cal F}_{\alpha_1}={\cal F}_{-\pi}={\cal F}_{\alpha_2+\alpha_1}$, if $\alpha_1+\alpha_2 =-\pi$. 

\medskip
\noindent The proof is complete. \qed

\begin{remark}\label{rem0} {\rm The result of item {\its ii} of the previous proof can also be obtained by applying the method of Sect.\ \ref{sect225} to Eq.\ (\ref{eq81}): $\alpha_1$ is fixed and $\alpha_2$ tends to $\pi-\alpha_1$.
  }
\end{remark}

Proposition \ref{prop2} can be extended to every real $\alpha_1$ and $\alpha_2$ as shown by the following proposition.

\setcounter{proposition}{1}
\begin{proposition}\label{prop2prime}$\!\!'$ For every $\alpha_1$ and every $\alpha_2$ in ${\mathbb R}$, we have
  ${\cal F}_{\alpha_2}\circ{\cal F}_{\alpha_1}={\cal F}_{\alpha_2+\alpha_1}$, and the product is commutative.
\end{proposition}

\noindent{\its Proof.}

\medskip \noindent {\its i.} Since ${\cal F}_\pi ={\cal P}$ (see Sect.\ \ref{sect225}), for $\alpha_1=\pi$ and $\alpha_2\in ]-\pi, \pi[$, according to Eq.\ (\ref{eq26}), we have ${\cal F}_{\alpha_2}\circ {\cal F}_\pi={\cal F}_{\alpha_2}\circ {\cal P}={\cal F}_{\alpha_2+\pi}={\cal P}\circ{\cal F}_{\alpha_2}$. The same result holds  for $\alpha_1=-\pi$ and $\alpha_2\in ]-\pi, \pi[$. 

    \medskip \noindent {\its ii.}  If $\alpha_1=\alpha_2=\pi$, we have ${\cal F}_{\alpha_2}\circ{\cal F}_{\alpha_1}={\cal P}\circ{\cal P}={\cal I}={\cal F}_{2\pi}={\cal F}_{\alpha_2+\alpha_1}$.  The same methods  applies to $\alpha_1=\alpha_2=-\pi$, and to $\alpha_1=-\alpha_2=\pi$. In each case, we obtain ${\cal F}_{\alpha_2}\circ{\cal F}_{\alpha_1}={\cal F}_{\alpha_2+\alpha_1}$.

    We conclude that Proposition \ref{prop2} holds when $\alpha_1$ and $\alpha_2$ belong to $[-\pi, \pi]$.

 \medskip \noindent{\its iii.} The transformations ${\cal F}_\alpha$ and ${\cal F}_{\alpha +2n\pi}$ ($n\in{\mathbb Z}$) have the same symbol and are then identical: ${\cal F}_\alpha ={\cal F}_{\alpha +2n\pi}$. For every $\alpha_j$ in ${\mathbb R}$ ($j=1,2$), there exists an integer $n_j$ such that $\alpha_j=\alpha'_j+2n_j\pi$, with $\alpha'_j\in [-\pi ,\pi ]$. Then
 \begin{equation}
   {\cal F}_{\alpha_2}\circ{\cal F}_{\alpha_1}= {\cal F}_{\alpha'_2}\circ{\cal F}_{\alpha'_1}=
   {\cal F}_{\alpha'_2+\alpha'_1}={\cal F}_{\alpha'_2+\alpha'_1+2(n_2+n_1)\pi}= {\cal F}_{\alpha_2+\alpha_1}\,,
   \end{equation}
and the product is commutative.

\smallskip
The proof is complete. \qed

\subsection{Hyperbolic fractional-order Fourier transformations of the first kind}\label{sect23}

\subsubsection{Kernel}
According to Eq.\ (\ref{eq27}), the kernel of ${\cal H}_\beta$ is
\begin{equation}
  K(\vec x,\vec y )=C_\beta\,\E^{-\I\pi x^2\coth\beta}\E^{-\I\pi y^2\coth\beta}\exp\left({2\I\pi\over \sinh\beta}\,\vec x\vec\cdot\vec y\right)\,,\label{eq55}
\end{equation}
where
\begin{equation}
  C_\beta={\I\E^{\beta}\over \sinh\beta}\,.\label{eq74}\end{equation}

\subsubsection{Symbol}
\begin{proposition}\label{prop3} {The hyperbolic fractional-order Fourier transformation of the first kind ${\cal H}_\beta$ is a Weyl pseudo-differential operator whose symbol is
\begin{equation}
  a(\vec x,\vec \xi )= \E^{\beta}\left(1-\tanh^2{\beta\over 2}\right)\exp \left(-2\I\pi x^2\tanh{\beta \over 2}\right)\exp \left(2\I\pi \xi^2\tanh{\beta \over 2}\right)\,.\label{eq59a}
\end{equation}
}
\end{proposition}

\noindent{\its Proof.} We write the kernel in the form
\begin{eqnarray}
  K\left(\vec x+{\vec t\over 2},\vec x-{\vec t\over 2}\right)\rap &=&\rap C_\beta
  \exp\left(-\I\pi \Bigl\|\vec x+{\vec t\over 2}\Bigr\|^2\coth\beta\right) \exp\left(-\I\pi \Bigl\|\vec x-{\vec t\over 2}\Bigr\|^2\coth\beta\right)\nonumber \\
  & & \hskip 3cm \times \;\;\exp\left[{2\I\pi\over \sinh\beta}\left(\vec x+{\vec t\over 2}\right)\vec\cdot \left(\vec x-{\vec t\over 2}\right)\right]\nonumber \\
  \rap &=&\rap C_\beta \exp (-2\I\pi x^2\coth\beta )\exp \left(-\I\pi {t^2\over 2}\coth\beta\right)\nonumber \\
  & &\hskip 3cm \times \;\;
  \exp\left({2\I\pi x^2\over \sinh\beta}\right)\exp\left(-{\I\pi t^2\over 2\sinh\beta}\right) \\
  \rap &=&\rap C_\beta\exp\left[-2\I\pi x^2\left(\coth\beta-{1\over \sinh\beta}\right)\right]
  \exp\left[-\I\pi {t^2\over 2}\left(\coth\beta+{1\over \sinh\beta}\right)\right]\,.\nonumber
  \end{eqnarray}
We use
\begin{equation}
  \coth\beta -{1\over \sinh\beta}={\cosh\beta -1\over \sinh\beta}={2\tanh^2(\beta /2)\over 2\tanh (\beta /2)}=\tanh{\beta \over 2}\,,\end{equation}
and
\begin{equation}
  \coth\beta +{1\over \sinh\beta}={\cosh\beta +1\over \sinh\beta}={1\over \tanh(\beta /2)}=\coth{\beta \over 2}\,,\end{equation}
and we obtain
\begin{equation}
  K\left(\vec x+{\vec t\over 2},\vec x-{\vec t\over 2}\right)=C_\beta\exp \left(-2\I\pi x^2\tanh{\beta \over 2}\right)\exp \left(-{\I\pi t^2\over 2}\coth{\beta \over 2}\right)\,.
\end{equation}
We use Eq.\ (\ref{eq19}) and apply the Fourier transformation  to $K$, considered as a function of the two-dimensional variable $\vec t$. We also use Eq.\ (\ref{eq3}) with  $A=-2/\coth (\beta /2)$, and we obtain
\begin{equation}
  \exp \left(-{\I\pi t^2\over 2}\coth{\beta \over 2}\right)\;\rightleftharpoons\; -2\I  \tanh{\beta\over  2}\exp\left(2\I\pi \xi^2\tanh{\beta\over 2}\right)\,.
\end{equation}
Since
\begin{equation}
  -2\I C_\beta \tanh {\beta \over 2}={2\, \E^{\beta}\over \sinh\beta}\tanh{\beta \over 2}=\E^{\beta}\left(1-\tanh^2{\beta\over 2}\right)\,,
\end{equation}
we obtain
\begin{equation}
  a(\vec x,\vec \xi )= \E^{\beta}\left(1-\tanh^2{\beta\over 2}\right)\exp \left(-2\I\pi x^2\tanh{\beta \over 2}\right)\exp \left(2\I\pi \xi^2\tanh{\beta \over 2}\right)\,,\label{eq61}
\end{equation}
which is Eq.\ (\ref{eq59a}). \qed

\subsubsection{Calculation of ${\cal H}_0$}
When $\beta =0$, we obtain $a(\vec x,\vec \xi )=1$, so that for every $f\in{\cal S}({\mathbb R}^2)$ we have $a^{\rm w}[f](\vec x)=f(\vec x)$, as in Sect.\ \ref{sect223}, and we conclude that ${\cal H}_0={\cal I}$.

\subsubsection{Composition of two hyperbolic fractional-order Fourier transformations of the first kind}

\begin{proposition}\label{prop4} For every $\beta_1$ and $\beta_2$ in ${\mathbb R}$, the product  ${\cal H}_{\beta_2}\circ{\cal H}_{\beta_1}$ is commutative and
  \begin{equation}
    {\cal H}_{\beta_2}\circ{\cal H}_{\beta_1}={\cal H}_{\beta_1+\beta_2}\,.\label{eq67}
  \end{equation}
  \end{proposition}

\noindent{\its Proof.}
We use the same method as in Set.\ \ref{sect324}, but for symbols $a_1$ and $a_2$ given by Eq.\ (\ref{eq61}), that is, for $j=1,2$,
\begin{equation}
  a_j(\vec x,\xi)=\E^{\beta_j} (1-\kappa_j^2)\,\exp (-2\I\pi \kappa_jx^2)\,\exp (2\I\pi \kappa_j\xi^2)\,,
  \end{equation}
where $\kappa_j=\tanh (\beta_j/2)$. Then ${\cal H}_{\beta_2}\circ{\cal H}_{\beta_1}=(a_2\#a_1)^{\rm w}$.

We remark that Eq.\ (\ref{eq43}) still holds formally, so that we introduce
\begin{eqnarray}
  I(\vec z',\vec \eta )\rap &=&\rap\int_{{\mathbb R}^4}\exp \bigl[4\I\pi (\vec x-\vec z')\vec\cdot\vec\tau \bigr] \,\exp \bigl[-4\I\pi \vec t'\vec\cdot (\vec \eta -\vec \xi )\bigr]\,a_1 (\vec t',\vec \tau)\, \D\vec \tau\, \D \vec t'\nonumber \\
 \rap &=&\rap\E^{\beta_1}(1-\kappa_1^2) \int_{{\mathbb R}^4}\exp \bigl[4\I\pi (\vec x-\vec z')\vec\cdot\vec\tau \bigr]\,\exp (2\I\pi \kappa_1\,\tau^2 )\,\D\vec \tau \nonumber \\
 \rap & &\rap \hskip 3cm \times \;\; \int_{{\mathbb R}^4}\exp \bigl[-4\I\pi \vec t'\vec\cdot (\vec \eta -\vec \xi )\bigr]\,\exp (-2\I\pi \kappa_1\,t'^2 )\,\D\vec t'\nonumber \\
\rap &=&\rap\E^{\beta_1}(1-\kappa_1^2) \left({1\over 2\kappa_1}\right)^2\,\exp\left(-{2\I\pi\over \kappa_1}\| \vec x-\vec z'\|^2\right)\,\exp\left({2\I\pi\over \kappa_1}\| \vec \xi-\vec \eta\|^2\right)\,.\label{eq63}
\end{eqnarray}

Then
\begin{eqnarray}
  (a_2\# a_1)(\vec x,\vec \xi)\rap &=&\rap 2^4\int_{{\mathbb R}^4} \!\!\!
  \exp (4\I\pi \vec z'\vec\cdot\vec \xi )\,\exp (-4\I\pi\vec x\vec\cdot\vec\eta )\,a_2(\vec z',\vec\eta )\, I (\vec z',\vec \eta )\,\D\vec z'\,\D \vec\eta \nonumber \\
\rap &=&\rap  {2^4\over 4\kappa_1^2}{\E^{\beta_1+\beta_2 }(1-\kappa_1^2)(1-\kappa_2^2)}\nonumber \\
  & & \hskip .5cm \times \;\;\int_{{\mathbb R}^2} \!\!\!
  \exp (4\I\pi \vec z'\vec\cdot\vec \xi ) \,\exp (-2\I\pi \kappa_2z'^2)\,\exp\left(-{2\I\pi\over \kappa_1}\| \vec x-\vec z'\|^2\right)\,\D \vec z'\nonumber \\
  & & \hskip .5cm \times \;\;\int_{{\mathbb R}^2} \!\!\!\exp (-4\I\pi\vec x\vec\cdot\vec\eta ) \,\exp (2\I\pi \kappa_2\eta^2)\exp\left({2\I\pi\over \kappa_1}\| \vec \xi-\vec \eta\|^2\right)\,\D\vec \eta \nonumber \\
\rap &=&\rap {2^4\over 4\kappa_1^2}{\E^{\beta_1+\beta_2 }(1-\kappa_1^2)(1-\kappa_2^2)}\nonumber \\
   & & \hskip .5cm \times \;\;\int_{{\mathbb R}^2} \!\!\!
  \exp \left[2\I\pi \left(2\vec z'\vec\cdot\vec \xi +{2\over \kappa_1}\vec z'\vec\cdot \vec x\right) \right] \,\exp \left(-{2\I\pi x^2\over \kappa_1}\right)\nonumber \\
  & &\hskip 6cm \times \;\;\exp\left[-2\I\pi z'^2\left(\kappa_2+{1\over \kappa_1}\right)\right]\,\D\vec z'\nonumber \\
  & & \hskip .5cm \times \;\;\int_{{\mathbb R}^2} \!\!\!\exp \left[-2\I\pi\left(2 \vec x\vec\cdot\vec\eta +{2\over \kappa_1}\vec\xi\vec\cdot\vec\eta \right)\right] \,\exp \left({2\I\pi \xi^2\over \kappa_1}\right)\nonumber \\
  & & \hskip 6cm \times \;\; \exp\left[2\I\pi \eta^2\left(\kappa_2+{1\over \kappa_1}\right)\right]\,\D\vec \eta \nonumber  \\
  \rap &=&\rap  {2^4\over 4\kappa_1^2}{\E^{\beta_1+\beta_2 }(1-\kappa_1^2)(1-\kappa_2^2)}\nonumber \\
   & &\hskip .5cm \times \;\; {-\I \kappa_1\over 2(\kappa_1\kappa_2+1)}\,\exp\biggl({2\I\pi \kappa_1\over \kappa_1\kappa_2+1}\Bigl\|{\vec x\over \kappa_1}+\vec \xi\Bigr\|^2\biggr)\,\exp\left(-{2\I\pi x^2\over \kappa_1}\right)\nonumber \\
  & &\hskip .5cm \times \;\; {\I \kappa_1\over 2(\kappa_1\kappa_2+1)}\,\exp\biggl({-2\I\pi \kappa_1\over \kappa_1\kappa_2+1}\Bigl\|\vec x+{\vec \xi\over \kappa_1}\Bigr\|^2\biggr)\,\exp\left({2\I\pi \xi^2\over \kappa_1}\right)\nonumber \\
  \rap &=&\rap \E^{\beta_1+\beta_2}{(1-\kappa_1^2)(1-\kappa_2^2)\over (\kappa_1\kappa_2+1)^2}
  \,\exp\left[{2\I\pi \kappa_1\over \kappa_1\kappa_2+1}\left({1\over \kappa_1^2 }-1\right) (x^2-\xi^2)\right]\nonumber \\
  & & \hskip 6cm \times \;\;\exp\left[{-2\I\pi\over \kappa_1} (x^2-\xi^2)\right]\nonumber \\
   \rap &=&\rap \E^{\beta_1+\beta_2}{(1-\kappa_1^2)(1-\kappa_2^2)\over (\kappa_1\kappa_2+1)^2}\,\exp\left[-2\I\pi (x^2-\xi^2){\kappa_1+\kappa_2\over \kappa_1\kappa_2+1}\right]\,.
\end{eqnarray}
We have
\begin{equation}
  {\kappa_1+\kappa_2\over \kappa_1\kappa_2+1}={\tanh \displaystyle{\beta_1\over 2}+\tanh{\beta_2\over 2}\over \tanh \displaystyle{\beta_1\over 2}\tanh{\beta_2\over 2}+1}=\tanh {\beta_1+\beta_2\over 2}\,,\end{equation}
and
\begin{equation}
  {(1-\kappa_1^2)(1-\kappa_2^2)\over (\kappa_1\kappa_2+1)^2}
   =1-{(\kappa_1+\kappa_2)^2\over (\kappa_1\kappa_2+1)^2}
  =1-\tanh^2{\beta_1+\beta_2\over 2}\,.
\end{equation}

Eventually, we obtain
\begin{equation}
  (a_2\# a_1)(\vec x,\vec \xi)=\E^{\beta_1+\beta_2}\left(1-\tanh^2{\beta_1+\beta_2\over 2}\right)
  \,\exp \left[-2\I\pi (x^2-\xi^2)\tanh{\beta_1+\beta_2\over 2}\right]\,,
\end{equation}
that is
\begin{equation}
  {\cal H}_{\beta_2}\circ{\cal H}_{\beta_1}=(a_2\# a_1)^{\rm w}={\cal H}_{\beta_1+\beta_2}\,,\end{equation}
which is Eq. (\ref{eq67}). 
The proof is complete. \qed

\subsection{Hyperbolic fractional-order Fourier transformations of the second kind}\label{sect24}

\subsubsection{Kernel}
According to Eq.\ (\ref{eq28}), the kernel of ${\cal K}_\beta$ is 
\begin{equation}
  K(\vec x,\vec y )=C_\beta\,\E^{\I\pi x^2\tanh\beta}\E^{-\I\pi y^2\tanh\beta}\exp\left({2\I\pi\over \cosh\beta}\,\vec x\vec\cdot\vec y\right)\,,\label{eq71}
\end{equation}
where
\begin{equation}
  C_\beta={\I\E^{\beta}\over \cosh\beta}\,.\label{eq92}\end{equation}

\subsubsection{Symbol}

\begin{proposition}\label{prop5}
{The hyperbolic fractional-order Fourier transformation of the second kind ${\cal K}_\beta$ is a Weyl pseudo-differential operator whose symbol is
\begin{equation}
  a(\vec x,\vec \xi )= 2\,\E^{\beta}\,\exp (2\I\pi x^2\cosh \beta )\exp (2\I\pi \xi^2\cosh \beta )\exp (4\I\pi \vec x\vec\cdot\vec\xi\,\sinh \beta )\,.\label{eq77}
\end{equation}
}
\end{proposition}

  \noindent{\its Proof.} We write the kernel of ${\cal K}_\beta$ in the form
\begin{eqnarray}
  K\left(\vec x+{\vec t\over 2},\vec x-{\vec t\over 2}\right)\rap &=&\rap C_\beta
  \exp\left(\I\pi \Bigl\|\vec x+{\vec t\over 2}\Bigr\|^2\tanh\beta\right) \exp\left(-\I\pi \Bigl\|\vec x-{\vec t\over 2}\Bigr\|^2\tanh\beta\right)\nonumber \\
  & & \hskip 3cm \times \;\;\exp\left[{2\I\pi\over \cosh\alpha}\left(\vec x+{\vec t\over 2}\right)\vec\cdot \left(\vec x-{\vec t\over 2}\right)\right]\nonumber \\
  \rap &=&\rap C_\beta \exp (2\I\pi \,\vec x\vec\cdot\vec t\tanh\beta )
  \exp\left({2\I\pi x^2\over \cosh\beta}\right)\exp\left(-{\I\pi t^2\over 2\cosh\beta}\right)\,. 
  \end{eqnarray}
We then apply the Fourier transformation  to $K$, considered as a function of the two-dimensional variable $\vec t$. First we use
\begin{equation}
  \exp \left(-{\I\pi t^2\over 2\cosh\beta }\right)\;\rightleftharpoons\; -{2\I  \cosh \beta}\exp (2\I\pi \xi^2\cosh \beta )\,.
\end{equation}
Let $g$ be defined by
\begin{equation}
  g(\vec x,\vec t)=   \exp (2\I\pi \,\vec x\vec\cdot\vec t\tanh\beta )\,\exp\left(-{\I\pi t^2\over 2\cosh\beta}\right)\,,
\end{equation}
so that  $g(\vec x,\vec t)\rightleftharpoons \widehat g(\vec x,\vec \xi)$ with
\begin{equation}
  \widehat g(\vec x,\vec \xi)= -{2\,\I  \cosh \beta}\exp \bigl(2\I\pi \|\vec \xi +\vec x\tanh\beta \|^2\cosh \beta \bigr)\,,
\end{equation}
in accordance with Eq.\ (\ref{eq3t}).

Eventually, the symbol of ${\cal K}_\beta$, considered as a Weyl pseudo-differential operator, is
\begin{eqnarray}
  a(\vec x,\vec \xi )\rap & =& \rap C_\beta \, \exp\left({2\I\pi x^2\over \cosh\beta}\right)\, \widehat g(\vec x,\vec \xi)\nonumber \\
  \rap &=&\rap 2\,\E^{\beta}\exp \left({2\I\pi x^2\over \cosh\beta}\right) \,\exp \bigl(2\I\pi \|\vec\xi +\vec x\tanh\beta \|^2\cosh \beta \bigr)\nonumber \\
  \rap &=&\rap 2\,\E^{\beta}\,\exp (2\I\pi x^2\cosh\beta ) \exp (2\I\pi \xi^2\cosh\beta )
  \,\exp(4\I\pi \vec x\vec\cdot\vec\xi \sinh\beta )\,,
\end{eqnarray}
which is Eq.\ (\ref{eq77}). \qed

\subsubsection{Calculation of ${\cal K}_0$}
When $\beta =0$, Eq.\ (\ref{eq28}) gives ${\cal K_0}[f](\vec x)=\I\widehat{f}(\vec x)$, that is, ${\cal K}_0=\I{\cal F}$. We check that the symbol of ${\cal K}_0$ is
\begin{equation}
  a(\vec x,\vec \xi )=2\,\E^{2\I\pi (x^2+\xi^2)}\,,\end{equation}
which leads to ${\cal K_0}=\I{\cal F}_{\pi /2}=\I{\cal F}$, according to the derivation developed in Sect.\ \ref{sect224}.

\subsubsection{Composition of two hyperbolic fractional-order Fourier transformations of the second kind}

\begin{proposition}\label{prop6} The product of two hyperbolic fractional-order Fourier transformations of the se\-cond order is given by
   \begin{equation} {\cal K}_{\beta_2}\circ{\cal K}_{\beta_1}=-\E^{2\beta_2}\,{\cal H}_{\beta_1-\beta_2}\circ{\cal P}\,,
  \label{eq87} \end{equation}
   where ${\cal P}$ is the parity operator. The product ${\cal K}_{\beta_2}\circ{\cal K}_{\beta_1}$ is not commutative.
\end{proposition}

\noindent{\its Proof.}
We will derive the result from the composition of kernels. We consider two fractional-order Fourier transformations of the second kind ${\cal K}_{\beta_1}$ and ${\cal K}_{\beta_2}$ with respective symbols $a_1$ and $a_2$ and kernels $K_1$ and $K_2$. The kernel of ${\cal K}_{\beta_2}\circ{\cal K}_{\beta_1}$ is $K$, 
and according to Eq.\ (\ref{eq22a}) we have, with $C_{\beta_j}=\I\E^{\beta_j}/\cosh\beta_j$ ($j=1,2$),
\goodbreak
\begin{eqnarray}
  K(\vec x,\vec y)\rap &=&\rap\int_{{\mathbb R}^2}K_2(\vec x,\vec z)\,K_1(\vec z,\vec y)\,\D\vec z\nonumber \\
  \rap &=&\rap C_{\beta_1} C_{\beta_2}\int_{{\mathbb R}^2}\E^{\I\pi x^2\tanh\beta_2}\E^{-\I\pi z^2\tanh\beta_2}\exp\left({2\I\pi\over \cosh\beta_2}\,\vec x\vec\cdot\vec z\right)\nonumber \\
  & &\hskip 3cm \times \;\;\E^{\I\pi z^2\tanh\beta_1}\E^{-\I\pi y^2\tanh\beta_1}\exp\left({2\I\pi\over \cosh\beta_1}\,\vec z\vec\cdot\vec y\right)\,\D\vec z\nonumber \\
  \rap &=&\rap C_{\beta_1} C_{\beta_2}\E^{\I\pi x^2\tanh\beta_2}\E^{-\I\pi y^2\tanh\beta_1}\nonumber \\
  & &\hskip 1cm \times \;\;
  \int_{{\mathbb R}^2}\E^{\I\pi z^2(\tanh\beta_1-\tanh\beta_2)}\exp\left[2\I\pi \vec z\vec\cdot \left({\vec x\over\cosh\beta_2}+{\vec y\over \cosh\beta_1}\right)\right]\,\D\vec z\,. \label{eq83}
\end{eqnarray}

\noindent {\its i.} When $\beta_1\ne \beta_2$, we have
\begin{eqnarray}
  K(\vec x,\vec y)
  \rap &=&\rap
  {\I C_{\beta_1} C_{\beta_2}\E^{\I\pi x^2\tanh\beta_2}\E^{-\I\pi y^2\tanh\beta_1}\over \tanh\beta_1-\tanh\beta_2}\,\exp \left[{-\I\pi \over \tanh\beta_1-\tanh\beta_2}\left\|{\vec x\over\cosh\beta_2}+{\vec y\over \cosh\beta_1}\right\|^2\right]
  \nonumber \\
  \rap &=&\rap {\I\,\E^{\beta_1+\beta_2}\over  \cosh\beta_2\sinh\beta_1-\sinh\beta_2\cosh\beta_1}\exp (\I\pi Lx^2)\exp(\I\pi M y^2) \exp(\I\pi N\vec x\vec\cdot\vec y)
   \nonumber \\
   \rap &=&\rap {\I\,\E^{\beta_2+\beta_1}\over \sinh(\beta_1-\beta_2)}\exp (\I\pi Lx^2)\exp(\I\pi M y^2) \exp(\I\pi N\vec x\vec\cdot\vec y)\,,\label{eq104}
\end{eqnarray}
with
\begin{eqnarray}
  L\rap &=&\rap \tanh\beta_2-{1\over (\tanh\beta_1-\tanh\beta_2)\cosh^2\beta_2}={\sinh\beta_2\cosh\beta_2\tanh\beta_1-\sinh^2\beta_2 -1\over
    (\tanh\beta_1-\tanh\beta_2)\cosh^2\beta_2}\nonumber \\
  \rap &=&\rap
 { \sinh\beta_2\tanh\beta_1-\cosh\beta_2 \over
   (\tanh\beta_1-\tanh\beta_2)\cosh\beta_2}={\tanh\beta_2\tanh\beta_1-1\over\tanh\beta_1-\tanh\beta_2}=-{1\over \tanh (\beta_1-\beta_2)} \nonumber \\
  \rap &=&\rap -\coth(\beta_1-\beta_2)\,,
\end{eqnarray}
\begin{equation}
  M= -\tanh\beta_1+{1\over (\tanh\beta_2-\tanh\beta_1)\cosh^2\beta_1}=-\coth(\beta_1-\beta_2)\,,
\end{equation}
\begin{eqnarray}
  N={-2\over (\tanh\beta_1-\tanh\beta_2)\cosh\beta_2\cosh\beta_1}\rap &=&\rap{-2\over \cosh\beta_2\sinh\beta_1-\sinh\beta_2\cosh\beta_1}\nonumber \\
  \rap & =&\rap {-2\over \sinh(\beta_1-\beta_2)}\,.
\end{eqnarray}
We obtain
\begin{equation}
  K(\vec x,\vec y)={\I\,\E^{\beta_1+\beta_2}\over \sinh(\beta_1-\beta_2)}\,\exp [-\I\pi(x^2+y^2)\coth (\beta_1-\beta_2)] \,\exp \left({-2\I\pi\over \sinh (\beta_1-\beta_2)}  \vec x\vec\cdot\vec y\right)\,.\label{eq108}
\end{equation}

Let $K_{12}$ be the kernel of the fractional transformation ${\cal H}_{\beta_1-\beta_2}$. If we compare Eq.\ (\ref{eq55}), written for $K_{12}$, with Eq.\  (\ref{eq108}), we obtain
$K(\vec x,\vec y)=\E^{2\beta_2}K_{12}(\vec x,-\vec y)$, and then
\begin{eqnarray}
  {\cal K}_{\beta_2}\circ{\cal K}_{\beta_1} [f](\vec x)=\E^{2\beta_2}\int_{{\mathbb R}^2}K_{12}(\vec x,-\vec y)\,f(\vec y)\,\D\vec y\rap &=&\rap -\E^{2\beta_2}\int_{{\mathbb R}^2}K_{12}(\vec x,\vec y)\,f(-\vec y)\,\D\vec y\nonumber \\
  \rap &=&\rap -\E^{2\beta_2}\int_{{\mathbb R}^2}K_{12}(\vec x,\vec y)\,\widetilde f(\vec y)\,\D\vec y\nonumber \\
   \rap &=&\rap -\E^{2\beta_2}{\cal H}_{\beta_1-\beta_2}\circ{\cal P}[f](\vec x)\,,
\end{eqnarray}
which is Eq.\ (\ref{eq87}).

The product  $ {\cal K}_{\beta_2}\circ{\cal K}_{\beta_1}$ is not commutative, because $\E^{2\beta_1}\,{\cal H}_{\beta_1-\beta_2}\ne \E^{2\beta_2}\,{\cal H}_{\beta_2-\beta_1}$.

\medskip

\noindent {\its ii.} When $\beta_1=\beta_2=\beta$, Eq.\ (\ref{eq83}) gives
\begin{eqnarray}
   K(\vec x,\vec y )\rap &=&\rap  {-\E^{2\beta}\over\cosh^2\beta}\E^{\I\pi x^2\tanh\beta}\E^{-\I\pi y^2\tanh\beta}
   \int_{{\mathbb R}^2}\exp\left[2\I\pi \vec z\vec\cdot \left({\vec x+\vec y\over\cosh\beta}\right)\right]\,\D\vec z \nonumber \\
\rap & = &\rap  -\E^{2\beta}\,\delta (\vec x+\vec y)\,,
\end{eqnarray}
and we conclude that ${\cal K}_\beta\circ{\cal K}_\beta=-\E^{2\beta}\,{\cal P}$ (see Sect.\ \ref{sect152}). Since ${\cal H}_0={\cal I}$, we may also write  ${\cal K}_\beta\circ{\cal K}_\beta=-\E^{2\beta}\,{\cal H}_0\circ{\cal P}$, so that Eq.\ (\ref{eq87}) holds when $\beta_1=\beta_2=\beta$.

\medskip
The proof is complete. \qed

\begin{remark}\label{rem1} {\rm When $\beta =0$, by Eq.\ (\ref{eq87}) we obtain ${\cal K}_0\circ{\cal K}_0=-{\cal P}$. From  ${\cal K}_0=\I{\cal F}_{\pi/2}$ we also obtain
    \begin{equation}
      {\cal K}_0\circ{\cal K}_0[f]=-{\cal F}_{\pi/2}\circ{\cal F}_{\pi/2} [f]=-\,\,\widehat{\!\!\widehat f}  =-\widetilde f=-{\cal P}[f]\,.
      \end{equation}
    }
\end{remark}

\begin{remark}{\rm The result given in item {\its ii} of the previous proof can be checked as follows. When $\beta_1=\beta_2=\beta$, we derive from Eq.\ (\ref{eq83})
\begin{eqnarray}
 {\cal K}_\beta\circ{\cal K}_\beta [f](\vec x)\rap &=&\rap \int_{{\mathbb R}^2} K(\vec x,\vec y )\,f(\vec y)\,\D \vec y\nonumber \\
    \rap &=&\rap {-\E^{2\beta}\over\cosh^2\beta}\E^{\I\pi x^2\tanh\beta}\int_{{\mathbb R}^2}\!\!\!\E^{-\I\pi y^2\tanh\beta}
  \,f(\vec y)\left\{\int_{{\mathbb R}^2}\!\!\!\exp\left[2\I\pi \vec z\vec\cdot \left({\vec x+\vec y\over\cosh\beta}\right)\right]\,\D\vec z\right\}\D\vec y\nonumber \\
  \rap &=&\rap {-\E^{2\beta}\over\cosh^2\beta}\E^{\I\pi x^2\tanh\beta}\int_{{\mathbb R}^2}\!\!\!\E^{-\I\pi y^2\tanh\beta}\, f(\vec y)\,\delta\left({\vec x+\vec y\over \cosh\beta}\right)\,\D\vec y\nonumber \\
  \ \rap &=&\rap -\E^{2\beta}\E^{\I\pi x^2\tanh\beta}\int_{{\mathbb R}^2}\!\!\!\E^{-\I\pi y^2\tanh\beta}\,f(\vec y)\,\delta (\vec x+\vec y)\,\D\vec y \nonumber \\
  \rap &=&\rap -\E^{2\beta}\,f (-\vec x)= -\E^{2\beta}\widetilde f(\vec x)= -\E^{2\beta}\,{\cal P}[f] (\vec x)= -\E^{2\beta}\,{\cal H}_0\circ{\cal P}[f] (\vec x)\,.
\end{eqnarray}

  }
  \end{remark}

\subsubsection{Symbol of the composed operator}
We now provide the symbol of ${\cal K}_{\beta_2}\circ{\cal K}_{\beta_1}$, denoted $a_{\rm c}$. If $\beta_1\ne \beta_2$, we write
\begin{equation}
  C={\I\,\E^{\beta_1+\beta_2}\over \sinh (\beta_1-\beta_2)} 
  \,,\end{equation}
and, since $L=M$, Eq.\ (\ref{eq104}) gives
\begin{eqnarray}
  K\left(\vec x +{\vec t\over 2},\vec x-{\vec t\over 2}\right)\rap &=&\rap  C\exp \left(\I\pi L \left\|\vec x +{\vec t\over 2}\right\|^2\right)\exp \left(\I\pi L \left\|\vec x -{\vec t\over 2}\right\|^2\right)\nonumber \\
  & &\hskip 4cm \exp\left[\I\pi N\left(\vec x +{\vec t\over 2}\right)\vec \cdot \left(\vec x-{\vec t\over 2}\right)\right]\nonumber \\
  \rap &=&\rap  C\exp (2\I\pi Lx^2)\,\exp(\I\pi Lt^2/2)\exp (\I\pi Nx^2)\exp (-\I\pi Nt^2/4)\nonumber \\
     \rap &=&\rap C\exp [\I\pi (2L+N)x^2]\,\exp[\I\pi (2L-N)t^2/4]\,.
\end{eqnarray}
Equation (\ref{eq19}) gives
\begin{equation}
  a_{\rm c}(\vec x,\vec \xi )= {4\,\I\,C\over 2L-N} \exp [\I\pi (2L+N)x^2]\exp \left(-{4\I\pi \xi^2\over 2L-N}\right)\,.
  \end{equation}
We have
\begin{eqnarray}
 2L+N=-2\coth (\beta_1-\beta_2)-{2\over \sinh (\beta_1-\beta_2)}\rap &=&\rap -2\,{\cosh (\beta_1-\beta_2)+1\over \sinh (\beta_1-\beta_2)}\nonumber \\ \rap &=&\rap-2\,\coth {\beta_1-\beta_2\over 2}\,,
\end{eqnarray}
\begin{eqnarray}
  2L-N=-2\coth (\beta_1-\beta_2)+{2\over \sinh (\beta_1-\beta_2)}\rap &=&\rap -2\,{\cosh (\beta_1-\beta_2)-1\over \sinh (\beta_1-\beta_2)}\nonumber \\ \rap &=&\rap 2\tanh {\beta_1-\beta_2\over 2}\,,
\end{eqnarray}
\begin{eqnarray}
  {4\,\I\,C\over 2L-N}={-2\,\E^{\beta_1+\beta_2}\over \sinh(\beta_1-\beta_2)\tanh \displaystyle{\beta_1-\beta_2\over 2}}
  \rap & =&\rap \E^{\beta_1+\beta_2}\left(1-\coth^2{\beta_1-\beta_2\over 2}\right)\nonumber \\
  \rap &=&\rap {-\E^{\beta_1+\beta_2}\over \sinh^2\displaystyle{\beta_1-\beta_2\over 2}}\,.
  \end{eqnarray}
Eventually, we obtain
\begin{equation}
  a_{\rm c}(\vec x,\vec \xi)={-\E^{\beta_1+\beta_2}\over \sinh^2\displaystyle{\beta_1-\beta_2\over 2}}\,\exp \left(-2\I\pi x^2\coth{\beta_1-\beta_2\over 2}\right)\,\exp \left(2\I\pi \xi^2\coth{\beta_1-\beta_2\over 2}\right)\,. \label{eq118n}
\end{equation}

For obtaining the symbol when $\beta_1=\beta_2=\beta$, we start from Equation (\ref{eq118n})---which holds for $\beta_1\ne\beta_2$---and  proceed as follows. Let $A=(1/2)\tanh [(\beta_1-\beta_2)/2]$, so that Eq.\ (\ref{eq118n}) becomes
\begin{equation}
  a_{\rm c}(\vec x,\vec \xi)={-\E^{\beta_1+\beta_2}\over 4 \cosh^2\displaystyle{\beta_1-\beta _2\over 2}}\,{1\over A^2}\,\exp \left(-{\I\pi x^2\over A}\right)\,\exp \left({\I\pi \xi^2\over A}\right)\,. 
\end{equation}
Then $A$ tends to $0$, if both $\beta_1$ and $\beta_2$ tend to $\beta$.
By Eq.\ (\ref{eq4}) we obtain
\begin{equation}
  \doublelimite{\beta_1\rightarrow\beta}{\beta_2\rightarrow\beta} {1\over A}\,\exp\left(-{\I\pi x^2\over A}\right)=-\I\delta_{\vec x}\,,\hskip .5cm\mbox{and}\;\;\;\doublelimite{\beta_1\rightarrow\beta}{\beta_2\rightarrow\beta} {1\over A}\,\exp\left({\I\pi \xi^2\over A}\right)=\I\delta_{\vec \xi}\,,
\end{equation}
and thus
\begin{equation}
  \doublelimite{\beta_1\rightarrow\beta}{\beta_2\rightarrow\beta}a_{\rm c}(\vec x,\vec \xi) =-{1\over 4}\,\E^{2\beta}\,\delta_{\vec x}\otimes\delta_{\vec \xi}= -{1\over 4}\,\E^{2\beta}\,\delta_{\vec x}\,\delta_{\vec \xi}
  =-\E^{2\beta}\,a_{\cal P}(\vec x,\vec\xi )\,,\end{equation}
where $a_{\cal P}$ is the symbol of the parity operator (see Sect.\ \ref{sect152}).

\subsection{Concluding remark}\label{sect25}
In his thesis, Harsoyo describes the (circular) fractional Fourier transformation of order $\alpha$ as a Weyl pseudo-differential operator \cite{Har}, and he provides the corresponding kernel and symbol. In the pre\-vious sections, we have extended his work to hyperbolic transformations. (We will develop explicit applications to optics in the second part of this article.)

Some recents publications refer to both (circular) fractional-order Fourier transformations and pseudo-differential operators \cite{Pra,Das}. 
    As far as we understand, in these articles,  the link between fractional Fourier transformations  and pseudo-differential operators differs significantly from that used in our present work. Whereas we show that fractional-order Fourier transformations are Weyl pseudo-differential operators, the mentioned articles introduce the notion of a fractional pseudo-differential operator by replacing the Fourier transform in the definition of a pseudo-differential operator with a (circular) fractional-order Fourier transform. More explicitly Eq.\ (\ref{eq5}) is replaced with
    \begin{equation}
      {\rm Op}_\alpha[a](\vec x)=\int_{{\mathbb R}^2}a(\vec x,\vec \xi )\,\overline{K_{\alpha}(\vec x,\vec \xi )}\,{\cal F}_\alpha[f](\vec \xi )\,\D\vec \xi\,,\end{equation}
    where $K_{\alpha}$ is the kernel of ${\cal F}_{\-\alpha}$. (We note that $\overline{K_{\alpha}(\vec x,\vec \xi )}$ is the kernel of ${\cal F}_{-\alpha}$.) The operator ${\rm Op}_\alpha[a]$ is called a ``fractional pseudo-differential'' operator \cite{Pra,Das}.

\section{Compositions of fractional-order transformations \\ of different kinds}\label{sect3}

\subsection{Circular transformations and hyperbolic transformations of the first kind}\label{sect31}

\subsubsection{Some specific cases}\label{sect311}

Before dealing with the general case, we examine some specific cases.  Since ${\cal H}_0={\cal I}$, for every $\alpha\in{\mathbb R}$, we have ${\cal F}_\alpha\circ{\cal H}_0={\cal H_0}\circ{\cal F}_\alpha ={\cal F}_\alpha$. Since ${\cal F}_{2n\pi}={\cal F}_0={\cal I}$, for every $n\in{\mathbb Z}$, we have ${\cal F}_{2n\pi}\circ{\cal H}_\beta={\cal H}_\beta\circ{\cal F}_{2n\pi}={\cal H}_\beta$, for every $\beta\in{\mathbb R}$. In particular ${\cal F}_0\circ{\cal H}_0={\cal I}$. Since ${\cal F}_{(2n+1)\pi}={\cal P}$, we have  ${\cal F}_{(2n+1)\pi}\circ{\cal H}_0 ={\cal P}$.

\subsubsection{Product ${\cal F}_\alpha\circ{\cal H}_\beta$ : kernel (for $\alpha\ne n\pi$ and $\beta\ne 0$)}\label{sect312}
We consider the circular transformation ${\cal F}_\alpha$ ($\alpha\ne n\pi$, $n\in{\mathbb Z}$) with symbol $a$ and kernel $K_a$, and the hyperbolic transformation of the first kind ${\cal H}_\beta$  ($\beta\ne 0$), with symbol $b$ and kernel $K_b$. 

The kernel of the operator ${\cal F}_\alpha\circ{\cal H}_\beta$ is
\begin{equation}
  K_{\cal {FH}}(\vec x,\vec y)=\int_{{\mathbb R}^2} K_a(\vec x,\vec z)K_b(\vec z,\vec y)\,\D \vec z\,,
\end{equation}
with $K_a$ and $K_b$ given by Eqs.\ (\ref{eq34}) and (\ref{eq55}).

We derive
\begin{eqnarray}
  K_{\cal {FH}}(\vec x,\vec y)\rap &=&\rap C_\alpha C_\beta\,\E^{-\I\pi x^2\cot\alpha}\int_{{\mathbb R}^2} \E^{-\I\pi z^2\cot\alpha}\exp\left({2\I\pi \vec x\vec\cdot\vec z\over \sin\alpha}\right) \nonumber \\
  & &\hskip 2cm \times\;\; \E^{-\I\pi z^2\coth\beta}\E^{-\I\pi y^2\coth\beta} \exp\left({2\I\pi \vec z\vec\cdot\vec y\over \sinh\beta}\right)\,\D\vec z \nonumber \\
  \rap &=&\rap C_\alpha C_\beta\,\E^{-\I\pi x^2\cot\alpha}\E^{-\I\pi y^2\coth\beta} \int_{{\mathbb R}^2} \E^{-\I\pi z^2(\cot\alpha +\coth\beta )}\nonumber \\
  & & \hskip 2cm \times \;\;  \exp\left[2\I\pi \vec z\vec\cdot\left({\vec x\over \sin\alpha}+{\vec y\over \sinh\beta}\right)\right]\,\D\vec z \nonumber \\
  \rap &=&\rap {-\I\,C_\alpha C_\beta\,\E^{-\I\pi x^2\cot\alpha}\E^{-\I\pi y^2\coth\beta}\over \cot\alpha +\coth\beta}\,\exp\left({\I\pi \over \cot\alpha +\coth\beta}\left\|{\vec x\over \sin\alpha}+{\vec y\over \sinh\beta}\right\|^2\right)\nonumber \\
  \rap & = &\rap C\, \E^{\I\pi Lx^2}\,\E^{\I\pi My^2}\,\E^{\I\pi N\vec x\vec\cdot\vec y}\,, \label{eq101}
\end{eqnarray}
with
\begin{equation}
  C={\I\,\E^{-\I\alpha+\beta}\over \sin\alpha \sinh\beta (\cot\alpha +\coth\beta)} =
  {\I\,\E^{-\I\alpha+\beta}\over \cos\alpha\sinh\beta +\sin\alpha\cosh\beta}\,,\label{eq124}
\end{equation}
\begin{eqnarray}L\rap &=&\rap -\cot\alpha +{1\over(\cot\alpha +\coth\beta)\sin^2\alpha}\nonumber \\
  \rap &=&\rap
  {-\cos^2\alpha -\cos\alpha\sin\alpha \coth\beta+1  \over(\cot\alpha +\coth\beta)\sin^2\alpha}
  ={1-\cot\alpha\coth\beta\over \cot\alpha +\coth\beta}\,,\label{eq125}
\end{eqnarray}
\begin{eqnarray}M\rap &=&\rap -\coth\beta +{1\over(\cot\alpha +\coth\beta)\sinh^2\beta}\nonumber \\
  \rap &=&\rap
  {-\cosh^2\beta -\cosh\beta \sinh\beta \cot\alpha+1  \over(\cot\alpha +\coth\beta)\sinh^2\beta}
  ={-1-\cot\alpha\coth\beta\over \cot\alpha +\coth\beta}\,,\label{eq126}
  \end{eqnarray}
\begin{equation}N = {2\over(\cot\alpha +\coth\beta)\sin\alpha\sinh\beta}=
  {2\over \cos\alpha\sinh\beta +\sin\alpha\cosh\beta}\,.\label{eq127}
  \end{equation}

\subsubsection{Product ${\cal F}_\alpha\circ{\cal H}_\beta$ : symbol (for $\alpha\ne n\pi$ and $\beta\ne 0$)}\label{sect313}
In accordance with Eqs.\ (\ref{eq19}) and (\ref{eq101}), we begin with
\begin{eqnarray}
  K_{\cal {FH}}\left(\vec x+{\vec t\over 2},\vec x-{\vec t\over 2}\right)\rap & = &\rap
  C\exp \left(\I\pi L\left\| \vec x+{\vec t\over 2}\right\|^2\right)
  \exp \left(\I\pi M\left\| \vec x-{\vec t\over 2}\right\|^2\right)\nonumber \\
  & &\hskip 3cm \times \;\; \exp\left[\I\pi N\left( \vec x+{\vec t\over 2}\right)\vec\cdot
    \left( \vec x-{\vec t\over 2}\right)\right] \nonumber \\
  \rap & = &\rap C \exp [\I\pi x^2(L+M+N)] \exp [\I\pi t^2(L+M-N)/4]\nonumber \\
   & &\hskip 3cm \times \;\;\exp [\I\pi  (L-M) \vec x\vec\cdot\vec t]\,.\label{eq106}
\end{eqnarray}
The corresponding symbol is
\begin{eqnarray}
  a_{\cal {FH}}(\vec x,\vec \xi )\rap &=&\rap \int_{{\mathbb R}^2} K_{\cal {FH}}\left(\vec x+{\vec t\over 2},\vec x-{\vec t\over 2}\right)\E^{2\I\pi \vec t\vec\cdot\vec \xi}\,\D\vec t \nonumber \\
  \rap &=&\rap  C  \exp [\I\pi x^2(L+M+N)]  \int_{{\mathbb R}^2}\exp [\I\pi t^2(L+M-N)/4]\nonumber \\
  & &\hskip 3cm \times \;\;\exp \left[2\I\pi \vec t\vec\cdot\left(\vec \xi +{L-M\over 2}\,\vec x \right)\right]\,\D\vec t\nonumber \\
   \rap &=&\rap  C  \exp [\I\pi x^2(L+M+N)]\,  I (\vec x,\vec \xi)\,,
\end{eqnarray}
with
\begin{eqnarray}
  I(\vec x,\vec \xi)\rap &=&\rap \int_{{\mathbb R}^2}\exp [\I\pi t^2(L+M-N)/4] \exp \left[2\I\pi \vec t\vec\cdot\left(\vec \xi +{L-M\over 2}\,\vec x \right)\right]\,\D\vec t\nonumber \\
  \rap &=&\rap {4\I\over L+M-N}\exp \left({-4\I\pi \over L+M-N}\left\|\vec \xi +{L-M\over 2}\vec x\right\|^2\right)\,. 
\end{eqnarray}
We obtain
\begin{eqnarray}
  a_{\cal {FH}}(\vec x,\vec \xi )\rap &=&\rap {4\I C\over L+M-N}\exp\left[\I\pi \left(L+M+N-{(L-M)^2\over (L+M-N)}\right)x^2\right]\nonumber \\
  & &\hskip 1cm \times \;\;\exp\left({-4\I\pi \xi^2\over L+M-N}\right)\,\exp\left(-4\I\pi\,{L-M\over L+M-N}\,\vec x \vec  \cdot\vec \xi \right)\,.\label{eq109}
\end{eqnarray}
For a more explicit expression, we use $M'$ and $N'$, such that
\begin{eqnarray}
  {1\over M'}\rap &=&\rap 
  L+M-N \nonumber \\
  \rap &=&\rap {1-\cot\alpha\coth\beta\over \cot\alpha+\coth\beta}- {1+\cot\alpha\coth\beta\over \cot\alpha+\coth\beta}
  -{2\over (\cot\alpha+\coth\beta)\sin\alpha\sinh\beta}\nonumber \\
  \rap &=&\rap {-2\cos\alpha\cosh\beta -2\over  (\cot\alpha+\coth\beta)\sin\alpha\sinh\beta}
  ={-2 (\cos\alpha\cosh\beta +1)\over   \cos\alpha\sinh\beta +\sin\alpha\cosh\beta}\,,
  \label{eq134}
\end{eqnarray}
\begin{eqnarray}
  N'\rap &=&\rap  {L-M\over L+M-N}\nonumber \\\rap &=& \rap {2\over \cot\alpha+\coth\beta}\,{(\cot\alpha +\coth\beta)\sin\alpha\sinh\beta\over
    -2\cos\alpha\cosh\beta -2}
  = {-\sin\alpha\sinh\beta\over
   1+ \cos\alpha\cosh\beta}\,.\label{eq133}
\end{eqnarray}
We also use $L'$, such that
\begin{eqnarray}
  -4L'\rap &=&\rap 
  L+M+N-{(L-M)^2\over L+M-N}
  ={(L+M)^2-N^2-(L-M)^2\over L+M-N}\nonumber \\
  \rap &=&\rap {4LM-N^2\over L+M-N}\,,\label{eq136}\end{eqnarray}
that is
\begin{eqnarray}
  L'(L+M-N)
  \rap &=&\rap {(1-\cot\alpha\coth\beta)(1+\cot\alpha\coth\beta )\over (\cot\alpha +\coth\beta )^2}
  +{1\over (\cot\alpha +\coth\beta)^2\sin^2\alpha\sinh^2\beta}\nonumber \\
   \rap &=&\rap {(1-\cot^2\alpha\coth^2\beta)\sin^2\alpha\sinh^2\beta +1\over (\cot\alpha +\coth\beta)^2\sin^2\alpha\sinh^2\beta}\nonumber \\
 \rap &=&\rap {\sin^2\alpha\sinh^2\beta -\cos^2\alpha\cosh^2\beta +1\over (\cos\alpha\sinh\beta+\sin\alpha\cosh\beta)^2} \nonumber \\
 \rap &=&\rap {\sin^2\alpha(1+\sinh^2\beta ) +\cos^2\alpha (1-\cosh^2\beta )\over (\cos\alpha\sinh\beta+\sin\alpha\cosh\beta)^2} \nonumber \\
 \rap &=&\rap {-\cos \alpha \sinh\beta+\sin \alpha \cosh \beta \over \cos\alpha\sinh\beta+\sin\alpha\cosh\beta} \,.
\end{eqnarray}
Taking into account the value of $L+M-N$ given by Eq.\ (\ref{eq134}), we obtain
\begin{equation}
  L'={\cos \alpha \sinh\beta-\sin \alpha \cosh \beta \over 2(1+\cos\alpha\cosh \beta)}\,.
\end{equation}
Eventually, by Eq.\ (\ref{eq124}), we obtain
\begin{equation}
  {4\I C\over L+M-N}={2\,\E^{-\I\alpha +\beta}\over 1+\cos\alpha \cosh\beta}\,.
  \end{equation}

The symbol of ${\cal F}_\alpha\circ{\cal H}_\beta$ takes the form
\begin{eqnarray}
  a_{\cal {FH}}(\vec x,\vec \xi)\rap &=&\rap {4\I C\over L+M-N}\, \exp(-4\I\pi L'x^2)\,\exp (-4\I\pi M'\xi^2)\,\exp (-4\I\pi N'\vec x\vec\cdot\vec\xi )\nonumber \\
\rap &=&\rap {2\,\E^{-\I\alpha +\beta}\over 1+\cos\alpha \cosh\beta}\, \exp(-4\I\pi L'x^2)\,\exp (-4\I\pi M'\xi^2)\,\exp (-4\I\pi N'\vec x\vec\cdot\vec\xi )
  \,.\label{eq114}
\end{eqnarray}

\subsubsection{Product ${\cal H}_\beta\circ{\cal F}_\alpha$. Non-commutativity}\label{noncom1}

We consider ${\cal F}_\alpha$ ($\alpha\ne n\pi$, $n\in{\mathbb Z}$) with symbol $a$ and kernel $K_a$, and  ${\cal H}_\beta$  ($\beta\ne 0$), with symbol $b$ and kernel $K_b$, as in Section \ref{sect312}. 
The kernel of the operator ${\cal H}_\beta\circ{\cal F}_\alpha$ is
\begin{eqnarray}
  K_{\cal {HF}}(\vec x,\vec y)\rap & =&\rap \int_{{\mathbb R}^2} K_b(\vec x,\vec z)K_a(\vec z,\vec y)\,\D \vec z\nonumber \\ 
  \rap &=&\rap C_\alpha C_\beta\,\E^{-\I\pi x^2\coth\beta}\int_{{\mathbb R}^2} \E^{-\I\pi z^2\coth\beta}\exp\left({2\I\pi \vec x\vec\cdot\vec z\over \sinh\beta}\right) \nonumber \\
  & &\hskip 2cm \times\;\; \E^{-\I\pi z^2\cot\alpha}\E^{-\I\pi y^2\cot\alpha} \exp\left({2\I\pi \vec z\vec\cdot\vec y\over \sin\alpha}\right)\,\D\vec z \nonumber \\
  \rap &=&\rap C_\alpha C_\beta\,\E^{-\I\pi x^2\coth\beta}\E^{-\I\pi y^2\cot\alpha} \int_{{\mathbb R}^2} \E^{-\I\pi z^2(\cot\alpha +\coth\beta )}\nonumber \\
  & & \hskip 2cm \times \;\;  \exp\left[2\I\pi \vec z\vec\cdot\left({\vec x\over \sinh\beta}+{\vec y\over \sin\alpha}\right)\right]\,\D\vec z \nonumber \\
  \rap &=&\rap {-\I\,C_\alpha C_\beta\,\E^{-\I\pi x^2\coth\beta}\E^{-\I\pi y^2\cot\alpha}\over \cot\alpha +\coth\beta}\,\exp\left({\I\pi \over \cot\alpha +\coth\beta}\left\|{\vec x\over \sinh\beta}+{\vec y\over \sin\alpha}\right\|^2\right)\nonumber \\
  \rap & = &\rap C\, \E^{\I\pi Mx^2}\,\E^{\I\pi Ly^2}\,\E^{\I\pi N\vec x\vec\cdot\vec y}\,, \label{eq101n}
\end{eqnarray}
where $C$, $L$, $M$ and $N$ are still given by Eqs.\ (\ref{eq124}--\ref{eq127}).

By comparing Eq.\ (\ref{eq101n}) with Eq.\ (\ref{eq101}), we obtain  $K_{\cal {HF}}(\vec x,\vec y)=K_{\cal {FH}}(\vec y,\vec x)$. Since $L\ne M$, we have $K_{\cal {FH}}(\vec y,\vec x)\ne K_{\cal {FH}}(\vec x,\vec y)$ and we conclude that  $K_{\cal {HF}}(\vec x ,\vec y)\ne K_{\cal {FH}}(\vec x,\vec y)$, which means that the product ${\cal F}_\alpha\circ{\cal H}_\beta$ is not commutative.

We eventually provide the symbol $a_{\cal{HF}}$ of the composed operator ${\cal H}_\beta\circ{\cal F}_\alpha$. We remark that the kernel $K_{\cal{HF}}$ is deduced from $K_{\cal{FH}}$ by exchanging $L$ and $M$ in Eq.\ (\ref{eq101}). According to Eqs.\ (\ref{eq134}) and (\ref{eq136}), $M'$ and $L'$ remain unchanged, whereas according to Eq.\ (\ref{eq133}), $N'$ is changed into $-N'$. From Eq.\ (\ref{eq114}) we deduce
\begin{equation}
a_{\cal{HF}}(\vec x,\vec \xi)={2\,\E^{-\I\alpha +\beta}\over 1+\cos\alpha \cosh\beta}\, \exp(-4\I\pi L'x^2)\,\exp (-4\I\pi M'\xi^2)\,\exp (4\I\pi N'\vec x\vec\cdot\vec\xi )\,,
\end{equation}
where $L'$, $M'$ and $N'$ are respectively given by Eqs.\ (\ref{eq136}), (\ref{eq134}) and (\ref{eq133}).


\subsection{Circular transformations and hyperbolic transformations \\ of the second kind}\label{sect32}

\subsubsection{Specific cases}\label{sect321}
For every $n\in{\mathbb Z}$, we have ${\cal F}_{2n\pi}\circ {\cal K}_\beta={\cal K}_\beta\circ{\cal F}_{2n\pi}={\cal K}_\beta$, for every $\beta\in{\mathbb R}$.

Since ${\cal K}_0=\I{\cal F}_{\pi /2}$, we have
  ${\cal F}_{\alpha}\circ {\cal K}_0={\cal K}_0\circ{\cal F}_{\alpha}=\I {\cal F}_{\alpha +(\pi /2)}$. In particular ${\cal K_0}\circ{\cal F}_{\pi /2}=\I {\cal F}_\pi =\I{\cal P}$, and   ${\cal K_0}\circ{\cal F}_{-\pi /2}=\I {\cal F}_0 =\I{\cal I}$.

\subsubsection{Product ${\cal F}_\alpha\circ{\cal K}_\beta$ : kernel (for $\alpha \ne \pi /2\;[\pi ]$ and $\beta \ne 0$)}\label{sect322}
We consider the circular transformation ${\cal F}_\alpha$ ($\alpha \ne \pi /2\; [\pi ]$) with symbol $a$ and kernel $K_a$, and the hyperbolic transformation of the second kind ${\cal K}_\beta$ ($\beta \ne 0$), with symbol $b$ and kernel $K_b$.

The kernel of the operator ${\cal F}_\alpha\circ{\cal K}_\beta$ is
\begin{equation}
  K_{\cal{FK}}(\vec x,\vec y)=\int_{{\mathbb R}^2} K_a(\vec x,\vec z)K_b(\vec z,\vec y)\,\D \vec z\,,
\end{equation}
where  $K_a$ and $K_b$ are given by Eqs.\ (\ref{eq34}) and (\ref{eq71}).

With $C_\alpha$ still given by Eq.\ (\ref{eq46a}), but $C_\beta$ now given by Eq.\ (\ref{eq92}), we
derive
\begin{eqnarray}
  K_{\cal{FK}}(\vec x,\vec y)\rap &=&\rap C_\alpha C_\beta\,\E^{-\I\pi x^2\cot\alpha}\int_{{\mathbb R}^2} \E^{-\I\pi z^2\cot\alpha}\exp\left({2\I\pi \vec x\vec\cdot\vec z\over \sin\alpha}\right) \nonumber \\
  & &\hskip 2cm \times\;\; \E^{\I\pi z^2\tanh\beta}\E^{-\I\pi y^2\tanh\beta} \exp\left({2\I\pi \vec z\vec\cdot\vec y\over \cosh\beta}\right)\,\D\vec z \nonumber \\
 \hskip 1cm  \rap &=&\rap C_\alpha C_\beta\,\E^{-\I\pi x^2\cot\alpha}\;\E^{-\I\pi y^2\tanh\beta} \int_{{\mathbb R}^2} \E^{\I\pi z^2(\tanh\beta -\cot\alpha)}\nonumber \\
   & & \hskip 2cm \times \;\;
  \exp\left[2\I\pi \vec z\vec\cdot\left({\vec x\over \sin\alpha}+{\vec y\over \cosh\beta}\right)\right]\,\D\vec z \nonumber \\
  \rap &=&\rap {\I\,C_\alpha C_\beta\,\E^{-\I\pi x^2\cot\alpha}\;\E^{-\I\pi y^2\tanh\beta}\over \tanh\beta -\cot\alpha}\,\exp\left({-\I\pi \over \tanh\beta -\cot\alpha}\left\|{\vec x\over \sin\alpha}+{\vec y\over \cosh\beta}\right\|^2\right)\nonumber \\
  \rap & = &\rap C \, \E^{\I\pi Lx^2}\,\E^{\I\pi My^2}\,\E^{\I\pi N\vec x\vec\cdot\vec y}\,, \label{eq116}
\end{eqnarray}
with
\begin{equation}
  C={-\I\E^{-\I\alpha+\beta}\over \sin\alpha \cosh\beta (\tanh\beta -\cot\alpha)}=
  {-\I\E^{-\I\alpha+\beta}\over \sin\alpha\sinh\beta -\cos\alpha\cosh\beta}\,, \label{eq117}
\end{equation}
\begin{eqnarray}L\rap &=&\rap -\cot\alpha -{1\over(\tanh\beta -\cot\alpha)\sin^2\alpha} \nonumber \\
  \rap &=&\rap 
  { -\cos\alpha\sin\alpha \tanh\beta+\cos^2\alpha -1  \over(\tanh\beta -\cot\alpha)\sin^2\alpha}\nonumber \\
  \rap &=&\rap {-1-\cot\alpha\tanh\beta\over \tanh\beta -\cot\alpha}\,,\label{eq118}
\end{eqnarray}
\begin{eqnarray}M\rap &=&\rap -\tanh\beta -{1\over(\tanh\beta -\cot\alpha)\cosh^2\beta}\nonumber \\
  \rap &=&\rap 
  {-\sinh^2\beta +\cosh\beta \sinh\beta \cot\alpha-1  \over(\tanh\beta -\cot\alpha)\cosh^2\beta}\nonumber \\
  \rap &=&\rap {-1+\cot\alpha\tanh\beta\over \tanh\beta -\cot\alpha}\,,\label{eq119}
  \end{eqnarray}
\begin{equation}N = {-2\over(\tanh\beta -\cot\alpha )\sin\alpha\cosh\beta}=
  {-2\over \sin\alpha\sinh\beta -\cos\alpha\cosh\beta}\,.\label{eq120}
\end{equation}

\subsubsection{Product ${\cal F}_\alpha\circ{\cal K}_\beta$ : symbol (for $\alpha \ne \pi /2\;[\pi ]$ and $\beta \ne 0$)}
Equation (\ref{eq106}) still holds formally, where $C$, $L$, $M$ and $N$ are now given by Eqs.\ (\ref{eq117}--\ref{eq120}). Then Eq.\ (\ref{eq109}) also holds, and eventually the symbol, denoted $a_{\cal FK}$, is given by the following equation, similar to the first line of Eq.\ (\ref{eq114})
\begin{equation}
  a_{\cal {FK}}(\vec x,\vec \xi)={4\I C\over L+M-N}\, \exp(-4\I\pi L'x^2)\,\exp (-4\I\pi M'\xi^2)\,\exp (-4\I\pi N'\vec x\vec\cdot\vec\xi )
  \,,\label{eq128}
\end{equation}
where $L'$, $M'$ and $N'$ correspond to the new values of $L$, $M$ and $N$, and where $C$ is given by Eq.\ (\ref{eq117}). Explicitly we have
\begin{eqnarray}
  {1\over M'}\rap &=&\rap 
  L+M-N \nonumber \\
  \rap &=&\rap {-1-\cot\alpha\tanh\beta\over \tanh\beta -\cot\alpha}+ {-1+\cot\alpha\tanh\beta\over \tanh\beta-\cot\alpha}
  +{2\over (\tanh\beta -\cot\alpha)\sin\alpha\cosh\beta}\nonumber \\
  \rap &=&\rap {-2\sin\alpha\cosh\beta +2\over  (\tanh\beta -\cot\alpha)\sin\alpha\cosh\beta}
       ={2 (\sin\alpha\cosh\beta -1)\over   \sin\alpha\sinh\beta -\cos\alpha\cosh\beta}\,,
\end{eqnarray}
\begin{eqnarray}
  N'\rap &=&\rap  {L-M\over L+M-N}\nonumber \\\rap &=& \rap {-2\cot\alpha \tanh\beta \over \tanh\beta -\cot\alpha}\,\;{(\tanh\beta -\cot\alpha)\sin\alpha\cosh\beta\over
    -2\sin\alpha\cosh\beta +2}
  = {\cos\alpha\sinh\beta\over
    \sin\alpha\cosh\beta -1}\,.
\end{eqnarray}
From Eq.\ (\ref{eq136}), which remains formally valid, we obtain
\begin{eqnarray}
  L'(L+M-N)
  \rap &=&\rap -LM+(N/2)^2\nonumber \\
  \rap &=&\rap {(1+\cot\alpha\tanh\beta)(\cot\alpha\tanh\beta -1)\over (\tanh\beta -\cot\alpha)^2}
  +{1\over (\tanh\beta -\cot\alpha )^2\sin^2\alpha\cosh^2\beta}\nonumber \\
  \rap &=&\rap {(\cot^2\alpha\tanh^2\beta  -1)\sin^2\alpha\cosh^2\beta +1\over (\tanh\beta -\cot\alpha)^2\sin^2\alpha\cosh^2\beta} \nonumber \\
  \rap &=&\rap {\cos^2\alpha\sinh^2\beta -\sin^2\alpha\cosh^2\beta +1\over (\sin\alpha\sinh\beta-\cos\alpha\cosh\beta)^2} \nonumber \\
   \rap &=&\rap {\cos^2\alpha (\sinh^2\beta +1) +\sin^2\alpha (1-\cosh^2\beta)\over (\sin\alpha\sinh\beta-\cos\alpha\cosh\beta)^2} \nonumber \\
   \rap &=&\rap {\cos^2\alpha \cosh^2\beta  -\sin^2\alpha \sinh^2\beta\over (\sin\alpha\sinh\beta-\cos\alpha\cosh\beta)^2} \nonumber \\
   \rap &=&\rap  {\cos\alpha\cosh\beta+\sin\alpha\sinh\beta\over  \cos\alpha\cosh\beta -\sin\alpha\sinh\beta}
   \,,
\end{eqnarray}
so that
\begin{equation}
  L'=-{\cos\alpha\cosh\beta +\sin\alpha\sinh\beta \over 2(1-\sin\alpha\cosh\beta )}\,.
\end{equation}
Since
\begin{equation}
  {4\I C\over L+M-N}={2\E^{-\I\alpha +\beta}\over 1-\sin\alpha \cosh\beta}\,,
  \end{equation}
the symbol of ${\cal F}_\alpha\circ{\cal K}_\beta$ takes the form
\begin{equation}
  a_{\cal{FK}}(\vec x,\vec \xi)={2\E^{-\I\alpha +\beta}\over 1-\sin\alpha \cosh\beta}\, \exp(-4\I\pi L'x^2)\,\exp (-4\I\pi M'\xi^2)\,\exp (-4\I\pi N'\vec x\vec\cdot\vec\xi )\,.\label{eq151}
\end{equation}

\subsubsection{Product ${\cal K}_\beta\circ{\cal F}_\alpha$. Non-commutativity}\label{noncom2}

We consider ${\cal F}_\alpha$ ($\alpha\ne \pi /2 \;[\pi ]$) with symbol $a$ and kernel $K_a$, and  ${\cal K}_\beta$  ($\beta\ne 0$), with symbol $b$ and kernel $K_b$, as in Section \ref{sect322}. 
The kernel of the operator ${\cal K}_\beta\circ{\cal F}_\alpha$ is
\begin{equation}K_{\cal {KF}}(\vec x,\vec y) =  \int_{{\mathbb R}^2} K_b(\vec x,\vec z)K_a(\vec z,\vec y)\,\D \vec z \,,\end{equation}
that is
\begin{eqnarray}
  K_{\cal {KF}}(\vec x,\vec y)
\rap &=&\rap C_\alpha C_\beta\,\E^{\I\pi x^2\tanh\beta}\int_{{\mathbb R}^2} \E^{-\I\pi z^2\tanh\beta}\exp\left({2\I\pi \vec x\vec\cdot\vec z\over \cosh\beta}\right) \nonumber \\
  & &\hskip 2cm \times\;\; \E^{-\I\pi z^2\cot\alpha}\E^{-\I\pi y^2\cot\alpha} \exp\left({2\I\pi \vec z\vec\cdot\vec y\over \sin\alpha}\right)\,\D\vec z \nonumber \\
  \rap &=&\rap C_\alpha C_\beta\,\E^{\I\pi x^2\tanh\beta}\E^{-\I\pi y^2\cot\alpha} \int_{{\mathbb R}^2}  \E^{-\I\pi z^2(\cot\alpha+\tanh\beta)}\nonumber \\
  & &\hskip 2cm \times\;\;
  \exp\left[2\I\pi \vec z\vec\cdot\left({\vec x\over \cosh\beta}+{\vec y\over \sin\alpha}\right)\right]\,\D\vec z \nonumber \\
  \rap &=&\rap {-\I\,C_\alpha C_\beta\,\E^{\I\pi x^2\tanh\beta}\E^{-\I\pi y^2\cot\alpha}\over \cot\alpha +\tanh\beta }
  \,\exp\left({\I\pi \over \cot\alpha +\tanh\beta}\left\|{\vec x\over \cosh\beta}+{\vec y\over \sin\alpha}\right\|^2\right)\nonumber \\
  \rap & = &\rap C''\,\exp({\I\pi L'' x^2})\,\exp ({\I\pi M'' y^2})\,\exp ({\I\pi N''\vec x\vec\cdot\vec y})\,,
 \label{eq154} \end{eqnarray}
with
\begin{equation}
  C''={\I\,\E^{-\I\alpha+\beta}\over \sin\alpha \cosh\beta (\cot\alpha +\tanh\beta)}=
  {\I\,\E^{-\I\alpha+\beta}\over \cos\alpha\cosh\beta +\sin\alpha\sinh\beta}\,, \label{eq117a}
\end{equation}
\begin{eqnarray}L''\rap &=&\rap \tanh\beta +{1\over(\cot\alpha +\tanh\beta)\cosh^2\beta}\nonumber \\
  \rap &=&\rap 
  {\sinh^2\beta +\cosh\beta \sinh\beta \cot\alpha+1  \over(\cot\alpha +  \tanh\beta)\cosh^2\beta}\nonumber \\
  \rap &=&\rap {1+\cot\alpha\tanh\beta\over\cot\alpha +  \tanh\beta }\,,\label{eq118a}
  \end{eqnarray}
\begin{eqnarray}M''\rap &=&\rap -\cot\alpha +{1\over(\cot\alpha +\tanh\beta )\sin^2\alpha} \nonumber \\
  \rap &=&\rap 
  { -\cos^2\alpha  -\cos\alpha\sin\alpha \tanh\beta+1  \over(\cot\alpha +\tanh\beta)\sin^2\alpha}\nonumber \\
  \rap &=&\rap {1-\cot\alpha\tanh\beta\over \cot\alpha +\tanh\beta}\,,\label{eq119a}
\end{eqnarray}
\begin{equation}N'' = {2\over(\cot\alpha +\tanh\beta)\sin\alpha\cosh\beta}=
  {2\over \cos\alpha\cosh\beta + \sin\alpha\sinh\beta}\,.\label{eq120a}
\end{equation}

The comparison of Eqs.\ (\ref{eq117a}--\ref{eq120a}) with  Eqs.\ (\ref{eq117}--\ref{eq120}) show that $K_{\cal {KF}}(\vec x,\vec y)\ne K_{\cal {FK}}(\vec x,\vec y)$, which means that the product ${\cal F}_\alpha\circ{\cal K}_\beta$ is not commutative.

For the symbol $a_{\cal {KF}}$, we remark that  Eq.\ (\ref{eq154})  is Eq.\ (\ref{eq101}) where $C$, $L$, $M$ and $N$ have been replaced with $C''$, $L''$, $M''$ and $N''$. According to the results given in Section \ref{sect313}, we obtain
\begin{equation}
  a_{\cal {KF}}(\vec x,\vec \xi)= C'''\, \exp(-4\I\pi L'''x^2)\,\exp (-4\I\pi M'''\xi^2)\,\exp (-4\I\pi N'''\vec x\vec\cdot\vec\xi )\,,
  \end{equation}
  where
  \begin{equation}
  C'''={4\I\,C''\over L''+M''-N''}\,,
  \end{equation}
   \begin{equation}
  L'''={-L''M''+(N''/2)^2\over L''+M''-N''}\,,\end{equation}
  \begin{equation}
  M'''={1\over L''+M''-N''}\,,
  \end{equation}
  \begin{equation}
  N'''={L''-M''\over L''+M''-N''}\,.
  \end{equation}
Explicitly, we obtain
  \begin{eqnarray}
  L''+M''-N''\rap &=&\rap {2\over \cot\alpha +\tanh\beta}-{2\over (\cot\alpha +\tanh\beta )\sin\alpha\cosh\beta} \nonumber \\
  \rap &=&\rap
{2(\sin\alpha\sinh\beta -1)\over (\cot\alpha +\tanh\beta )\sin\alpha\cosh\beta }=
{2(\sin\alpha\sinh\beta -1)\over  \cos\alpha\cosh\beta +\sin\alpha\sinh\beta}\,,
  \end{eqnarray}
 and then
  \begin{equation}
 M'''={\cos\alpha\cosh\beta +\sin\alpha\sinh\beta \over 2(\sin\alpha\sinh\beta -1)}\,,
 \end{equation}
 \begin{equation}
 N'''={2\cot\alpha\tanh\beta\over \cot\alpha +\tanh\beta}\,\;{(\cot\alpha +\tanh\beta)\sin\alpha\cosh\beta \over 2(\sin\alpha\sinh\beta -1)}=
 {\cos\alpha\sinh\beta\over \sin\alpha\sinh\beta -1}\,,
 \end{equation}
 \begin{eqnarray}
 L'''(L''+M''-N'')\rap &=&\rap -L''M''+(N''/2)^2\nonumber \\
 \rap &=&\rap -{(1-\cot^2\alpha \tanh^2\beta)\over (\cot\alpha +\tanh\beta)^2}+{1\over  (\cot\alpha +\tanh\beta)^2\sin^2\alpha\cosh^2\beta}\nonumber \\
 \rap &=&\rap {-\sin^2\alpha\cosh^2\beta +\cos^2\alpha \sinh^2\beta+1\over  (\cot\alpha +\tanh\beta)^2\sin^2\alpha\cosh^2\beta}\nonumber \\
 \rap &=&\rap {\sin^2\alpha (1-\cosh^2\beta1)+\cos^2\alpha (1+\sinh^2\beta) \over  (\cos\alpha\cosh\beta +\sin\alpha\sinh\beta)^2}\nonumber \\
  \rap &=&\rap {-\sin^2\alpha \sinh^2\beta+\cos^2\alpha \cosh^2\beta \over  (\cos\alpha\cosh\beta +\sin\alpha\sinh\beta)^2}\nonumber \\
   \rap &=&\rap {\cos\alpha \cosh\beta -\sin\alpha \sinh\beta\over  \cos\alpha\cosh\beta +\sin\alpha\sinh\beta}\,,
 \end{eqnarray}
 so that
 \begin{equation}L'''={\cos\alpha \cosh\beta -\sin\alpha \sinh\beta \over 2(\sin\alpha\sinh\beta -1)}\,.
 \end{equation}


\subsection{Hyperbolic transformations of the first and  the second kind}\label{sect33}

\begin{proposition}\label{prop7} For every $\beta_1$ and $\beta_2$ in ${\mathbb R}$, we have
  \begin{itemize}
  \item ${\cal H}_{\beta_2}\circ{\cal K}_{\beta_1}=\E^{2\beta_2}\,{\cal K}_{\beta_1-\beta_2}$ ;
     \item ${\cal K}_{\beta_2}\circ{\cal H}_{\beta_1}={\cal K}_{\beta_1+\beta_2}$.
    \end{itemize}
\end{proposition}

The proof of Proposition \ref{prop7} has been given in a recent publication \cite{PPF5}.

\begin{remark} {\rm The composition of two hyperbolic transformations of the first and the second kind is not commutative, because   ${\cal H}_{\beta_2}\circ{\cal K}_{\beta_1}=\E^{2\beta_2}\,{\cal K}_{\beta_1-\beta_2}\ne {\cal K}_{\beta_1+\beta_2}={\cal K}_{\beta_1}\circ{\cal H}_{\beta_2}$, for $\beta_2\ne 0$.
    }
  \end{remark}

\subsection{Fractional Fourier transformation algebra}\label{sect34}

Circular and hyperbolic fractional Fourier transformations obey the following algebra (${\cal F}$ denotes the ``standard'' Fourier transformation, and ${\cal P}$ the parity operator).

\begin{enumerate}
\item ${\cal F}_0={\cal I}$ ; ${\cal F}_{\pi /2}={\cal F}$  ;
  ${\cal F}_\alpha^{-1}={\cal F}_{-\alpha}$ ; 
${\cal F}_{\alpha_2}\circ{\cal F}_{\alpha_1}={\cal F}_{\alpha_2+\alpha_1}$ ;  ${\cal F}_{\pi + \alpha}={\cal P}\circ{\cal F}_\alpha$ ;  ${\cal F}_{2\pi +\alpha}={\cal F}_{\alpha}$ ;
\item ${\cal H}_0={\cal I}$ ; ${\cal H}_\beta^{-1}={\cal H}_{-\beta}$ ;
 ${\cal H}_{\beta_2}\circ{\cal H}_{\beta_1}={\cal H}_{\beta_2+\beta_1}$ ;
\item  ${\cal K}_0= \I{\cal F}$ ; ${\cal K}_\beta^{-1}=-\E^{-2\beta}{\cal K}_\beta\circ {\cal P}$ ; ${\cal K}_{\beta_2}\circ{\cal K}_{\beta_1}=-\E^{2\beta_2}\,{\cal H}_{\beta_2-\beta_1}\circ{\cal P}$ ;
\item ${\cal K}_{\beta_2}\circ{\cal H}_{\beta_1}={\cal K}_{\beta_2+\beta_1}$ ;
  ${\cal H}_{\beta_2}\circ{\cal K}_{\beta_1}=\E^{2\beta_2}\,{\cal K}_{\beta_1-\beta_2}$ .
\end{enumerate}

\subsection{Identity and parity operators generally  cannot be the product of two fractional-order Fourier transformations of different kinds}\label{sect35}

\subsubsection{Motivation}\label{sect351}
From the perspective of applications in optics, we examine the possibility that the product of two fractional-order Fourier transformations results in either the identity or the  parity operator. If the transfer from an emitter to a receiver is described by the identity or the parity operator, then the optical field at the receiver is indeed the image of the field at the emitter. This  may occur, 
for instance, if the emitter and the receiver are located in different propagation media, separated by a refracting surface, a mirror or a lens.  We may say that  optical imaging consists in physically realizing optical-field transfers described by the identity or parity operators.

Since ${\cal F}_\alpha\circ{\cal F}_{-\alpha}={\cal I}={\cal H}_{\beta}\circ{\cal H}_{-\beta}$, for every real $\alpha$ and $\beta$, imaging may result from the composition of two circular fractional-order Fourier transformations or from the composition of two hyperbolic transformations of the first kind. Also, since ${\cal K}_\beta\circ{\cal K}_{\beta}=-\E^{2\beta}{\cal P}$, the composition of two hyperbolic transformations of the second kind may correspond to an imaging.

But may the composition of two fractional-order Fourier transformations of different kinds result in the identity or the parity operator?

The answer is affirmative, if we consider the following ``trivial'' cases (see Sections \ref{sect311} and \ref{sect321}):
\begin{enumerate}
\item Composition of a circular transformation ${\cal F}_\alpha$ with a hyperbolic transformation of the first kind ${\cal H}_\beta$:
  \begin{itemize}
  \item for $\alpha =0\; [2\pi ]$ and $\beta =0$, we obtain  ${\cal F}_\alpha\circ{\cal H}_0={\cal F}_0\circ{\cal H}_0={\cal I}$;
    \item for $\alpha =\pi \; [2\pi ]$ and $\beta =0$, we obtain ${\cal F}_\alpha\circ{\cal H}_0={\cal F}_\pi\circ{\cal H}_0={\cal P}$.
    \end{itemize}
\item
   Composition of a circular transformation ${\cal F}_\alpha$ with a hyperbolic transformation of the second kind ${\cal K}_\beta$:
  \begin{itemize}
  \item for $\alpha =\pi /2 \; [2\pi ]$ and $\beta =0$,  we obtain ${\cal F}_{\alpha}\circ{\cal K}_0={\cal F}_{\pi /2}\circ{\cal K}_0=\I\,{\cal P}$;
    \item for $\alpha =-\pi /2 \; [2\pi ]$ and $\beta =0$, we obtain ${\cal F}_{\alpha}\circ{\cal K}_0={\cal F}_{-\pi /2}\circ{\cal K}_0=\I\,{\cal I}$.
    \end{itemize}
\end{enumerate}

However, these trivial examples do not correspond to relevant or realistic configurations in optics. For example the composition ${\cal F}_0\circ{\cal H}_0={\cal I}$ is physically realized with a refracting spherical cap when the emitter and its image are merged with the cap (the refracting cap is its own image).   The result is valid, but it is of limited interest.

Once these trivial examples are set aside, the answer is negative, as shown by Theorem \ref{th1} in the next section. Consequently, the description of optical imaging can be addressed using fractional-order Fourier transformations by considering only compositions of transformations of the same kind. This will be implemented in Section \ref{sect6}.

\subsubsection{The main theorem}
\begin{theorem}\label{th1} Except  for the trivial cases mentioned in Sect.\ \ref{sect351}, the composition of two fractional-order Fourier transformations can  be proportional to neither the identity operator nor the parity operator, unless both transformations are of the same kind. 
\end{theorem}

\noindent {\its Proof.} The proof relies on comparing the symbols of the composed transformations with those of the identity and  the parity operator.

The symbol of the identity is $a_{\cal I}(\vec x,\vec \xi)=1$ (see Section \ref{sect152}), and the symbol of the parity operator is $a_{\cal P}(\vec x,\vec \xi)=(1/4)\,\delta ({\vec x})\,\delta({\vec\xi})$ (see Section \ref{sect223}).

We consider circular fractional-order Fourier transformations ${\cal F}_\alpha$ and hyperbolic transformations of the first kind ${\cal H}_\beta$ and of the second kind ${\cal K}_\beta$. 

\medskip
\noindent {\its i. Products ${\cal H}_{\beta_2}\circ{\cal K}_{\beta_1}$ and ${\cal K}_{\beta_2}\circ{\cal H}_{\beta_1}$.}

\smallskip
\noindent We remark first that ${\cal K}_\beta\ne \kappa {\cal I}$ ($\kappa \in {\mathbb R}$) for every $\beta$, because the symbol of ${\cal K}_\beta$ is
\begin{equation}
  a(\vec x,\vec \xi )= 2\,\E^{\beta}\,\exp (2\I\pi x^2\cosh \beta )\exp (2\I\pi \xi^2\cosh \beta )\exp (4\I\pi \vec x\vec\cdot\vec\xi\,\sinh \beta )\,,\label{eq77b}
\end{equation}
according to Eq.\ (\ref{eq77}). Then $a(\vec x,\vec \xi)$ cannot be a constant, namely $a(\vec x,\vec \xi)\ne \kappa\, a_{\cal I}(\vec x,\vec \xi )$, and ${\cal K}_\beta\ne \kappa {\cal I}$, for every $\beta$.
Since ${\cal H}_{\beta_2}\circ{\cal K}_{\beta_1}=\E^{2\beta_2}\,{\cal K}_{\beta_1-\beta_2}$ and ${\cal K}_{\beta_2}\circ{\cal H}_{\beta_1}={\cal K}_{\beta_2+\beta_1}$ (see Sect.\ \ref{sect33}), we conclude that ${\cal H}_{\beta_2}\circ{\cal K}_{\beta_1}\ne \kappa{\cal I}$ and ${\cal K}_{\beta_2}\circ{\cal H}_{\beta_1}\ne \kappa {\cal I}$.

For finite $\beta$, we have $a(\vec x,\vec \xi )\ne \kappa \,a_{\cal P}(\vec x,\vec \xi)$ ($\kappa\in{\mathbb R}$), so that ${\cal K}_\beta\ne \kappa{\cal P}$. We conclude   that ${\cal H}_{\beta_2}\circ{\cal K}_{\beta_1}\ne \kappa{\cal P}$ and ${\cal K}_{\beta_2}\circ{\cal H}_{\beta_1}\ne \kappa {\cal P}$.

\medskip
\noindent {\its ii. Products ${\cal F}_{\alpha}\circ{\cal H}_{\beta}$ and ${\cal H}_{\beta}\circ{\cal F}_{\alpha}$.}

\smallskip
\noindent Since trivial cases are set aside, we assume $\alpha \ne 0 \;[\pi]$ and $\beta \ne 0$.

According to Eq.\ (\ref{eq114}), the symbol of ${\cal F}_\alpha\circ{\cal H}_\beta$ is
\begin{equation}
  a_{\cal {FH}}(\vec x,\vec \xi)={2\,\E^{-\I\alpha +\beta}\over 1+\cos\alpha \cosh\beta}\, \exp(-4\I\pi L'x^2)\,\exp (-4\I\pi M'\xi^2)\,\exp (-4\I\pi N'\vec x\vec\cdot\vec\xi )\,,\label{eq159}
\end{equation}
with
\begin{equation}
L'={\cos\alpha\sinh\beta -\sin\alpha\cosh\beta\over 2(1+\cos\alpha\cosh\beta)}\,,
\end{equation}
\begin{equation}
M'=-{\cos\alpha\sinh\beta +\sin\alpha\cosh\beta\over 2(1+\cos\alpha\cosh\beta )}\,,
\end{equation}
\begin{equation}
N'=-{\sin\alpha\sinh\beta\over 1+\cos\alpha\cosh\beta}\,.
\end{equation}

For $a_{\cal {FH}}(\vec x,\vec \xi )$ to be a constant, we should have $L'=M'=N'=0$. If $\alpha\ne 0\;[\pi ]$ and $\beta \ne 0$, then $L'=0$ requires
$\tan\alpha =\tanh\beta$, and $M'=0$ requires $\tan\alpha =-\tanh\beta$ ; the two conditions are incompatible, if $\alpha\ne 0 \;[\pi ]$ and $\beta\ne 0$. Moreover $N'\ne 0$ for  $\alpha\ne 0\;[\pi ]$ and $\beta \ne 0$. We conclude that
$a_{\cal {FH}}(\vec x,\vec \xi )$ cannot be proportional  to $a_{\cal I}(\vec x,\vec \xi)$ and ${\cal F}_\alpha\circ{\cal H}_\beta$ cannot be proportional to the identity operator.

For $a_{\cal {FH}}(\vec x,\vec \xi )$ to be proportional to $a_{\cal P}(\vec x,\vec \xi )=(1/4)\,\delta (\vec x)\,\delta (\vec \xi)$, the quantity $1+\cos\alpha \cos\beta$ should tend to $0$, according to Eq.\ (\ref{eq4}). In the limit, that is obtained for $\alpha =\pi\, [2\pi ]$ and $\beta =0$, but these values correspond to a trivial case, as defined in Section \ref{sect351}. Then ${\cal F}_\alpha\circ{\cal H}_\beta$ 
cannot be proportional to ${\cal P}$, if  $\alpha \ne \pi\, [2\pi ]$ and $\beta \ne 0$.

We now examine the composition ${\cal H}_\beta\circ {\cal F}_\alpha$, whose kernel if $K_{\cal{HF}}$. Since ${\cal F}_\alpha\circ{\cal H}_\beta$ is not proportional to the identity operator, as shown above, we have $K_{\cal {FH}}(\vec x,\vec y)\ne \kappa \,\delta (\vec x-\vec y)$ (with $\kappa\in {\mathbb R}$).  Since $K_{\cal{HF}}(\vec x,\vec y)= K_{\cal{FH}}(\vec y,\vec x)$ (see Section \ref{noncom1}), we obtain  $K_{\cal{HF}}(\vec x,\vec y)\ne \kappa \,\delta (\vec y-\vec x)=\kappa\,\delta (\vec x-\vec y)$, which means that ${\cal H}_\beta\circ {\cal F}_\alpha$ is not proportional to the identity. The same reasoning holds if we replace the operator identity ${\cal I}$ with the parity operator ${\cal P}$, that is, $\delta (\vec x-\vec y)$ with $\delta (\vec x+\vec y)$ (which is the kernel of ${\cal P}$---see Section \ref{sect152}).

\medskip
\noindent {\its iii. Products ${\cal F}_{\alpha}\circ{\cal K}_{\beta}$ and ${\cal K}_{\beta}\circ{\cal F}_{\alpha}$.}

\smallskip
\noindent We assume $\alpha \ne \pi /2 \;[\pi ]$ and $\beta\ne 0$.

The symbol of ${\cal F}_\alpha\circ{\cal K}_\beta$ is given by Eq.\ (\ref{eq151}) as
\begin{equation}
  a_{\cal{FK}}(\vec x,\vec \xi)={2\E^{-\I\alpha +\beta}\over 1-\sin\alpha \cosh\beta}\, \exp(-4\I\pi L'x^2)\,\exp (-4\I\pi M'\xi^2)\,\exp (-4\I\pi N'\vec x\vec\cdot\vec\xi )\,,\label{eq157}
\end{equation}
where
\begin{equation}
L'=-{\cos\alpha\cosh\beta +\sin\alpha\sinh\beta\over 2(1-\sin\alpha\cosh\beta )}\,,
\end{equation}
\begin{equation}
M'={\sin\alpha\sinh\beta -\cos\alpha\cosh\beta\over 2(1-\sin\alpha\cosh\beta )}\,,
\end{equation}
\begin{equation}
N'=-{\cos\alpha\sinh\beta\over 1-\sin\alpha\cosh\beta}\,.
\end{equation}

For $\alpha\ne \pi /2\; [\pi ]$ and $\beta\ne 0$, we obtain $L'=0$, if $\cot\alpha =-\tanh\beta$, and $M'=0$, if $\cot\alpha =\tanh\beta$. The two conditions are incompatible. We conclude that ${\cal F}_{\alpha}\circ{\cal K}_{\beta}$  cannot be proportional to the identity operator for $\alpha\ne \pi /2\; [\pi ]$ and $\beta\ne 0$.

For $a_{\cal FH}(\vec x,\vec \xi )$ to be proportional to $a_{\cal P}(\vec x,\vec \xi )=(1/4)\,\delta (\vec x)\,\delta (\vec \xi)$, the quantity $1-\sin\alpha \sin\beta$ should tend to $0$, according to Eq.\ (\ref{eq4}). In the limit, we would have $\alpha =\pi /2 \;[\pi ]$ and $\beta =0$, which corresponds to a trivial case.  Then ${\cal F}_\alpha\circ{\cal K}_\beta$ cannot be proportional to ${\cal P}$, if  $\alpha \ne \pi\, [2\pi ]$ and $\beta \ne 0$.

The same reasoning holds for the composition ${\cal K}_{\beta}\circ {\cal F}_{\alpha}$ on the basis of the results obtained in Section \ref{noncom2}.

\smallskip
The proof of the theorem is complete. \qed


\newpage
{\Large
\centerline{\sc Part II -- Applications to diffraction and optical imaging}\label{part2}
}

\bigskip
\noindent
In this part we develop the mathematical representation of diffraction phenomena by fractional-order Fourier transformations, with emphasis on geometrical properties of field tranfers from sphe\-ri\-cal emitters to receivers (Sect.\ \ref{sect43}). Some  essential properties of imaging by optical centered systems are directly deduced from basic considerations (Sect.\ \ref{sect5}). Another approach is based on applying Theorem \ref{th1}---proved in Part I---to imaging by a refracting spherical cap and completes some findings given in a previous article \cite{PPF5} (Sect.\ \ref{sect6}).

The notation in this part is that we use in Fourier optics \cite{PPF4,PPF5}.

\section{Elements of fractional Fourier optics}\label{sect4}

\subsection{Field transfer by diffraction}\label{sect41}

\subsubsection{General transfer: Fresnel diffraction}

\begin{figure}[h]
\centering
    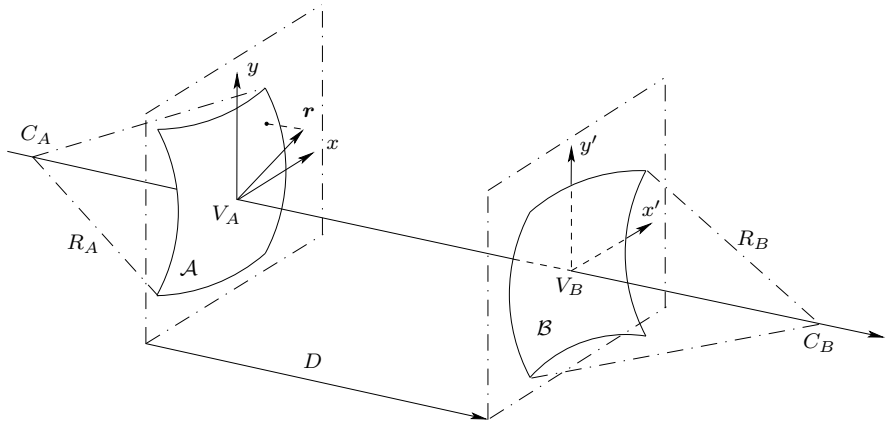
    \caption{\small Elements for representing diffraction from a spherical emitter ${\cal A}$ to a spherical receiver ${\cal B}$ at a distance $D$. Light propagates from left to right. Algebraic measures $D=\overline{V_AV_B}$, $R_A=\overline{V_AC_A}$, etc. are positive if taken along light propagation.  For example above: $R_B=\overline{V_BC_B}>0$, whereas $R_A<0$.\label{fig1}}
\end{figure}

The following approach to diffraction is based on the metaxial-optics theory of G. Bonnet \cite{PPF4,GB1,GB2}. We examine light propagation from a spherical cap  ${\cal A}$ (vertex $V_A$, center of curvature $C_A$,  radius  of curvature $R_A=\overline{V_AC_A}$), called the emitter, to another spherical cap ${\cal B}$ (vertex $V_B$, center $C_B$, radius $R_B=\overline{V_BC_B}$), called the receiver, located at a distance $D=\overline{V_AV_B}$ from ${\cal A}$ (Fig.\ \ref{fig1}). Spherical caps generally are aerial surfaces having a common axis and  on which lightwaves generate  electromagnetic fields. An intermediate spherical cap ${\cal C}$, located between ${\cal A}$ and ${\cal B}$, may be regarded  both like a receiver and like an emitter, since it receives light from ${\cal A}$ and re-emits it towards ${\cal B}$. Spherical caps may be real or virtual as usual in optics. The propagation medium is isotropic and homogeneous; lightwaves are monochromatic (wavelength $\lambda$, in the propagation medium).

In a scalar theory of diffraction, only the electric field is considered, since lightwaves are locally-plane waves: the electric field, the magnetic field and the Poynting vector form a rectangular trihedron at every point in space. The electric field is also called ``optical field.''

Locating a point on a spherical cap, say ${\cal A}$, is achieved through the Cartesian coordinates of its projection on the plane tangent to ${\cal A}$ at its vertex $V_A$ (Fig.\ \ref{fig1}). The corresponding variable is the two-dimensional vector $\vec r=(x,y)$. We denote $r=\| \vec r\|=(x^2+y^2)^{1/2}$ and $\D\vec r=\D x\,\D y$. The Euclidean scalar product of vectors $\vec r$ and $\vec r'$ is $\vec r\vec\cdot\vec r'$. The field amplitude at point $\vec r$ on ${\cal A}$ is denoted $U_A(\vec r)$. We do not write the time dependence.

For $R_A$, $R_B$ and $D$ not zero, the field transfer from ${\cal A}$ to ${\cal B}$ is expressed by \cite{PPF2,PPF4,GB1,GB2}
\begin{eqnarray}
U_B(\vec r')\rap &=&\rap {\I\over \lambda D}\exp\left[-{\I\pi\over \lambda}\left({1\over R_B}+{1\over D}\right)r'^2\right]\nonumber \\
& & \hskip 1.5cm \times \;\;
\int_{{\mathbb R}^2}\exp\left[-{\I\pi\over \lambda}\left({1\over D}-{1\over R_A}\right)r^2\right]\,\exp\left({2\I\pi \over \lambda D}\,\vec r\vec\cdot\vec r'\right)\,U_A(\vec r)\,\D\vec r\,,\label{eq2.1}
\end{eqnarray}
where a multiplicative factor $\exp (-2\I\pi D/\lambda)$ has been omitted. In general Eq.\ (\ref{eq2.1}) is associated with a Fresnel diffraction phenomenon.

\subsubsection{Fraunhofer diffraction. Fourier sphere}
If $D=R_A=-R_B$, Eq.\ (\ref{eq2.1}) becomes
\begin{equation}
U_B(\vec r')={\I\over \lambda D}
\int_{{\mathbb R}^2}\exp\left({2\I\pi \over \lambda D}\,\vec r\vec\cdot\vec r'\right)\,U_A(\vec r)\,\D\vec r ={\I\over\lambda D}\,\widehat{U}_A\left({\vec r'\over \lambda D}\right)\,,\label{eq2.1b}
\end{equation}
and ${\cal B}$ is called the Fourier sphere of ${\cal A}$, denoted as ${\cal F}$ (the context should avoid confusion with the standard Fourier transformation, also denoted as ${\cal F}$). The diffraction phenomenon is a Fraunhofer phenomenon.

We say that the field amplitude on ${\cal F}$ is the optical Fourier transform of the field amplitude on ${\cal A}$. The passage from $U_A$ to $U_F$ ($U_F\equiv U_B$ above) is achieved in three steps:
\begin{enumerate}
\item  Fourier transformation:  $U_A(\vec r)\,\rightleftharpoons\,\widehat U_A(\vec F)$, where $\vec F$ is a spatial frequency, the conjugate variable of $\vec r$;
\item Changing $\vec F$ to  $\vec r'/(\lambda D)$, where $\vec r'$ is the spatial variable on ${\cal F}$;
\item Multiplication by $\I/(\lambda D)$ (which corresponds to a phase shift and a mitigation).
\end{enumerate}

We recall that in Eq.\ (\ref{eq2.1b}) a constant phase factor $\exp (-2\I\pi D/\lambda )$ is omitted, and that time dependence is not written.

\begin{figure}[h]
  \centering
    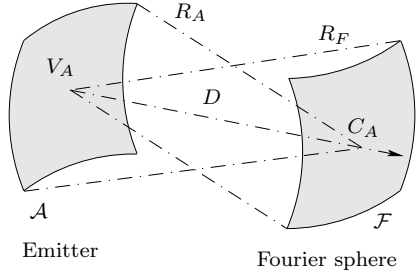
    \caption{\small Fraunhofer diffraction phenomenon. Spherical caps ${\cal A}$ and ${\cal F}$ are confocal: the vertex of  one is the curvature center of the other ($R_A=D=-R_F$). The field transfer from ${\cal A}$ to its Fourier sphere ${\cal F}$ is expressed by an ``optical'' Fourier transform---see Eq.\ (\ref{eq2.1b}). \label{fig1bis}}
\end{figure}

\subsection{Field-transfer representation by a fractional-order Fourier transformation}\label{sect42}

There is a similarity between Eq.\ (\ref{eq2.1}) and Eqs.\ (\ref{eq26}--\ref{eq28}): quadratic phase factors in front of the integrals and inside, Fourier kernels. Hence the idea of writing Eq.\ (\ref{eq2.1}) as a fractional-order Fourier transform. This can be done as follows \cite{PPF4,PPF5}.

Let $J$ be defined by
\begin{equation}
J={(R_A-D)(R_B+D)\over D(D-R_A+R_B)}\,.\label{eq2.2}\end{equation}
We assume $D\ne 0$ and $D\ne R_A-R_B$ ($D=0$ and $D= R_A-R_B$ may be examined as limit cases). 

Then we have three cases.
\begin{enumerate}
\item[{\its i}.]{\its Circular transformation:} $J>0$.

We define $\alpha\in ]-\pi ,\pi [$ by
\begin{equation}\cot^2\alpha =J\,,\hskip 1cm \alpha D>0\,,\hskip 1cm {DR_A\over R_A-D} \cot\alpha >0\,.\label{eq2.4}\end{equation}
We also define auxiliary parameters $\varepsilon_A$ and $\varepsilon_B$ by
\begin{equation}
\varepsilon_A={D\over R_A-D}\cot\alpha\,,\hskip 1.5cm  \varepsilon_B={D\over R_B+D}\cot\alpha\,,\end{equation}
which are shown to be such that $\varepsilon_AR_A>0$ and $\varepsilon_BR_B>0$.

Finally we choose the following reduced variables $\vec \rho$ on ${\cal A}$ and $\vec \rho '$ on ${\cal B}$
\begin{equation}
\vec\rho ={\vec r\over \sqrt{\lambda \varepsilon_AR_A}}\,,\hskip 1.5cm
\vec\rho ' ={\vec r'\over \sqrt{\lambda \varepsilon_BR_B}}\,,\label{eq2.5}
\end{equation}
and the following reduced  field-amplitudes
\begin{equation}
u_A(\vec \rho )=\sqrt{\lambda\varepsilon_AR_A}\;U_A(\sqrt{\lambda \varepsilon R_A}\,\vec \rho)\,,\hskip 1cm
u_B(\vec \rho ')=\sqrt{\lambda\varepsilon_BR_B}\;U_B(\sqrt{\lambda \varepsilon R_B}\,\vec \rho ')\,.\label{eq2.6}\end{equation}
Equation (\ref{eq2.1}) becomes \cite{PPF4,PPF5}
\begin{equation}u_B(\vec \rho ')=\E^{\I\alpha}{\cal F}_\alpha [u_A](\vec \rho ')\,.\label{eq2.161}\end{equation}
The field transfer from ${\cal A}$ to ${\cal B}$ is expressed through a circular fractional Fourier transformation of order $\alpha$.

\item[{\its ii}.]{\its Hyperbolic transformation of the first kind:} $J<-1$.

We define $\beta\in{\mathbb R}$ by
\begin{equation}\coth^2\beta =-J\,,\hskip 1cm \beta D>0\,.\label{eq2.7}\end{equation}
We also define auxiliary parameters $\varepsilon_A$ and $\varepsilon_B$ by
\begin{equation}
\varepsilon_A=\frak{s}{D\over R_A-D}\coth\beta\,,\hskip 1.5cm  \varepsilon_B=\frak{s}{D\over R_B+D}\coth\beta\,,\label{eq2.8}\end{equation}
where $\frak{s}$ denotes the sign of $R_A(R_A-D)$. 

If reduced variables and reduced field-amplitudes are chosen as in Eqs.\ (\ref{eq2.5}) and (\ref{eq2.6}), Eq.~(\ref{eq2.1}) takes the form \cite{PPF5}
\begin{equation}u_B(\vec \rho ')=\frak{s}\E^{-\frak{s}\beta}{\cal H}_{\frak{s}\beta} [u_A](\frak{s}\vec \rho ')\,.\label{eq2.164}\end{equation}
The field transfer from ${\cal A}$ to ${\cal B}$ is expressed through a hyperbolic fractional Fourier transformation of the first kind of order $\frak{s}\beta$.

\item[{\its iii}.]{\its Hyperbolic transformation of the second kind:} $-1<J<0$.

We define $\beta\in{\mathbb R}$ by
\begin{equation}\coth^2\beta =-{1\over J}\,,\hskip 1cm \beta D>0\,.\end{equation}
We also define auxiliary parameters $\varepsilon_A$ and $\varepsilon_B$ by
\begin{equation}
\varepsilon_A=\frak{s}{D\over R_A-D}\,{1\over \coth\beta}\,,\hskip 1.5cm  \varepsilon_B=-\frak{s}{D\over R_B+D}\,{1\over \coth\beta}\,,\end{equation}
where $\frak{s}$ denotes the sign of $R_A(R_A-D)$. 

If reduced variables and reduced field-amplitudes are chosen as in Eqs.\ (\ref{eq2.5}) and (\ref{eq2.6}), then Eq.\ (\ref{eq2.1}) takes the form \cite{PPF5}
\begin{equation}u_B(\vec \rho ')=\E^{-\frak{s}\beta}{\cal K}_{\frak{s}\beta} [u_A](\vec \rho ')\,,\hskip .5cm\mbox{if}\;\;\;D>0\,,\label{eq2.167}\end{equation}
\begin{equation}u_B(\vec \rho ')=\E^{-\frak{s}\beta}{\cal K}_{\frak{s}\beta} [\widetilde u_A](\vec \rho ')=\E^{-\frak{s}\beta}\,{\cal P}\circ{\cal K}_{\frak{s}\beta} [u_A](\vec \rho ') \,,\hskip .5cm\mbox{if}\;\;\;D<0\,.\label{eq2.168}\end{equation}
The field transfer from ${\cal A}$ to ${\cal B}$ is expressed through a hyperbolic fractional Fourier transformation of the second kind of order $\frak{s}\beta$.
\end{enumerate}

Proofs of Eqs.\ (\ref{eq2.161}), (\ref{eq2.164}),   (\ref{eq2.167}) and (\ref{eq2.168}) have been given in a previous publication \cite{PPF5}.

\begin{definition}[Three classes of field transfers by diffraction]\label{def1} A field transfer by diffraction from a spherical emitter 
to a receiver is said to be  circular (or, respectively, hyperbolic of the first kind or second kind)
 if it is represented by a fractional-order Fourier transformation  that is circular (or, respectively,  hyperbolic of the first or second kind).
\end{definition}

\subsection{Geometrical characterization of a field transfer by diffraction}\label{sect43}

\subsubsection{Analyzing sequences of vertices and  centers of curvature}
The following proposition allows us to determine without calculation which kind of fractional-order Fourier transform corresponds to the field transfer from a given spherical emitter ${\cal A}$ to a given spherical receiver ${\cal B}$.

\begin{proposition}[Geometrical characterization]\label{propgeo} Let ${\cal A}$ (vertex $V_A$, center of curvature $C_A$) be a spherical emiter and let ${\cal B}$ (vertex $V_B$, center of curvature $C_B$) be a receiver at a distance $D=\overline{V_AV_B}$. Then the field transfer from ${\cal A}$ to ${\cal B}$ is represented by
\begin{enumerate}
\item[i.] A circular fractional-order Fourier transform if, and only if, the indices of the vertices and  of the centers  of curvature of ${\cal A}$ and ${\cal B}$ are sequentially arranged along the optical axis, in the direction of light propagation, according to the pattern $ABAB$ or $BABA$. Example: $(V_AC_BC_AV_B)$ and circular permutations.
\item[ii.] A hyperbolic fractional-order Fourier transform of the first kind if, and only if, (a) the indices of vertices and  of the centers  of curvature  of ${\cal A}$ and ${\cal B}$ are sequentially arranged along the optical axis, in the direction of light propagation, according to $AABB$, up to a circular permutation, and (b) the vertices or the centers of curvature are adjacent. Example:  $(V_AC_AC_BV_B)$ and circular permutations.
\item[iii.] A hyperbolic fractional-order Fourier transform of the second kind if, and only if, the indices of the vertices and  of the centers of curvature of ${\cal A}$ and ${\cal B}$ are sequentially arranged  along the optical axis, in the direction of light propagation, according to the pattern $AABB$, up to a circular permutation,  and (b) the vertices alternate with the  centers  of curvature.
Example:  $(C_AV_BC_BV_A)$ and circular permutations.
\end{enumerate}
\end{proposition}

\noindent{\its Proof.} We begin with a general analysis of sequences of vertices and centers of curvature of ${\cal A}$ and ${\cal B}$ along the optical axis.

For each configuration of ${\cal A}$ and ${\cal B}$, there is a corresponding value of $J$, given by Eq.\ (\ref{eq2.2}).
Let ${\cal S}$ and ${\cal T}$ be two  spherical caps tangent to ${\cal B}$ (their common vertex  is $V_B$): ${\cal S}$ is centred at $V_A$ and  ${\cal T}$ at $C_A$.  We have $R_S=\overline{V_BV_A}=-D$, and $R_T=\overline{V_BC_A}=R_A-D$. Let us introduce the following (algebraic) curvatures
\begin{equation}
{\frak C_B}={1\over R_B}\,,\hskip 1cm {\frak C_S}={1\over R_S} \,,\hskip 1cm {\frak C_T}={1\over R_T}\,,\end{equation}
so that
\begin{equation}
J={(R_A-D)(R_B+D)\over D(D-R_A+R_B)}=-{R_T(R_B-R_S)\over R_S(R_B-R_T)}=-{\displaystyle{1\over R_S}-{1\over R_B}\over \displaystyle{1\over R_T}-{1\over R_B}}=-{\frak{C}_S-\frak{C}_B\over \frak{C}_T-\frak{C}_B}\,.\end{equation}

\begin{figure}[h]
\begin{center}
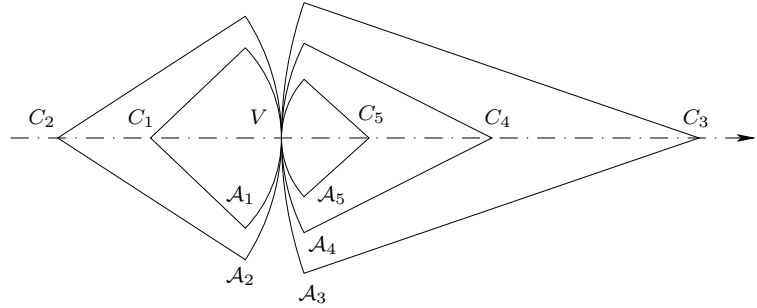
\caption{\small A sequence of spherical caps with increasing curvatures:
$\frak{C}_{A_1}<\frak{C}_{A_2}<0<\frak{C}_{A_3}<\frak{C}_{A_4}<\frak{C}_{A_5}$. \label{fig3-0}}
\end{center}
\end{figure}

For practical purposes, comparisons of curvatures can be handled graphically as shown in Fig.\ \ref{fig3-0} for
a sequence of tangent spherical caps ${\cal A}_j$ ($j=1,\dots,5$) with increasing curvatures  $\frak{C}_{A_j}=1/R_{A_j}$ ($R_{A_j}=\overline{VC_{A_j}\!\!}\,\,$). 
Algebraic measures are positive if taken along the direction of light propagation (indicated by the arrow in Fig.\ \ref{fig3-0}).

To prove Proposition \ref{propgeo}, we examine the following three cases.

\medskip
\noindent{\its i. Circular fractional-order Fourier transformation: $J>0$.}

\begin{figure}
\centering
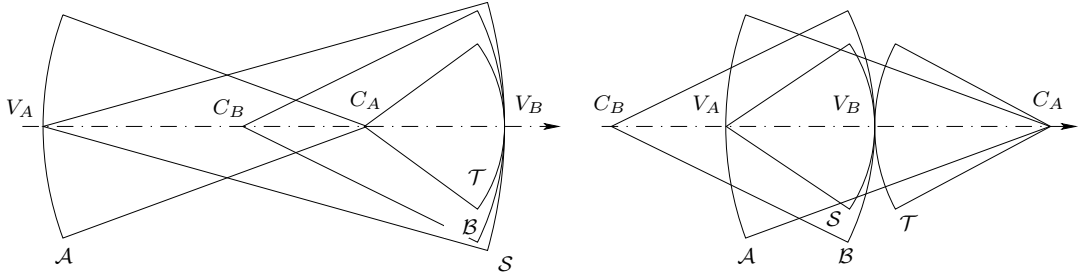
\caption{\small Two configurations for which  $J>0$.  The receiver ${\cal B}$ is located between ${\cal S}$ and ${\cal T}$. The vertices $V_A$ and $V_B$ and the centers of curvature $C_A$ and $C_B$ are arranged along the axis according to  $(V_A C_B C_A V_B)$, on the left diagram, and according to $(C_BV_AV_BC_A)$, on the right one. \label{fig3-1}}
\end{figure}

The condition $J>0$ is equivalent to
\begin{equation}
{\frak{C}_S-\frak{C}_B\over \frak{C}_T-\frak{C}_B}<0\,,\end{equation}
that is
\begin{equation}
\min (\frak{C}_S,\frak{C}_T)<\frak{C}_B <\max(\frak{C}_S,\frak{C}_T)\,.\label{eq162}\end{equation}
Equation (\ref{eq162}) indicates that ${\cal B}$  lies geometrically between ${\cal S}$ and ${\cal T}$. Figure \ref{fig3-1} shows two examples: vertices and centers of curvature are arranged along the axis according to  $(V_AC_BC_AV_B)$, on the left diagram, and to $(C_BV_AV_BC_A)$, on the right one. Other configurations are obtained by circular permutations, as we shall show.

We consider the arrangements (or orderings) I of Table  \ref{tab1} (denoted I.1, I.2, and so on), which are circular permutations of $(V_AC_BC_AV_B)$, and the arrangements II of Table \ref{tab2} (denoted II.1, II.2, and so on), which are circular permutations of $(C_BV_AV_BC_A)$. The corresponding values of $J$ are denoted $J_{{\rm I}.1}$,  $J_{{\rm I}.2}$, and so on. We know that $(V_AC_BC_AV_B)$ and $(C_BV_AV_BC_A)$ correspond to $J_{{\rm I}.1}>0$ and $J_{{\rm II}.1}>0$, respectively, because the vertices and centers of ${\cal A}$ and ${\cal B}$ are arranged such that ${\cal B}$  lies geometrically  between ${\cal S}$ and ${\cal T}$, in both cases. We have to prove that all values of $J$ corresponding to arrangements 
listed in  Tables \ref{tab1} and \ref{tab2} are positive. Although graphically checking each arrangement would suffice to prove that, we instead adopt a different method.

We start with $(V_AC_BC_AV_B)$ (arrangement I.1, for which $J_{{\rm I}.1}>0$), but we read it from right to left, that is, we consider the arrangement $(V_BC_AC_BV_A)$, which is the arrangement II.3. Clearly ${\cal B}$ remains between ${\cal S}$ and ${\cal T}$, and we conclude that  $J_{{\rm II}.3}>0$. The same method shows that $J_{{\rm I}.3}>0$, because  I.3 is none other  than II.1, read from right to left.

Now, the arrangement  $(C_AC_BV_AV_B)$, which is arrangement II.2, is deduced from I.1 by interchanging $V_A$ and $C_A$ without changing ${\cal B}$. Caps ${\cal S}$ and ${\cal T}$ are interchanged, but ${\cal B}$, which lies  between them for arrangement I.1 ($J_{{\rm I}.1}>0$) remains between them, so that $J_{{\rm II}.2}>0$.  If we read II.2 from right to left, we obtain the arrangement I.2, which corresponds then to $J_{{\rm I}.2}>0$.

Finally, we exchange $V_A$ and $C_A$ in I.3 and obtain II.4. Since I.3 corresponds to $J_{{\rm I}.3}>0$, we deduce that II.4
 correspond to $J_{{\rm II}.4}>0$. We also obtain $J_{{\rm I}.4}>0$, because I.4 is II.4, read from right to left.

In conclusion, all the configurations given by Tables \ref{tab1} and \ref{tab2} correspond to positive values of $J$, that is, to field transfers expressed by circular fractional-order Fourier transformations. The direct part of item {\its i} (necessary condition) is proved.

\begin{table}[h]
\centering
\begin{minipage}{5cm}
\hskip .64cm
\begin{tabular}{cc}
\hline
I.1&$(V_A\;C_B\;C_A\;V_B)$\cr
I.2&$(V_B\;V_A\;C_B\;C_A)$\cr
I.3&$(C_A\;V_B\;V_A\;C_B)$\cr
I.4&$(C_B\;C_A\;V_B\;V_A)$\cr
\hline
\end{tabular}
\caption{\small Arrangements I\label{tab1}}
\end{minipage}
\hskip 1cm
\begin{minipage}{5cm}
\hskip .62cm
\begin{tabular}{cc}
\hline
II.1&$(C_B\;V_A\;V_B\;C_A)$\cr
II.2&$(C_A\;C_B\;V_A\;V_B)$\cr
II.3&$(V_B\;C_A\;C_B\;V_A)$\cr
II.4&$(V_A\;V_B\;C_A\;C_B)$\cr
\hline
\end{tabular}
\caption{\small Arrangements II\label{tab2}}
\end{minipage}
\end{table}

\medskip
\noindent{\its ii. Hyperbolic fractional-order Fourier transformation of the first kind: $J<-1$.}

\begin{figure}[h]
\centering
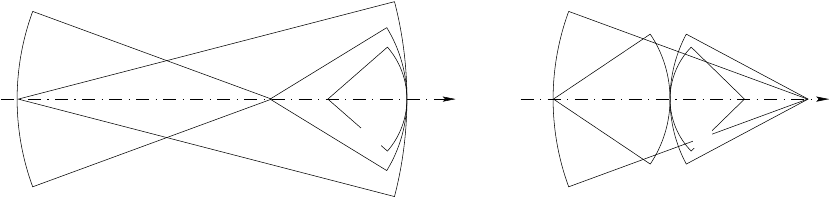
\caption{\small Two configurations for which  $J<-1$. The receiver ${\cal B}$ should not lie between ${\cal S}$ and ${\cal T}$. The vertices $V_A$ and $V_B$ and the centers $C_A$ and $C_B$ are ordered along the axis according to $(V_A  C_A C_B V_B)$ and to $(V_AV_BC_BC_A)$. \label{fig3-2}}
\end{figure}

The condition $J<-1$ is equivalent to
\begin{equation}
{\frak{C}_S-\frak{C}_B\over \frak{C}_T-\frak{C}_B}>1\,.\label{eq163}\end{equation}
which can be obtained as follows:
\begin{itemize}
\item If $\frak{C}_S> \frak{C}_T$, then ${\frak{C}_S-\frak{C}_B> \frak{C}_T-\frak{C}_B}$. Therefore, Inequality (\ref{eq163}) requires $\frak{C}_T-\frak{C}_B>0$, implying   $\frak{C}_B< \frak{C}_T<\frak{C}_S$. An example is shown in the left diagram of Fig. \ref{fig3-2}.
\item If $\frak{C}_S< \frak{C}_T$ then ${\frak{C}_S-\frak{C}_B< \frak{C}_T-\frak{C}_B}$. Therefore, Inequality (\ref{eq163}) requires $\frak{C}_T-\frak{C}_B<0$, implying  $\frak{C}_S< \frak{C}_T<\frak{C}_B$. Such a case is illustrated in the right diagram of Fig. \ref{fig3-2}.
\end{itemize}

Table \ref{tab3} shows the arrangements that are circular permutations of  $(V_AC_AC_BV_B)$ and Table \ref{tab4} those of  $(V_AV_BC_BC_A)$. A graphical checking proves that all these arrangements correspond to $J<-1$ (explicitly: $J_{{\rm III}.1}<-1$, $J_{{\rm III}.2}<-1$,\dots , $J_{{\rm IV}.1}<-1$, \dots), because  they are such that $\frak{C}_B< \frak{C}_T<\frak{C}_S$, if  $\frak{C}_T<\frak{C}_S$, and such that $\frak{C}_S< \frak{C}_T<\frak{C}_B$, if  $\frak{C}_S<\frak{C}_T$.  We notice that the arrangements listed in Table \ref{tab4} are those listed in Table \ref{tab3}, read from right to left. For example the arrangement IV.1 is III.4, read from left to right.

In conclusion all configurations given by Tables \ref{tab3} and \ref{tab4} correspond to field transfers expressed by hyperbolic fractional-order Fourier transformations of the first kind. The direct part of item {\its ii} (necessary condition) is proved.

\begin{table}[h]
\centering
\begin{minipage}{5cm}
\hskip .55cm
\begin{tabular}{cc}
\hline
III.1&$(V_A\;C_A\;C_B\;V_B)$\cr
III.2&$(V_B\;V_A\;C_A\;C_B)$\cr
III.3&$(C_B\;V_B\;V_A\;C_A)$\cr
III.4&$(C_A\;C_B\;V_B\;V_A)$\cr
\hline
\end{tabular}
\caption{\small Arrangements III\label{tab3}}
\end{minipage}
\hskip 1cm
\begin{minipage}{5cm}
\hskip .55cm
\begin{tabular}{cc}
\hline
IV.1&$(V_A\;V_B\;C_B\;C_A)$\cr
IV.2&$(C_A\;V_A\;V_B\;C_B)$\cr
IV.3&$(C_B\;C_A\;V_A\;V_B)$\cr
IV.4&$(V_B\;C_B\;C_A\;V_A)$\cr
\hline
\end{tabular}
\caption{\small Arrangements IV\label{tab4}}
\end{minipage}
\end{table}

\medskip
\noindent{\its iii. Hyperbolic fractional-order Fourier transformation of the second kind: $-1<J<0$.}

\begin{figure}[b]
\centering
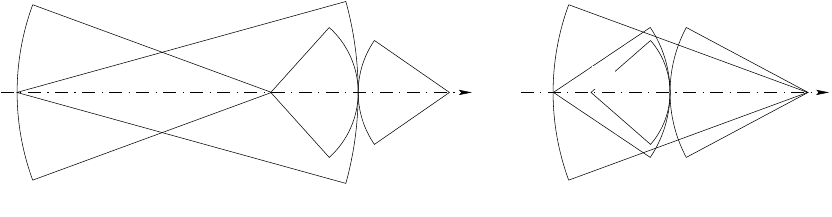
\caption{\small Two configurations to  obtain $-1<J<0$. The receiver ${\cal B}$ is not  between ${\cal S}$ and ${\cal T}$. The vertices $V_A$ and $V_B$ and the centers $C_A$ and $C_B$ are ordered along the axis according to  $(V_A  C_A V_B C_B)$ and to $(V_AC_BV_BC_A)$. \label{fig3-3}}
\end{figure}

The condition $-1<J<0$ is equivalent to
\begin{equation}
0<{\frak{C}_S-\frak{C}_B\over \frak{C}_T-\frak{C}_B}<1\,.\label{eq164}\end{equation}
Then $\frak{C}_S-\frak{C}_B$ and $\frak{C}_T-\frak{C}_B$ have the same sign. Inequality (\ref{eq164}) is obtained as follows:
\begin{itemize}
\item If $\frak{C}_S> \frak{C}_T$, then ${\frak{C}_S-\frak{C}_B> \frak{C}_T-\frak{C}_B}$. Therefore; Inequality (\ref{eq164}) requires $\frak{C}_S-\frak{C}_B<0$ and also $\frak{C}_T-\frak{C}_B<0$, implying  $\frak{C}_T< \frak{C}_S<\frak{C}_B$. An example is shown in the left diagram of Fig.~\ref{fig3-3}.
\item If $\frak{C}_S< \frak{C}_T$  then ${\frak{C}_S-\frak{C}_B< \frak{C}_T-\frak{C}_B}$. Therefore, Inequality (\ref{eq164}) requires $\frak{C}_T-\frak{C}_B>0$, so that  $\frak{C}_S-\frak{C}_B>0$, and then $\frak{C}_B< \frak{C}_S<\frak{C}_T$, as shown in the right diagram of Fig.\ \ref{fig3-3}.
\end{itemize}

Table \ref{tab5} shows the arrangements that are circular permutations of  $(V_AC_AV_BC_B)$ and Table \ref{tab6} those of  $(V_AC_BV_BC_A)$. A graphical checking proves that all these arrangements correspond to $-1<J<0$, according to the conditions mentioned above.   We notice that the arrangements listed in Table \ref{tab5} are those  listed in Table \ref{tab6}, read from right to left.

In conclusion all configurations given by Tables \ref{tab5} and \ref{tab6} correspond to field transfers represented by hyperbolic fractional-order Fourier transformations of the second kind. The direct part of item {\its iii} (necessary condition) is proved.

\begin{table}[h]
\centering
\begin{minipage}{5cm}
\hskip .55cm
\begin{tabular}{cc}
\hline
V.1&$(V_A\;C_A\;V_B\;C_B)$\cr
V.2&$(C_B\;V_A\;C_A\;V_B)$\cr
V.3&$(V_B\;C_B\;V_A\;C_A)$\cr
V.4&$(C_A\;V_B\;C_B\;V_A)$\cr
\hline
\end{tabular}
\caption{\small Arrangements V\label{tab5}}
\end{minipage}
\hskip 1cm
\begin{minipage}{5cm}
\hskip .54cm
\begin{tabular}{cc}
\hline
VI.1&$(V_A\;C_B\;V_B\;C_A)$\cr
VI.2&$(C_A\;V_A\;C_B\;V_B)$\cr
VI.3&$(V_B\;C_A\;V_A\;C_B)$\cr
VI.4&$(C_B\;V_B\;C_A\;V_A)$\cr
\hline
\end{tabular}
\caption{\small Arrangements VI\label{tab6}}
\end{minipage}

\end{table}

\medskip
\noindent{\its iv. Converse.}

Eventually, we remark that there are $4!=24$ arrangements of four objects. Tables \ref{tab1}--\ref{tab6} provide an exhaustive list of the arrangements of $\{V_A, C_A, V_B, C_B\}$. 
Then converse parts of items {\its i--iii} (sufficient conditions) are proved.

The proof of Proposition \ref{propgeo} is complete. \qed

\subsubsection{Application to the stability of optical resonators \cite{PPF6,PPF7,PPF8}}
We briefly mention some properties of optical resonators  \cite{PPF6,PPF7,PPF8}. A resonator is stable if, an only  if, the field transfer from a mirror to the other is a circular fractional-order Fourier transformation. If not, the resonator is said to be unstable. There are two kinds of unstable resonators:  resonators for which the field transfer from a mirror to the other is described by a hyperbolic fractional-order Fourier transformation of the first kind; resonators for which the field transfer is described by a hyperbolic fractional-order Fourier transformation of the second kind.

According to Proposition \ref{propgeo}, the stability of an optical resonator can be quickly determined by examinating the arrangement of vertices and centers of curvature of  the mirrors,  along the optical axis.

\subsection{Huygens--Fresnel principle}\label{sect44}
\subsubsection{Composition of impulse responses}\label{sect441}
We denote $\cal{G}_{BA}$ the operator ``field transfer'' from ${\cal A}$ to ${\cal B}$.

Let ${\cal C}$ be an intermediate spherical cap between ${\cal A}$ and ${\cal B}$. The Huygens--Fresnel principle states that the field transfer from ${\cal A}$ to ${\cal B}$ can be decomposed into a transfer from ${\cal A}$ to ${\cal C}$ followed by a transfer
from ${\cal C}$ to ${\cal B}$, that is
\begin{equation}
\cal{G}_{BA}=\cal{G}_{BC}\circ\cal{G}_{CA}\,.\label{eq2.15}\end{equation}

Equation (\ref{eq2.15}) also holds if ${\cal C}$ is outside the interval between ${\cal A}$ and ${\cal C}$. For example ${\cal C}$ may be on the left of ${\cal A}$. It is thus a virtual receiver: the field transfer from ${\cal A}$ to ${\cal C}$ is virtual (the ``distance''---algebraic measure---from ${\cal A}$ to ${\cal C}$ is negative: $\overline{V_AV_C}<0$).

Let us write Eq.\ (\ref{eq2.1}) in the form
\begin{equation}
U_B(\vec r')=\int_{{\mathbb R}^2}h_{BA}(\vec r',\vec r)\,U_A(\vec r)\,\D\vec r\,,\label{eq2.16}
\end{equation}
where
\begin{equation}
h_{BA}(\vec r',\vec r)={\I\over\lambda D}\exp\left[-{\I\pi\over\lambda}\left({\| \vec r'-\vec r\|^2\over D}+{r'^2\over R_B}-{r^2\over R_A}\right)\right]\label{eq2.17}
\end{equation}
is the impulse response of the diffraction phenomenon from ${\cal A}$ to ${\cal B}$ (it corresponds to the operator ${\cal G}_{BA}$).

If $D'$ is the algebraic measure from $V_A$ to $V_C$ ($D'=\overline{V_AV_C}$) and $\vec r''$ the spatial variable on ${\cal C}$, the field transfer from ${\cal A}$ to ${\cal C}$ is expressed in the form
\begin{equation}
U_C(\vec r'')=\int_{{\mathbb R}^2}h_{CA}(\vec r'',\vec r)\,U_A(\vec r)\,\D\vec r\,,\label{eq2.18}
\end{equation}
where
\begin{equation}
h_{CA}(\vec r'',\vec r)={\I\over\lambda D'}\exp\left[-{\I\pi\over\lambda}\left({\| \vec r''-\vec r\|^2\over D'}+{r''^2\over R_C}-{r^2\over R_A}\right)\right]\,.\label{eq2.19}
\end{equation}

Similarly, if $D''=\overline{V_CV_B}$, the field transfer from ${\cal C}$ to ${\cal B}$ takes the form
\begin{equation}
U_B(\vec r')=\int_{{\mathbb R}^2}h_{BC}(\vec r',\vec r'')\,U_C(\vec r'')\,\D\vec r''\,,\label{eq2.20}
\end{equation}
where
\begin{equation}
h_{BC}(\vec r',\vec r'')={\I\over\lambda D''}\exp\left[-{\I\pi\over\lambda}\left({\| \vec r'-\vec r''\|^2\over D''}+{r'^2\over R_B}-{r''^2\over R_C}\right)\right]\,.\label{eq2.21}
\end{equation}

According to Eq.\ (\ref{eq2.15}), we must have
\begin{eqnarray}
U_B(\vec r')\rap &=&\rap \int_{{\mathbb R}^2}h_{BC}(\vec r',\vec r'')\left(\int_{{\mathbb R}^2}\,h_{CA}(\vec r'',\vec r)\,U_A(\vec r)\,\D\vec r\right)\,\D \vec r''\nonumber \\
 \rap &=&\rap \int_{{\mathbb R}^2\times {\mathbb R}^2}\!\!\!\!\!\!\! h_{BC}(\vec r',\vec r'')
 \,h_{CA}(\vec r'',\vec r)\,U_A(\vec r)\,\D\vec r''\,\D \vec r
\,,\label{eq2.22}
\end{eqnarray}
that is
\begin{equation}
\int_{{\mathbb R}^2}h_{BA}(\vec r',\vec r)\,U_A(\vec r)\,\D\vec r=\int_{{\mathbb R}^2\times {\mathbb R}^2}\!\!\!\!\!\!\! h_{BC}(\vec r',\vec r'')
 \,h_{CA}(\vec r'',\vec r)\,U_A(\vec r)\,\D\vec r''\,\D \vec r
\,.\label{eq2.23}
\end{equation}

We prove in the  \ref{appenB} that Eq.\ (\ref{eq2.23}) holds with $h_{BA}$, $h_{BC}$, and $h_{CA}$ as in Eqs\ (\ref{eq2.17}), (\ref{eq2.21}), and (\ref{eq2.19}).

Equation (\ref{eq2.23}) gives the composition law of impulse responses $h_{BC}$ and $h_{CA}$ corresponding to the composition ${\cal G}_{BC}\circ{\cal G}_{CA}$, in accordance with the Huygens--Fresnel principle.

\subsubsection{Compositions of fractional transformations}

If field transfers by diffraction are expressed by means of fractional-order Fourier transformations, then compliance with the Huygens-Fresnel principle requires the composition of the appropriate transformations. As far as we know, only compositions involving  circular fractional Fourier transformations have been examined \cite{PPF4,Fog}. It has been shown that an additional condition, which relates to the curvature of the intermediate cap,  must also to be taken into account. The question remains whether the Weyl calculus can be applied  to compose fractional transformations of any kind in a way that removes the previous condition while still  obeying the  Huygens-Fresnel principle.  
 We will not address that issue in the present article, but will instead limit our analysis to  imaging by a refracting spherical cap, and more generally, by a centered system.

\section{Application to geometrical imaging  by a centered system}\label{sect5}

By a centered system we mean an objective lens (with foci) or an afocal system that forms an image of a spherical cap centered on its axis. The system 
consists of a sequence of refracting spherical caps that separate propagation media with distinct refractive indices (the system can also include mirrors).

In this section, based on results of Sections \ref{sect42} and \ref{sect43}, we establish the essential properties of geometrical imaging: double conjugation of vertices and centers of curvature;  longitudinal magnification for finite segments (equivalent to Bonnet's radius-magnification law);  conservation of  the class of a diffraction phenomenon;  preservation of the arrangement of luminous points on the optical axis, up to a circular permutation.

\subsection{Coherent geometrical imaging}\label{sect51}
Let us consider a centered system  (also referred as  ``a lens'' for brevity). The object is a spherical cap ${\cal A}$ and we assume that the lens  forms a coherent geometrical image ${\cal A}'$ of ${\cal A}$. By geometrical image, we mean that diffraction effects due to lens pupils are not taken into account. We also assume the lens is aberration-free. We take as an experimental fact that ${\cal A}'$ is a spherical cap in the image space (in the limits of the metaxial approximation). Finally, 
by coherent image, we mean that the field amplitude on ${\cal A}'$ (including the phase) is identical to the field amplitude on ${\cal A}$, up to a lateral-magnification factor $m_{\rm v}$, that is
\begin{equation}
U_{A'}(\vec r')={1\over m_{\rm v}}\,U_A\left({\vec r'\over m_{\rm v}}\right)\,.\end{equation}
(The factor $1/m_{\rm v}$ in front of $U_A$ is necessary for the conservation of power in the imaging process; it is such that $\int_{{\cal A}'}|U_{A'}(\vec r')|^2\,\D \vec r'=\int_{\cal A}|U_{A}(\vec r)|^2\,\D \vec r$.)

\subsection{Basic laws of coherent geometrical imaging}\label{sect52}

We begin with a vocabulary question. Let $A$ be a point light source located on the axis of a centered optical system. If we assume that the  system is free from aberrations, the image of $A$, formed by the system, is a luminous point $A'$, also located on the axis.  We now anticipate  and introduce a common term used in geometrical optics:  we say that point $A'$ is the {\its conjugate point} of $A$ through the system, and that $A$ and $A'$---understood as geometrical points, not necessarily luminous---are {\its conjugate points}. (This is precisely why one speaks of the {\its conjugation formula}.) To obtain a luminous point at $A'$ (in the image space of a lens), we must place  a luminous point at its conjugate point $A$ (in the object space).

\begin{proposition}[Longitudinal magnification for finite segments]\label{proplong} Let $V$ and $C$ be two points on the axis of a  centered system ${\cal L}$. If ${\cal L}$ is a lens with foci, points $V$ and $C$ are assumed not to lie at the foci. Let $V'$ and $C'$ be the conjugate points of $V$ and $C$ through ${\cal L}$;  let $m_{\rm  v}$ be the lateral magnification between $V$ and $V'$ and $m_{\rm c}$ the lateral magnification between $C$ and $C'$. Then, the longitudinal magnification between finite segments $\overline{VC}$ and $\overline{V'C'}$ is
\begin{equation}
m_{\rm r}={\overline{V'C'}\over \overline{VC}}={n'\over n}\,m_{\rm v}\,m_{\rm c} \,,\end{equation}
where $n$ is the refractive index of the object space and $n'$ that of the image space.
\end{proposition}

\noindent {\its Proof.}
Let ${\cal A}$ (vertex $V_A\equiv V$, center of curvature $C_A\equiv C$) be a spherical emitter in the object space of the lens ${\cal L}$ and let ${\cal F}$ be its Fourier sphere (vertex $V_F\equiv C$, center $C_F\equiv V$). Let ${\cal A}'$ and ${\cal F}'$ be their coherent images through ${\cal L}$. The distance (algebraic measure) from ${\cal A}$ to ${\cal F}$ is $D=\overline{VC}$.

From
\begin{equation}
U_{A'}(\vec r')={1\over m_{\rm v}}\,U_A\left({\vec r'\over m_{\rm v}}\right)\,,\end{equation}
we deduce (2--dimensional Fourier transformation)
\begin{equation}
\widehat U_{A'}(\vec F)=m_{\rm v}\,\widehat U_A (m_{\rm v}\vec F)\,,\end{equation}
where $\vec F$ is a spatial frequency, the conjugate\footnote{Here, ``conjugate'' refers to the duality between a vector space and its dual.} variable of $\vec r'$.

Equation (\ref{eq2.1b}) gives
\begin{equation}
U_F(\vec s)={\I\over \lambda D}\,\widehat{U}_A\left({\vec s\over \lambda D}\right)\,,\end{equation}
and since ${\cal F}'$ is the coherent image of ${\cal F}$ (their vertices are $C'$ and $C$), we have
\begin{equation}
U_{F'}(\vec s')={1\over m_{\rm c}}\,U_{F}\left({\vec s'\over m_{\rm c}}\right)
={\I\over \lambda Dm_{\rm c}}\,\widehat U_A\left({\vec s'\over \lambda D m_{\rm c}}\right)
={\I\over \lambda Dm_{\rm v}m_{\rm c}}\,\widehat U_{A'}\left({\vec s'\over \lambda D m_{\rm v} m_{\rm c}}\right)\,.\label{eq2.188}\end{equation}
Equation (\ref{eq2.188}) shows that the field amplitude on ${\cal F}'$ is proportional to the Fourier transform of the field amplitude on ${\cal A}'$. Then ${\cal F}'$ is the Fourier sphere of ${\cal A}'$, which means that $C'$ is the curvature center of ${\cal A}'$ and $V'$ the center of curvature of ${\cal F}'$. The distance from ${\cal A}'$ to ${\cal F}'$ is $D'=\overline{V'C'}$.

Since ${\cal F}'$ is the Fourier sphere of ${\cal A}'$, we may write
\begin{equation}
U_{F'}(\vec s')={\I\over \lambda 'D'}\,\widehat U_{A'}\left({\vec s'\over \lambda' D'}\right)
\,,\label{eq2.189}
\end{equation}
which is Eq.\ (\ref{eq2.1b}) applied to ${\cal A}'$ and ${\cal F}'$. 
The comparison of Eqs.\ (\ref{eq2.188}) and (\ref{eq2.189}) leads to
\begin{equation}
m_{\rm r}={\overline{V'C'}\over \overline{V'C'}}={D'\over D}={\lambda \over \lambda '}\,m_{\rm v}\,m_{\rm c}={n'\over n}\,m_{\rm v}\,m_{\rm c}\,.\end{equation}
The proof is complete. \qed

\bigskip
In the proof of Proposition \ref{proplong} we have proved the following proposition.

\begin{proposition}\label{prop10} Let ${\cal A}'$ be the image of the spherical emitter ${\cal A}$, formed by the centered system ${\cal L}$, and let ${\cal F}'$ be the image of the Fourier sphere ${\cal F}$ of ${\cal A}$. Then ${\cal F}'$ is the Fourier sphere of ${\cal A}'$. 
\end{proposition}

Since $V$ and $C$ are arbitrary points on the optical axis of ${\cal L}$, in the proof of Proposition \ref{proplong} the emitter ${\cal A}$  is an arbitrary spherical cap, so that we can deduce that coherent geometrical imaging by ${\cal L}$ is governed by the following proposition:

\begin{proposition}[Coherent geometrical imaging]\label{prop11} Let ${\cal A}'$ be the image of ${\cal A}$ through the centered system ${\cal L}$. Then
\begin{itemize}
\item {\rm Double conjugation.}  The vertex $V'$ of ${\cal A}'$ is the conjugate point of the vertex $V$ of ${\cal A}$, and the center of curvature $C'$ of  ${\cal A}'$ is the conjugate point of the center of curvature $C$ of ${\cal A}$.
\item {\rm Radius-magnification law of Bonnet}. Let $m_{\rm v}$ be the lateral magnification at vertices $V$ and $V'$, and let $m_{\rm c}$ be the lateral magnification at centers of curvature $C$ and $C'$. Then the radius magnification between ${\cal A}$ and ${\cal A}'$ is
\begin{equation}
m_{\rm r}={R_{A'}\over R_A}={\overline{V'C'}\over \overline{VC}}={n'\over n}\,m_{\rm v}m_{\rm c}\,.
\end{equation}
The radius-magnification law  is no more than the longitudinal-magnification law applied to radii of curvature of two conjugate spherical caps.
\item  Field amplitudes on ${\cal A}$ and ${\cal A}'$ are such that
\begin{equation}
U_{A'}(\vec r')={1\over m_{\rm v}}\,U_A\left({\vec r'\over m_{\rm v}}\right)\,,\end{equation}
up to a constant phase factor.
\end{itemize}
\end{proposition}

\begin{remark}[Afocal system]\label{rem51}   {\rm If ${\cal L}$ is an afocal system, the lateral magnification is the same, regardless of the position of the object. Then $m_{\rm v}=m_{\rm c}=m$ and $m_{\rm r}=(n'/n)m^2>0$, for every object ${\cal A}$.}
\end{remark}

\subsection{Imaging a diffraction phenomenon}\label{sect53}
By imaging a diffraction phenomenon we mean the following. An emitter ${\cal A}$ and a receiver ${\cal B}$ lie in the object space of a centered system ${\cal L}$.  Their images by ${\cal L}$ are ${\cal A}'$ and ${\cal B}'$. We say that the diffraction phenomenon from ${\cal A}'$ to ${\cal B}'$ is the image of the diffraction phenomenon from ${\cal A}$ to ${\cal B}$. 

\begin{theorem}[Part 1]\label{th2} Let ${\cal A}$ and ${\cal B}$ be two spherical caps in the object space of a centred system ${\cal L}$, and let ${\cal A}'$ and ${\cal B}'$ be their respective coherent images formed by ${\cal L}$. Then the field transfers from ${\cal A}$ to ${\cal B}$ and from ${\cal A}'$ to ${\cal B}'$ are  expressed by fractional-order Fourier transformations of the same kind. If these transfers are related to parameters $J$ and $J'\!$, respectively, then $J'=J$.

\end{theorem}

\noindent{\its Proof.}  We use the parameter $J$ defined in Sect.\ \ref{sect42}.  We  write
\begin{equation}
J={(R_A-D)(R_B+D)\over D(D-R_A+R_B)}={\overline{V_BC_A}\cdot\overline{V_AC_B}\over \overline{V_AV_B}\cdot\overline{C_AC_B}}\,,\end{equation}
for the field transfer from ${\cal A}$ to ${\cal B}$,
and \begin{equation} J'={\overline{V_{B'}C_{A'}}\cdot\overline{V_{A'}C_{B'}}\over \overline{V_{A'}V_{B'}}\cdot\overline{C_{A'}C_{B'}}}\,,
\end{equation}
for the field transfer from ${\cal A}'$ to ${\cal B}'$.

We denote $m_{{\rm v}A}$ the lateral magnification at vertices for caps ${\cal  A}$ and ${\cal A}'$, and $m_{{\rm c}A}$ the lateral magnification at their centers of curvature.
The radius-magnification law leads us to write
\begin{eqnarray}
J'={\overline{V_{B'}C_{A'}}\cdot\overline{V_{A'}C_{B'}}\over \overline{V_{A'}V_{B'}}\cdot\overline{C_{A'}C_{B'}}}
\rap &=&\rap {(n'/n)\cdot m_{{\rm v}B}\cdot m_{{\rm c}A}\cdot\overline{V_BC_A}\cdot (n'/n)\cdot m_{{\rm v}A} \cdot m_{{\rm c}B}\cdot \overline{V_AC_B}
\over (n'/n)\cdot m_{{\rm v}A}\cdot m_{{\rm v}B}\cdot \overline{V_AV_B}\cdot(n'/n)\cdot m_{{\rm c}A}\cdot m_{{\rm c}B}\cdot\overline{C_AC_B} }\nonumber \\
\rap &=&\rap{\overline{V_BC_A}\cdot\overline{V_AC_B}\over \overline{V_AV_B}\cdot\overline{C_AC_B}}=J\,.
\end{eqnarray}
Then we have  $J'=J>0$, or $J'=J<-1$, or $-1<J'=J<0$, and we conclude that the two transfers are of the same kind (that is, both circular, or both hyperbolic of the first kind, or both hyperbolic of the second kind, according to Definition \ref{def1}). \qed

\medskip
We complete Theorem \ref{th2} as follows.
\setcounter{theorem}{1}
\begin{theorem}[Part 2]\label{th2bis} Moreover:
\begin{enumerate}
\item[{\its i.}] Circular transformations. If the field transfers from ${\cal A}$ to ${\cal B}$ and from ${\cal A}'$ to ${\cal B}'$ are represented by ${\cal F}_\alpha$ and ${\cal F}_{\alpha '}$, then  $\alpha '=\alpha$ ; or $\alpha '=\alpha -\pi$ (with $\alpha >0$); or $\alpha '=\alpha +\pi$ (with $\alpha <0$). (We recall that both $\alpha$ and $\alpha '$ belong to $]-\pi, \pi [$.)
\item[{\its ii.}] Hyperbolic transformations of the first kind. If the field transfers are represented by ${\cal H}_{\frak{s}\beta}$ and ${\cal H}_{\frak{s'}\beta '}$, then $\frak{s'}\beta '=\frak{s}\beta$, where $\frak{s}$ is the sign of $R_A(R_A-D)$ and $\frak{s'}$ the sign of $R_{A'}(R_{A'}-D')$.
\item[{\its iii.}] Hyperbolic transformations of the second kind. If the field transfers are represented by ${\cal K}_{\frak{s}\beta}$ and ${\cal K}_{\frak{s'}\beta '}$, then $\frak{s'}\beta '=\frak{s}\beta$.
\end{enumerate}
\end{theorem}

\noindent{\its Proof.}

\noindent{\its i.} We begin with transformations ${\cal F}_\alpha$ (from ${\cal A}$ to ${\cal B}$) and ${\cal F}_{\alpha '}$ (from ${\cal A}'$ to ${\cal B}'$). If $D=\overline{V_AV_B}$ and $D'=\overline{V_{A'}V_{B'}}$,  according to Eq.\ (\ref{eq2.4}), we have
\begin{equation}
0<{DR_A\over R_A-D}\cot\alpha ={\overline{V_AV_B}\cdot\overline{V_AC_A}\over \overline{V_BC_A}}\cot\alpha\,,\label{eq222}\end{equation}
\begin{equation}
0<{D'R_{A'}\over R_{A'}-D'}\cot\alpha '={\overline{V_{A'}V_{B'}}\cdot\overline{V_{A'}C_{A'}}\over \overline{V_{B'}C_{A'}}}\cot\alpha '\,.\label{eq223}
\end{equation}
We denote $m_{{\rm v}B}$ the lateral magnification at vertices between caps ${\cal B}$ and ${\cal B}'$ and $m_{{\rm c}B}$ the lateral magnification at their centers of curvature.
Therefore
\begin{eqnarray}
{\overline{V_{A'}V_{B'}}\cdot\overline{V_{A'}C_{A'}}\over \overline{V_{B'}C_{A'}}}\rap &=&\rap
{(n'/n)m_{{\rm v}A}m_{{\rm v}B}\cdot\overline{V_AV_B}\cdot (n'/n) m_{{\rm v}A}m_{{\rm c}A}\cdot\overline{V_AC_A}
\over (n'/n) m_{{\rm v}B}m_{{\rm c}A}\cdot \overline{V_BC_A}}\nonumber \\
\rap &=&\rap
{n'\over n}\;{m_{{\rm v}A}^2\cdot\overline{V_AV_B}\cdot\overline{V_AC_A}
\over \overline{V_BC_A}}\,,
\end{eqnarray}
and Eqs. (\ref{eq222}) and (\ref{eq223}) imply  that $\cot\alpha$ and $\cot \alpha '$ have the same sign. Since $\cot^2\alpha '=\cot^2\alpha$, because $J'=J$ (Theorem \ref{th2} Part 1,), we conclude that $\cot\alpha '=\cot\alpha$.

 The longitudinal-magnification law gives
\begin{equation}D'=\overline{V_{A'}V_{B'}}={n'\over n}\;m_{{\rm v}A}m_{{\rm v}B}\overline{V_AV_B}
={n'\over n}m_{{\rm v}A}m_{{\rm v}B}D\;\,,\end{equation}
and $DD'$ has the sign of $m_{{\rm v}A}m_{{\rm v}B}$.

Moreover $\alpha D>0$ and $\alpha 'D'>0$, so that $\alpha\alpha' DD'>0$, and we conclude that $\alpha\alpha '$ has the sign of  $m_{{\rm v}A}m_{{\rm v}B}$. Since $\alpha$ and $\alpha'$ belong to $]-\pi, \pi [$, we obtain
\begin{enumerate}
\item $\alpha '=\alpha$, if  $m_{{\rm v}A}m_{{\rm v}B}>0$.
\item $\alpha '=\alpha +\pi$, if $\alpha <0$ and $m_{{\rm v}A}m_{{\rm v}B}<0$.
\item $\alpha '=\alpha -\pi$, if $\alpha >0$ and $m_{{\rm v}A}m_{{\rm v}B}<0$.
\end{enumerate}

\medskip
\noindent{\its ii.} We now consider hyperbolic transformations of the first kind, ${\cal H}_{\frak{s}\beta}$ and  ${\cal H}_{\frak{s'}\beta '}$, where $\frak{s}$ is the sign of $R_A(R_A-D)$ and $\frak{s'}$ is the sign of $R_{A'}(R_{A'}-D')$.

We have
\begin{eqnarray}
R_{A'}(R_{A'}-D')=\overline{V_{A'}C_{A'}}\cdot \overline{V_{B'}C_{A'}}
\rap &=&\rap {n'^2\over n^2} m_{{\rm v}A}m_{{\rm c}A}\overline{V_{A}C_{A}}\cdot m_{{\rm v}B}m_{{\rm c}A}\overline{V_{B}C_{A}}\nonumber \\
\rap &=&\rap  {n'^2\over n^2} m_{{\rm v}A}m_{{\rm v}B}({m_{{\rm c}A}})^2\;\overline{V_{A}C_{A}}\cdot \overline{V_{B}C_{A}}
\nonumber \\
\rap &=& \rap {n'^2\over n^2} m_{{\rm v}A}m_{{\rm v}B}({m_{{\rm c}A}})^2\;R_A(R_A-D)\,,
\end{eqnarray}
which means that $\frak{s'}$ is the sign of $m_{{\rm v}A}m_{{\rm v}B}\,\frak{s}$. The sign of $D'$ is the sign of $m_{{\rm v}A}m_{{\rm v}B}\,D$, because $D'=(n'/n)\,m_{{\rm v}A}m_{{\rm v}B}\,D$. Then the sign of  $\frak{s'}D'$ is the sign of $\frak{s}D$. Since $\beta D>0$ and $\beta 'D'>0$,  then $\frak{s'}\beta'$ has the sign of 
 $\frak{s}\beta$,  and $\coth \frak{s}\beta$ and $\coth \frak{s'}\beta '$ have the same sign. Since $\coth^2 \frak{s'}\beta '=\coth^2 \frak{s}\beta$ (because $J'=J$, Theorem \ref{th2} Part 1), we conclude that  $\coth \frak{s'}\beta '=\coth \frak{s}\beta$, that is, $\frak{s'}\beta '=\frak{s}\beta$.

\medskip
\noindent{\its iii.} For hyperbolic transformations of the second kind, ${\cal K}_{\frak{s}\beta}$ and  ${\cal K}_{\frak{s'}\beta '}$, the reasoning is the same as in item ${\its ii}$ above.  \qed

\subsection{Transforming a sequence of points on the optical axis}\label{sect54}

\begin{theorem}\label{th3} Let ${\cal L}$ be a centered system and let $A_1$, $A_2$, $A_3$, and $A_4$ be four points, in the object space, sequentially arranged along the system axis according to $(A_1A_2A_3A_4)$ (taken in the direction of light propagation). Then, in the image space, their respective conjugate points  $A'_i$  under ${\cal L}$ are sequentially arranged   along the axis according to  $(A'_1A'_2A'_3A'_4)$, up to a circular permutation.
\end{theorem}

\begin{table}[b]
\vskip -.15cm
\centering
\begin{tabular}{ccccccccc}
\hline
\rule{0mm}{3.7mm}\par
&Arrangement & $\left.\overline{V_AV_B}\right.$ &$\overline{V_AC_A}$ & $\overline{V_BC_A}$ & $\cot\alpha$ & $\alpha $ &  Range of $\alpha$ & $\overline{V_BC_B}$\cr
\hline
I.1&$(V_A\;C_B\;C_A\;V_B)$& $>0$ & $>0$ & $<0$& $ <0$ & $>0$ & $\pi /2<\alpha <\pi$& $<0$\cr
I.2&$(V_B\;V_A\;C_B\;C_A)$ & $<0$& $>0$ & $>0$ & $<0$ &  $<0$ & $-\pi /2<\alpha <0$ & $>0$ \cr
I.3&$(C_A\;V_B\;V_A\;C_B)$ & $<0$ & $<0$ & $<0$ & $<0$  &$<0$ &$-\pi /2<\alpha <0$ & $>0$\cr
I.4&$(C_B\;C_A\;V_B\;V_A)$ & $<0$ &$<0$ & $<0$ & $<0$ &  $<0$ & $-\pi /2<\alpha <0$ & $<0$\cr
II.1 & $(C_B\;V_A\;V_B\;C_A)$ & $>0$ &  $>0$ & $>0$ & $>0$ &  $>0$ & $0<\alpha <\pi /2$ & $<0$\cr
II.2 & $(C_A\;C_B\;V_A\;V_B)$ & $>0$ &  $<0$ & $<0$ & $>0$ &  $>0$ & $0<\alpha <\pi /2$ & $<0$\cr
II.3 & $(V_B\;C_A\;C_B\;V_A)$ & $<0$ &  $<0$ & $>0$ & $>0$ &  $<0$ & $-\pi <\alpha <-\pi /2$ & $>0$\cr
II.4 & $(V_A\;V_B\;C_A\;C_B)$ & $>0$ &  $>0$ & $>0$ & $>0$ &  $>0$ & $0<\alpha <\pi /2$ & $>0$\cr
\hline
\end{tabular}
\caption{\small Determination of the sign and range of $\alpha$ for the arrangements corresponding to field transfers described by circular fractional Fourier transformations. The sign of  $\alpha$ is given by  $\alpha \,\overline{V_AV_B}>0$, and the sign of $\cot\alpha$ by Eq. (\ref{eq227}). The column on the right, which gives the sign of $\overline{V_BC_B}$, is useful for analysing afocal systems (see Prop.\ \ref{propofocal}). 
The table is also used for determining the sign and range of $\alpha '$, after  $V_A$, $C_A$, $V_B$ and $C_B$ have been replaced by  $V_{A'}$, $C_{A'}$,  $V_{B'}$ and  $C_{B'}$. In the imaging process of  a centered system, the orders $\alpha$ and $\alpha '$ must comply with Theorem \ref{th2} Part 2.
\label{tab7}}
\end{table}

\noindent{\its Proof.} We will consider points $A_i$ as vertices $V_A$ and $V_B$ and centers  of curvature $C_A$ and $C_B$ of an emitter ${\cal A}$ and a receiver ${\cal B}$ (the arrangement of vertices and centers is not yet specified at this point).  Due to the  double-conjugation law (Prop.\ \ref{prop11}), the conjugate points $A'_i$  constitute the vertices ($V_{A'}$ and $V_{B'}$)  and centers of curvature  ($C_{A'}$ and $C_{B'}$) of an emitter ${\cal A}'$ and a receiver ${\cal B}'$  that are the coherent images of ${\cal A}$ and ${\cal B}$ under ${\cal L}$.

\smallskip
\noindent{\its i.}
Let us assume first that, in the object space, the  configuration is that of the left diagram in Fig.\ \ref{fig3-1}, i.e.  $(A_1A_2A_3A_4)\equiv (V_AC_BC_AV_B)$.
We have to prove that, in the image space, the arrangement of the points $A'_i$ 
is $(A'_1A'_2A'_3A'_4)$, up to a circular permutation, with $(A'_1A'_2A'_3A'_4)\equiv (V_{A'}C_{B'}C_{A'}V_{B'})$. 

According to Proposition \ref{propgeo}, the field transfer from ${\cal A}$ to ${\cal B}$ is described by a circular fractional Fourier transform; its order is $\alpha$. And according to Theorem \ref{th2}, the field transfer from ${\cal A}'$ to ${\cal B}'$ is also expressed by a circular fractional Fourier transform; its  order is $\alpha '$. The orders $\alpha$ and $\alpha '$ are connected according to Theorem \ref{th2} Part 2.

In Table \ref{tab7}, we examine the various possibilities for the sign and domain of $\alpha$, according to the considered arrangement. The sign of $\cot\alpha$ is given by Eq.\ (\ref{eq2.4}), that is,
\begin{equation}
0<{DR_A\over R_A-D}\cot\alpha ={\overline{V_AV_B}\cdot \overline{V_AC_A}\over \overline{V_BC_A}}\cot\alpha\,,\label{eq227}\end{equation}
and the sign of $\alpha$ by $0<\alpha D =\alpha \,\overline{V_AV_B}$.

Table \ref{tab7} is also used to determine the sign and domain of $\alpha '$, after replacing $V_A$, $C_A$, $V_B$ and $C_B$ by $V_{A'}$, $C_{A'}$,  $V_{B'}$ and  $C_{B'}$, respectively. We denote I$'.j$ the arrangement derived from I.$j$ by changing $V_A$ to   $V_{A'}$, $C_A$ to $C_{A'}$ and so on.

The arrangements ${\rm I}.1=(V_AC_BC_AV_B)$ and  I$'.\ell$ ($\ell =1,2,3,4$) are compatible with Theorem \ref{th2} Part 2, that is, with $\alpha =\alpha '$,  or $\alpha '=\alpha -\pi$. For example, let us consider I.1  and  ${\rm I}'.2$. The arrangement I.1 corresponds to $\pi /2<\alpha <\pi$, and   ${\rm I}'.2=(V_{B'} V_{A'} C_{B'} C_{A'})$ to   $-\pi /2<\alpha ' <0$, so that the values of $\alpha$ and $\alpha '$ make it possible to achieved the condition $\alpha '=\alpha -\pi$, in accordance with Theorem \ref{th2}. That means that imaging a diffraction phenomenon for which the  arrangement of vertices and centers is ${\rm I}.1=(V_AC_BC_AV_B)$ may lead to the arrangement ${\rm I}'.2=(V_{B'} V_{A'} C_{B'} C_{A'})$, in the image space, which actually occurs when $m_{{\rm v}A}\,m_{{\rm v}B} <0$, $m_{{\rm v}A}\,m_{{\rm c}A} >0$ and  $m_{{\rm v}B}\,m_{{\rm c}A} >0$.

On the other hand,  the arrangement I.1 cannot be transformed into an arrangement II$'$ by imaging, because it  corresponds to $\pi /2 <\alpha <\pi$,  whereas  the arrangements II$'$ correspond to either $0<\alpha '<\pi /2$ or $-\pi <\alpha ' < -\pi /2$. Thus, we can obtain neither $\alpha '=\alpha$ nor $\alpha '=\alpha -\pi$ (since  $\alpha >0$). This means that transforming  arrangement I.1 into an arrangement II$'$ by imaging would violate Theorem \ref{th2} Part 2.

In conclusion, we have proved that, by imaging, the arrangement I.1 can be transformed into an  arrangement ${\rm I}'.\ell$, not into an arrangement ${\rm II}'.\ell$. This means that if $(A_1A_2A_3A_4)\equiv(V_AC_BC_AV_B)$, then the conjugate points $A'_i$ are arranged according to $(A'_1A'_2A'_3A'_4)$, up to a circular permutation, with $A'_1\equiv V_{A'}$,  $A'_2\equiv C_{B'}$, $A'_3\equiv C_{A'}$ and $A'_4\equiv V_{B'}$.

\medskip
\noindent{\its ii.} The same reasoning can be applied  to  $(A_1A_2A_3A_4)\equiv(C_BV_AV_BC_A)={\rm II}.1$. It follows that the arrangement II.1 can be transformed into ${\rm II}'.\ell$ ($\ell =1,2,3,4$)  by imaging, but not into  ${\rm I}'.\ell$.

More generally, comparison of the ranges of $\alpha$ and $\alpha '$ in Table \ref{tab7}  shows that each arrangement ${\rm I}.k$ can be transformed into an arrangement  ${\rm I}'.\ell$ by imaging, but not into an arrangement ${\rm II}'.\ell$, in accordance with Theorem \ref{th2}. Conversely, each arrangement ${\rm II}.k$ can be transformed into an arrangement  ${\rm II}'.\ell$, but not into an arrangement ${\rm I}'.\ell$. Since each ${\rm I}.k$ is a circular permutation of ${\rm I}.\ell$ for every $\ell$, it follows that, for every $k$ ($k=1,2,3,4$), if $(A_1A_2A_3A_4)\equiv{\rm I}.k$, the conjugate points $A'_j$ are arranged according to $(A'_1A'_2A'_3A'_4)$, up to a circular permutation. The same result is valid if  $(A_1A_2A_3A_4)\equiv{\rm II}.k$.

\medskip
\noindent{\its iii.} We now assume that the arrangements of vertices $V_A$ and $V_B$ and centers of curvature $C_A$ and $C_B$ that correspond to field transfers by diffraction from ${\cal A}$ to ${\cal B}$, described by hyperbolic fractional Fourier transformations of the first kind and order $\frak{s}\beta$, according to Proposition \ref{propgeo}. They are the arrangemets III$.k$ and IV$.k$  listed in Table \ref{tab8}.

According to Eq.\ (\ref{eq2.7}), for every arrangement III or IV, the sign of $\beta$ (which is also the sign of $\coth\beta$) is given by $\beta D>0$, that is, by $\beta\,\overline{V_AV_B}>0$. The sign of $\frak{s}$ is the sign of $R_A(R_A-D)$, that is, the sign of $\overline{V_AC_A}\cdot\overline{V_BC_A}$. The sign of  $\frak{s}\beta$ is then deduced.

According to Theorem \ref{th2}, the field transfer from ${\cal A}'$ to ${\cal B}'$, the coherent images of ${\cal A}$ and ${\cal B}$ formed by ${\cal L}$, is also described by a hyperbolic fractional Fourier transformation of the first kind, whose order is $\frak{s'}\beta '$, with $\frak{s'}\beta '=\frak{s}\beta $.
Table \ref{tab8} also provides the sign of  $\frak{s'}\beta '$, after $V_A$, $C_A$, $V_B$ and $C_B$ have been changed to $V_{A'}$,  $C_{A'}$,  $V_{B'}$ and  $C_{B'}$.

We note that the condition  $\frak{s'}\beta '=\frak{s}\beta $ can be achieved only if an arrangement ${\rm III}.k$ is transformed into an arrangement ${\rm III}'.\ell$ by imaging, or if an arrangement ${\rm IV}.k$ is transformed into an arrangement ${\rm IV}'.\ell$.

\begin{table}[h]
\centering
\begin{tabular}{ccccccccc}
\hline
\rule{0mm}{3.7mm}
&Arrangement & $\overline{V_AV_B}$ & $\overline{V_AC_A}$ & $\overline{V_BC_A}$& $\frak{s}$ &$\coth\beta $ & $\frak{s}\beta$ & $\overline{V_BC_B}$ \cr
\hline
III.1&$(V_A\;C_A\;C_B\;V_B)$& $>0$ & $>0$ & $<0$& $ <0$ & $>0$ &$<0$  & $<0$  \cr
III.2&$(V_B\;V_A\;C_A\;C_B)$& $<0$ & $>0$ & $>0$& $ >0$ & $<0$ &$<0$  & $>0$ \cr
III.3&$(C_B\;V_B\;V_A\;C_A)$& $<0$ & $>0$ & $>0$& $ >0$ & $<0$ &$<0$  & $<0$\cr
III.4&$(C_AC_B\;V_B\;V_A)$& $<0$ & $<0$ & $<0$& $ >0$ & $<0$ &$<0$    & $<0$\cr
IV.1&$(V_A\;V_B\;C_B\;C_A)$& $>0$ & $>0$ & $>0$& $ >0$ & $>0$ & $>0$  & $>0$\cr
IV.2&$(C_A\;V_A\;V_B\;C_B)$& $>0$ & $<0$ & $<0$& $ >0$ & $>0$ & $>0$  & $>0$ \cr
IV.3&$(C_B\;C_A\;V_A\;V_B)$& $>0$ & $<0$ & $<0$& $ >0$ & $>0$ & $>0$  & $<0$\cr
IV.4&$(V_B\;C_B\;C_A\;V_A)$& $<0$ & $<0$ & $>0$& $ <0$ & $<0$ &$>0$   & $>0$\cr
\hline
\end{tabular}
\caption{\small Determination of the sign of $\frak{s}\beta$ for the arrangements corresponding to field transfers described by hyperbolic fractional Fourier tranformations of the first kind. The table  also provides the sign of  $\frak{s'}\beta '$, after $V_A$, $C_A$, $V_B$ and $C_B$ have been changed to $V_{A'}$,  $C_{A'}$,  $V_{B'}$ and  $C_{B'}$. Since $\frak{s}'\beta '=\frak{s}\beta$ by imaging under a centered system, the sign of $\frak{s}\beta$ is preserved; we deduce that an arrangement ${\rm III}.k$ can only be transformed into an arrangement ${\rm III}'.\ell$ by imaging, and an arrangement ${\rm IV}.k$ into an arrangement ${\rm IV}'.\ell$.
\label{tab8}}
\end{table}

\bigskip
\noindent{\its iv.} The arrangements of $V_A$, $C_A$, $V_B$ and $C_B$ for which the field transfers from ${\cal A}$ to ${\cal B}$ correspond to hyperbolic fractional Fourier transformation of the second kind are arrangements V and VI listed in Table \ref{tab9}. The analysis is exactly that of item {\its iii} above. The result is that, by imaging under an optical system, arrangements $V$ are transformed into arrangements V$'$, and arrangements VI into arrangements VI$'$. 

\medskip
The theorem is proved. \qed

\begin{table}
\centering
\begin{tabular}{ccccccccc}
\hline\rule{0mm}{3.7mm}
&Arrangement & $\overline{V_AV_B}$ & $\overline{V_AC_A}$ & $\overline{V_BC_A}$& $\frak{s}$ &$\coth\beta $ & $\frak{s}\beta$ & $\overline{V_BC_B}$ \cr
\hline
V.1&$(V_A\;C_A\;V_B\;C_B)$& $>0$ & $>0$ & $<0$& $ <0$ & $>0$ &$<0$   & $>0$\cr
V.2&$(C_B\;V_A\;C_A\;V_B)$& $>0$ & $>0$ & $<0$& $ <0$ & $>0$ &$<0$   & $<0$ \cr
V.3&$(V_B\;C_B\;V_A\;C_A)$& $<0$ & $>0$ & $>0$& $ >0$ & $<0$ &$<0$   & $>0$\cr
V.4&$(C_AV_B\;C_B\;V_A)$& $<0$ & $<0$ & $<0$& $ >0$ & $<0$ &$<0$     & $>0$\cr
VI.1&$(V_A\;C_B\;V_B\;C_A)$& $>0$ & $>0$ & $>0$& $ >0$ & $>0$ & $>0$ & $<0$\cr
VI.2&$(C_A\;V_A\;C_B\;V_B)$& $>0$ & $<0$ & $<0$& $ >0$ & $>0$ & $>0$ & $<0$\cr
VI.3&$(V_B\;C_A\;V_A\;C_B)$& $<0$ & $<0$ & $>0$& $ <0$ & $<0$ & $>0$ & $>0$\cr
VI.4&$(C_B\;V_B\;C_A\;V_A)$& $<0$ & $<0$ & $>0$& $ <0$ & $<0$ &$>0$  & $<0$ \cr
\hline
\end{tabular}
\caption{\small Determination of the sign of $\frak{s}\beta$ for the arrangements corresponding to field transfers described by hyperbolic fractional Fourier tranformations of the second  kind.  \label{tab9}}
\end{table}

\bigskip
Theorem \ref{th3} takes a specific form when the optical system ${\cal L}$ is afocal, as shown by the following proposition.

\begin{proposition}\label{propofocal} Let ${\cal L}$ be an afocal system.
Let $A_1$, $A_2$, $A_3$, and $A_4$ be four points, in the object space,  sequentially arranged along the system axis according to $(A_1A_2A_3A_4)$. In the image space,  their respective conjugate points $A'_i$ under ${\cal L}$ are sequentially arranged according to $(A'_1A'_2A'_3A'_4)$.
\end{proposition}

\noindent{\its Proof}. According to Remark \ref{rem51}, the longitudinal magnification $m_{\rm r}$ is a strictly positive constant for every position of the object. Then, for every $i$ and $j$ in $\{ 1,2,3,4\}$  ($i\ne j$), we have $\overline{A'_iA'_j}=m_{\rm r}\,\overline{A_iA_j}$, so that $\overline{A_iA_j}$ and $\overline{A'_iA'_j}$ have the same sign.

We associated points $A_i$ with the vertices $V_A$ and $V_B$ and the centers of curvature $C_A$ and $C_B$ of an emitter ${\cal A}$ and a receiver ${\cal B}$. Then points $A'_j$ are the vertices and centers of curvature of the images ${\cal A}'$ and ${\cal B}'$ of ${\cal A}$ and ${\cal B}$.

Let us assume that the field transfer from ${\cal A}$ to ${\cal B}$ is represented by a circular fractional Fourier transform of order $\alpha$, so that the field transfer from ${\cal A}'$ to ${\cal B}'$ is represented by a circular fractional Fourier transform of order $\alpha '$. If $(A_1A_2A_3A_4)={\rm I}.k$, then $(A'_1A'_2A'_3A'_4)={\rm I}'.\ell$, according to Theorem~\ref{th3}. Since $\overline{A_iA_j}$ and $\overline{A'_iA'_j}$ have the same sign, because ${\cal L}$ is an afocal system, we conclude that necessarily: (a)  $\overline{V_AV_B}$ and $\overline{V_{A'}V_{B'}}$ have the same sign; (b)   $\overline{V_AC_A}$ and $\overline{V_{A'}C_{A'}}$ have the same sign; (c)   $\overline{V_BC_A}$ and $\overline{V_{B'}C_{A'}}$ have the same sign;  (d)   $\overline{V_BC_B}$ and $\overline{V_{B'}C_{B'}}$ have the same sign. If we examine Table \ref{tab7}, we conclude that  if $(A_1A_2A_3A_4)={\rm I}.k$ ($k=1,2,3,4$), then $(A'_1A'_2A'_3A'_4)={\rm I}'.k$, which means that the points $A'_i$ are arranged in the same way as points $A_i$. That is Proposition \ref{propofocal}. The same reasoning shows that  $(A'_1A'_2A'_3A'_4)={\rm II}'.k$, if $(A_1A_2A_3A_4)={\rm II}.k$.

The same result is obtained from examining Tables \ref{tab8} and \ref{tab9} when field transfers from ${\cal A}$ to ${\cal B}$ and from ${\cal A}'$ to ${\cal B}'$  are described by hyperbolic fractional Fourier transformations of the first and second kind. The proof is complete. \qed

\bigskip
So far, we have used four points on the optical axis of the lens to introduce two spherical caps, that is, two vertices and two centers of curvature. The following theorem, concerning  three points, is equivalent to Theorem \ref{th3}.

\setcounter{theorem}{2}
\begin{theorem}\label{th3prime}$\!\!\!${\bf $'$} Let $A_1$, $A_2$ and $A_3$ be three points on the optical axis of a centered system ${\cal L}$, ordered accor\-ding to $(A_1A_2A_3)$, in the direction of light propagation. Their conjugate points under ${\cal L}$ are arranged according to $(A'_1A'_2A'_3)$, up to a circular permutation.
\end{theorem}

\noindent{\its Proof.}

\noindent {\its i. Direct implication: Theorem \ref{th3} implies Theorem \ref{th3}$\,'$.} We introduce an additional point $B$, so that points $A_j$ and $B$ are ordered according to $(A_1A_2A_3B)$. According to Theorem  \ref{th3}, the order of their conjugate points is a circular permutation of $(A'_1A'_2A'_3B')$, as given on the left part  of Table \ref{tab10}.   Then we read the table ignoring point $B$ (right part of Table \ref{tab10}) and we obtain that $A'_1$, $A'_2$, $A'_3$ are arranged according to a circular permutation of  $(A'_1A'_2A'_3)$.

\begin{table}[h]
\centering
\begin{tabular}{cc}
Four points & Ignoring $B'$\cr
\hline
$(A'_1\;A'_2\;A'_3\;B')$& $(A'_1\;A'_2\;A'_3)$ \cr
$(B'\;A'_1\;A'_2\;A'_3)$& $(A'_1\;A'_2\;A'_3)$\cr
$(A'_3\;B'\;A'_1\;A'_2)$& $(A'_3\;A'_1\;A'_2)$\cr
$(A'_2\;A'_3\;B'\;A'_1)$& $(A'_2\;A'_3\;A'_1)$\cr
\hline
\end{tabular}
\caption{\small The arrangements of $(A'_1A'_2A'_3)$ are ``sub-arrangements'' of $(A'_1A'_2A'_3 B')$.\label{tab10}}
\end{table}

\noindent {\its ii. Converse implication.} We assume that the conjugate points $P'$, $Q'$ and $R'$ of three arbitrary points $P$, $Q$ and $R$ (on the lens axis)  are arranged according to a circular permutation of $(P'Q'R')$.
Let $A_1$, $A_2$, and $A_3$ be arranged in the object space as $(A_1A_2A_3)$, and such that their conjugate points $A'_1$, $A'_2$ and $A'_3$ are arranged according to $(A'_1 A'_2A'_3)$, $(A'_3 A'_1A'_2)$  or $(A'_2 A'_3A'_1)$. Let $B$ be in the object space, such that we have the arrangement $(A_1A_2A_3B)$. 
Then $B'$ cannot be between $A'_2$ and $A'_3$ (taken is this order), because $(A'_2B'A_3)$ is not a circular permutation of $(A_2A_3B)$ (if $B'$ were between  $A'_2$ and $A'_3$, it would be inconsistent with the assumption made above regarding arbitrary points $P$, $Q$, $R$).
And $B'$ cannot be between $A'_1$ and $A'_2$ (in this order), because $(A'_1B'A_2)$ is not a circular permutation of $(A'_1A'_2B')$. The only possibilities are  those given in Table \ref{tab10} (left part). Then the conjugate points of $A_1$, $A_2$, $A_3$, $B$, taken in this order, form the arrangement $(A'_1A'_2A'_3 B')$, up to a circular permutation; this is Theorem \ref{th3}.

The proof is complete. \qed

\begin{remark}\label{rem52}{\rm If the optical system is afocal, Theorem \ref{th3}$'$ takes a specific form, according to Proposition \ref{propofocal}: the conjugate points $A'_1$, $A'_2$, and $A'_3$ under ${\cal L}$ are necessarily arranged according to $(A_1A_2A_3)$. 
}
\end{remark}

\section{Imaging as result of the composition of two fractional-order transformations}\label{sect6}

In this section we deduce the laws of coherent geometrical imaging by a refracting spherical cap from the composition of two fractional-order Fourier transformations, considered as Weyl pseudo-diifferential operators. One transformation is associated with the field transfer by diffraction from the emitter to the  refracting cap; the other with the field transfer from the refracting cap to the image. An important consequence of Theorem \ref{th1} is that only compositions of fractional Fourier transformations of the same kind have to be taken into account. The findings are extended to an arbitrary optical centered system.

\subsection{Condition for a refracting spherical cap to form an image}\label{sect61}

\begin{figure}
  \centering
    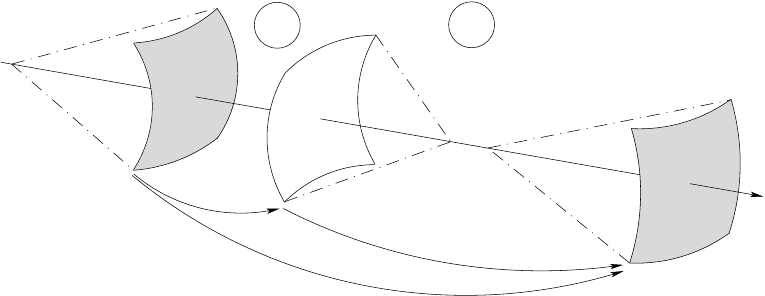
    \caption{\small The field transfer from ${\cal A}$ to ${\cal A}'$ is the composition of the field transfers from ${\cal A}$ to ${\cal D}$ and 
    from ${\cal D}$ to ${\cal A}'$: ${\cal G}_{A'A}={\cal G}_{A'D}\circ{\cal G}_{DA}$. The spherical cap ${\cal A}'$ is the coherent image of ${\cal A}$ through  the spherical refracting cap ${\cal D}$ if ${\cal G}_{A'A}$ is the identity or the parity operator, up to a scaling factor.   \label{fig2bis}}
\end{figure}

We refer to Fig.\ \ref{fig2bis}, which represents a refracting spherical cap ${\cal D}$ (vertex $V_D$, center of curvature $C_D$, radius of curvature $R_D=\overline{V_DC_D}$). The refractive indices are $n$ (object space) and $n'$ (image space). The object is the spherical cap ${\cal A}$ and its coherent image is ${\cal A}'$. We recall that by ``coherent image'' we mean that the field amplitude on ${\cal A}'$ is equal to the field amplitude on ${\cal A}$, up to a lateral magnification $m_{\rm v}$, that is
\begin{equation}
U_{A'}(\vec r')={1\over m_{\rm v}}\,U_A\left({\vec r'\over m_{\rm v}}\right)\,.\end{equation}

The field transfer from ${\cal A}$ to its image can decompose into the field transfer ${\cal G}_{DA}$ from ${\cal A}$ to ${\cal D}$ and the field transfer ${\cal G}_{A'D}$ from ${\cal D}$ to ${\cal A}'$. If both transfer are expressed by fractional-order Fourier transformations, their composition is expressed by the product of the corresponding transformations and the result should be the identity or the parity operator.

The field transfer from ${\cal A}$ to ${\cal D}$ can be expressed by a circular fractional Fourier transformation ${\cal F}_\alpha$, or a hyperbolic fractional transformation of the first kind ${\cal H}_\beta$ or of second kind ${\cal K}_\beta$. The same thing holds for the field transfer from ${\cal D}$ to ${\cal A}'$ with transformations ${\cal F}_{\alpha '}$, ${\cal H}_{\beta '}$ or ${\cal K}_{\beta '}$. A priori, the possible compositions are
\begin{enumerate}
\item ${\cal F}_{\alpha'}\circ{\cal F}_\alpha$ ; ${\cal H}_{\beta '}\circ{\cal H}_\beta$ ; ${\cal K}_{\beta '}\circ{\cal K}_\beta$ ;
\item  ${\cal H}_{\beta '}\circ {\cal F}_\alpha$ ;  ${\cal K}_{\beta '}\circ {\cal F}_\alpha$ ; ${\cal F}_{\alpha'}\circ {\cal H}_\beta$, ${\cal F}_{\alpha'}\circ {\cal K}_\beta$ ;
\item  ${\cal K}_{\beta '}\circ{\cal H}_\beta$ ; ${\cal H}_{\beta '}\circ{\cal K}_\beta$.
\end{enumerate}

An important consequence of Theorem  \ref{th1} is that only the three compositions listed in  item 1 above can be multiple of the identity or the parity operator. Then only these three cases have to be considered when dealing with imaging through a refracting cap: we have to compose fractional-order Fourier transformations of the same kind.

An ``optical'' version of Theorem \ref{th1} is:
\begin{theorem}\label{th4} If image formation by a refracting spherical cap is mathematically obtained by composing two fractional-order Fourier transformations, those transformations necessarily are of the same kind (apart from trivial cases).
\end{theorem}

\subsection{Imaging by a refracting spherical cap by composition of two circular fractional-order Fourier transformations}\label{sect62}

Imaging obtained by composing two hyperbolic fractional-order Fourier transformations of the same kind has been derived in a recent article \cite{PPF5}. To complete the previous publication, in the present article we will limit ourselves to the composition of two circular transformations.

\subsubsection{Composition of transformations. Conjugation law}

\begin{figure}
  \centering
    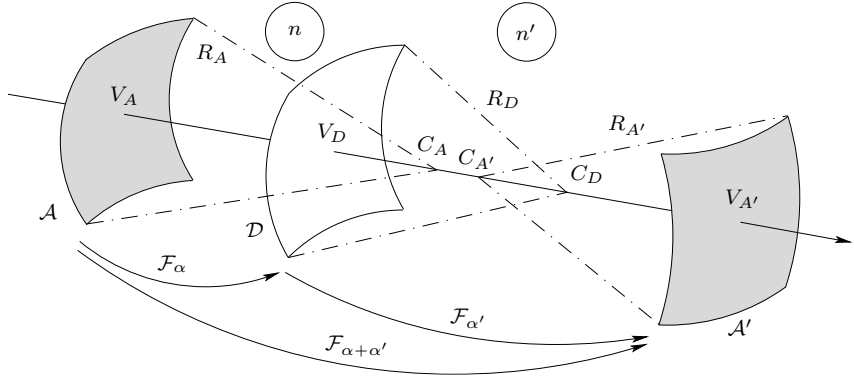
    \caption{\small The field transfer from ${\cal A}$ to ${\cal A}'$ is the composition of the field transfers from ${\cal A}$ to ${\cal D}$ and  from ${\cal D}$ to ${\cal A}'$ and is represented by ${\cal F}_{\alpha '}\circ{\cal F}_{\alpha}={\cal F}_{\alpha +\alpha '}$. The spherical cap ${\cal A}'$ is the coherent image of ${\cal A}$ through  the spherical refracting cap ${\cal D}$ if $\alpha +\alpha '=0$, or $\alpha +\alpha '=\pi$.   \label{fig2}}
\end{figure}

The configuration is shown in Fig.\ \ref{fig2}. 
Spatial variables are $\vec r$ on ${\cal A}$, $\vec s$ on ${\cal D}$ and $\vec r'$ on ${\cal A}'$. Field amplitudes are $U_A$, $U_D$ and $U_{A'}$.

We use notation of geometrical optics, that is, we use $d=-D=\overline{V_DV_A}$ for the distance from ${\cal D}$ to ${\cal A}$, and $d'=D'=\overline{V_DV_{A'}}$ for the distance from ${\cal D}$ to ${\cal A}'$.

Since $D=-d$, the parameter $J$ associated with the field transfer from ${\cal A}$ to ${\cal D}$  is
\begin{equation}
  J={(R_A+D)(R_D-d)\over d(d+R_A-R_D)}\,,
\end{equation}
and we assume $J>0$.  The order of the corresponding circular fractional-order Fourier transformation is $\alpha$, such that
\begin{equation}
 \cot^2\alpha =J\,,\hskip .7cm  \alpha d<0\,,\hskip .7cm  -\pi <\alpha <\pi\,, \hskip .7cm {-R_A d\over R_+dA}\cot\alpha >0\,.\end{equation}
We introduce
\begin{equation}
  \varepsilon_A={-d\over R_A+d}\cot\alpha\,,\hskip .5cm \mbox{and}\;\;\;  \varepsilon_D={-d\over R_D-d}\cot\alpha\,.
\end{equation}
Reduced variables on ${\cal A}$ and ${\cal D}$ are
\begin{equation}
  \vec \rho={\vec r\over\sqrt{\lambda \varepsilon_AR_A}}\,,\hskip 1cm
  \vec \sigma={\vec s\over\sqrt{\lambda \varepsilon_DR_D}}\,,\label{eq187}
\end{equation}
and the reduced field amplitudes are
\begin{equation}
  u_A(\vec \rho)=\sqrt{\lambda\varepsilon_AR_A}\,U_A\left(\sqrt{\lambda \varepsilon_AR_A}\,\vec \rho\right)\,,
  \hskip 1cm
  u_D(\vec \sigma)=\sqrt{\lambda\varepsilon_DR_D}\,U_D\left(\sqrt{\lambda \varepsilon_DR_D}\,\vec \sigma\right)\,.
\end{equation}
According to Eq. (\ref{eq2.161}), the field transfer from ${\cal A}$ to ${\cal D}$ takes the form
\begin{equation}
  u_D=\E^{\I\alpha}\,{\cal F}_\alpha[u_A]\,.\end{equation}

For the field transfer from ${\cal D}$ to ${\cal A}'$, we have
\begin{equation}
  J'={(R_D-d')(R_{A'}+d')\over d'(d'-R_D+R_{A'})}\,,\end{equation}
and we assume $J'>0$, so that the field transfer from ${\cal D}$ to ${\cal A}'$ is a circular fractional Fourier transformation whose order is $\alpha '$, such that
  \begin{equation}
\cot^2\alpha '=J'\,,\hskip .7cm \alpha ' d\primespe>0\,,\hskip .7cm -\pi <d\primespe <\pi\,,\hskip .7cm {R_Dd'\over R_D-d'}\cot\alpha >0\,.\end{equation} 
  Auxiliary parameters are
  \begin{equation}
    \varepsilon'_D={d'\over R_D-d'}\cot\alpha '\,,\hskip 1cm  \varepsilon_{A'}={d'\over R_{A'}+d\primespe}\cot\alpha '\,.\end{equation}
  Reduced variables on ${\cal D}$ and ${\cal A}'$ are
  \begin{equation}
     \vec \sigma '={\vec s\over\sqrt{\lambda '\varepsilon'_DR_D}}\,, \hskip 1cm
  \vec \rho '={\vec r'\over\sqrt{\lambda '\varepsilon_{A'}R_{A'}}}\,,\label{eq193}
\end{equation}
and the reduced field amplitudes are
\begin{equation}
   u'_D(\vec \sigma ')=\sqrt{\lambda '\varepsilon'_DR_D}\,U_D\left(\sqrt{\lambda \varepsilon'_DR_D}\,\vec \sigma \right)\,,\hskip .3cm
  u_{A'}(\vec \rho ')=\sqrt{\lambda '\varepsilon_{A'}R_{A'}}\,U_{A'}\left(\sqrt{\lambda  '\varepsilon_{A'}R_{A'}}\,\vec \rho\right)\,.
\end{equation}
The field transfer from ${\cal D}$ to ${\cal A}'$ takes the form
\begin{equation}
  u_{A'}=\E^{\I\alpha '}\,{\cal F}_{\alpha '}[u'_D]\,.\end{equation}

The composition of ${\cal F}_\alpha$ and ${\cal F}_{\alpha '}$ makes sense if $u'_D\equiv u_D$, that is, if reduced variables $\vec \sigma$ and $\vec \sigma '$ on ${\cal D}$ are identical, so that $u'_D(\vec \sigma )=u_D(\vec\sigma )$.
Then we must have $\lambda \varepsilon_D=\lambda '\varepsilon'_D$. Since  $n\lambda =n'\lambda '$, we obtain $\varepsilon_D/n=\varepsilon'_D/n'$, that is
\begin{equation}
  -{1\over n}\,{d\over R_D-d}\,\cot\alpha ={1\over n'}\,{d'\over R_D-d'}\,\cot\alpha '\,.
\end{equation}

The spherical cap ${\cal A}'$ is the image of ${\cal A}$ if $\alpha +\alpha '=0$, or if $\alpha +\alpha '=\pm\pi$. In both cases we have $\cot\alpha '=-\cot\alpha$ so that
  \begin{equation}
    n{R_D-d\over R_Dd}= n'{R_D-d'\over R_Dd'}\,,\label{eq196}
  \end{equation}
  which is usually written in the form
  \begin{equation}
    {n'\over d'}={n\over d}+{n'-n\over R_D}\,.\label{eq219}\end{equation}
Equation (\ref{eq219}) is the conjugation formula of the refracting spherical cap. It indicates that vertices of ${\cal A}$ and ${\cal A}'$ are conjugate points.

\subsubsection{Conjugation of centers of curvature}
We denote $q=\overline{V_DC_A}=d+R_A$ and $q'=\overline{V_DC_{A'}}=d'+R_{A'}$. Since $\cot\alpha '=-\cot\alpha$, we have
$J'=J$, and from the definitions of $J$ and $J'$ we obtain
\begin{equation}
  {R_D-d'\over d'}\;{q'\over q'-R_D}=J'=J= {R_D-d\over d}\;{q\over q-R_D}\,,
\end{equation}
and taking into account Eq.\ (\ref{eq196}), we deduce
\begin{equation}
  n'{q'-R_D\over q'R_D}= n{q-R_D\over qR_D}\,,
\end{equation}
that is
\begin{equation}
  {n'\over q'}={n\over q}+{n'-n\over R_D}\,.\label{eq200}\end{equation}
  Equation (\ref{eq200}) shows that the centers of curvature of ${\cal A}$ and ${\cal A}'$
  are conjugate points.

  \subsubsection{Lateral magnification}
  The lateral magnification at vertices is $m_{\rm v}$, defined by $\vec r'=m_{\rm v}\vec r$.  If the imaging is the identity, then $\vec \rho '=\vec \rho$. If the imaging is the parity operator, then $\vec \rho '=-\vec\rho$. According to Eqs. (\ref{eq187}) and (\ref{eq193}), we have
  \begin{equation}
    \vec r' =\sqrt{\lambda '\varepsilon_{A'}R_{A'}}\,\vec \rho'=\pm {\sqrt{\lambda '\varepsilon_{A'}R_{A'}}\over \sqrt{\lambda \varepsilon_{A}R_{A}}}\vec r\,,\end{equation}
  so that
  \begin{equation}
    m_{\rm v}^2={\lambda '\varepsilon_{A'}R_{A'}\over \lambda \varepsilon_{A}R_{A}}\,.\end{equation}
  We derive
   \begin{eqnarray}
     m_{\rm v}^2={\lambda '\varepsilon_{A'}R_{A'}\over \lambda \varepsilon_{A}R_{A}}
     \rap & =&\rap -{n\over n'}\,{d'\over R_{A'}+d'}\,{R_A+d\over d} \;{R_{A'}\over R_A}\;{\cot\alpha'\over \cot\alpha}\nonumber \\
     \rap & =&\rap {nd'\over n'd}\;{R_{A'}\over R_A}\;{R_A+d\over R_{A'}+d'}
     = {nd'\over n'd} \,{q'-d'\over q-d}\;{q\over q'}\,.\label{eq204}\end{eqnarray}
   From Eqs.\ (\ref{eq219}) and (\ref{eq200}) we obtain
   \begin{equation}
     n'{q'-d'\over q'd'}={n'-n\over R_D}=n{q-d\over qd}\,,\end{equation}
   and Eq.\ (\ref{eq204}) becomes
   \begin{equation}
     m_{\rm v}^2={n^2d'^2\over n'^2d^2}\,.\end{equation}

   If $\alpha '=-\alpha$, then  $d\primespe/d>0$, because $\alpha d<0$ and $\alpha 'd'>0$;  then $m_{\rm v}=nd'/n'd$. If  $\alpha +\alpha '=\pi$, then $\alpha>0$ and $\alpha '>0$, so that $d\primespe/d<0$ and $m_{\rm v}=-nd'/n'd$.

\subsubsection{Radius-magnification law (longitudinal magnification)}

The lateral magnification at curvature centers is $m_{\rm c}= nq'/n'q$. The radius magnification is
   \begin{equation}
     m_{\rm r}={R_{A'}\over R_A}={q'-d'\over q-d}={nq'd'\over n'qd}={n'\over n}\,m_{\rm v}\,m_{\rm c}\,.
   \end{equation}

   Two arbitrary points $V$ and $C$ on the axis of ${\cal L}$, in the object space, can be considered as the vertex and the center of curvature of a spherical emitter ${\cal A}$ whose radius is $R_A=\overline{VC}$. Their conjugate points $V'$ and $C'$ are the vertex and the  center of curvature  of ${\cal A}'$, the image of ${\cal A}$.  The radius magnification between $R_A$ and $R_{A'}$ may be interpreted as the  longitudinal magnification between $\overline{VC}$ and $\overline{V'C'}$, as in Proposition \ref{proplong}.

\subsection{Imaging by a centered optical system}\label{sect63}

\subsubsection{Coherent imaging of a spherical emitter. Double conjugation}

Since a centred system is the succession of refracting spherical caps or mirrors, the result obtained for a refracting sphere can be extended to an arbitrary centred system \cite{PPF4}. If ${\cal A}'$ is the coherent geometrical image of the spherical cap ${\cal A}$, then
\begin{enumerate}
\item 
The vertex of ${\cal A}'$ is the  conjugate of the vertex of ${\cal A}$, with lateral magnification $m_{\rm v}$.
\item The center of curvature of ${\cal A}'$ is the conjugate of the center of curvature of ${\cal A}$, with lateral magnification $m_{\rm c}$.
\item  The radius-magnification law is ${R_{A'}/R_A}=m_{\rm r}=(n'/n)m_{\rm v}m_{\rm c}$ ($n$ and $n'$ are the refractive indices of the object space and the image space, respectively).
\end{enumerate}

\subsubsection{System with foci: conjugation formula}

Equation (\ref{eq219}) is usually written
\begin{equation}
{n'\over d\primespe}={n\over d}+{n'\over f'}\,,\label{eq229}\end{equation}
where $f'$ is the image focal length of the refracting cap. It is such that
\begin{equation}
f'={n'R_D\over n'-n}\,.\end{equation}
($R_D$ is assumed to be not infinite.)
The object focal length is $f$ such that
\begin{equation}
f={nR_D\over n-n'}=-{n\over n'}f'\,.\end{equation}

One actually proves that Eq.\ (\ref{eq229}) holds for every system with foci (including mirrors). Algebraic measures $d$ and $d'$ have their origins at the principal (or unit) points of the lens (on the optical axis), usually denotes as $H$ and $H'$.  If $A'$ is the image of a luminous point $A$ located on the axis, then $d=\overline{HA}$ and $d'=\overline{H'A'}$. If $F'$ denotes the image focus, the image focal-length is $f'=\overline{H'F'}$: it is strictly positive for a convergent lens, and strictly negative for a divergent lens.

\subsubsection{A classical proof of Theorem 3}

We begin with a centered system with foci. 
Equation (\ref{eq229}) can be written
\begin{equation}
d'={n' d\over \displaystyle{n'\over f'} d+n}={n' d\over n\left(1-\displaystyle{d\over f}\right)}\,,\label{eq232}\end{equation}
which shows that the mapping $d\,\longmapsto\,d\primespe$ is  a homographic function, say $h$. 
\begin{figure}[h]
\centering
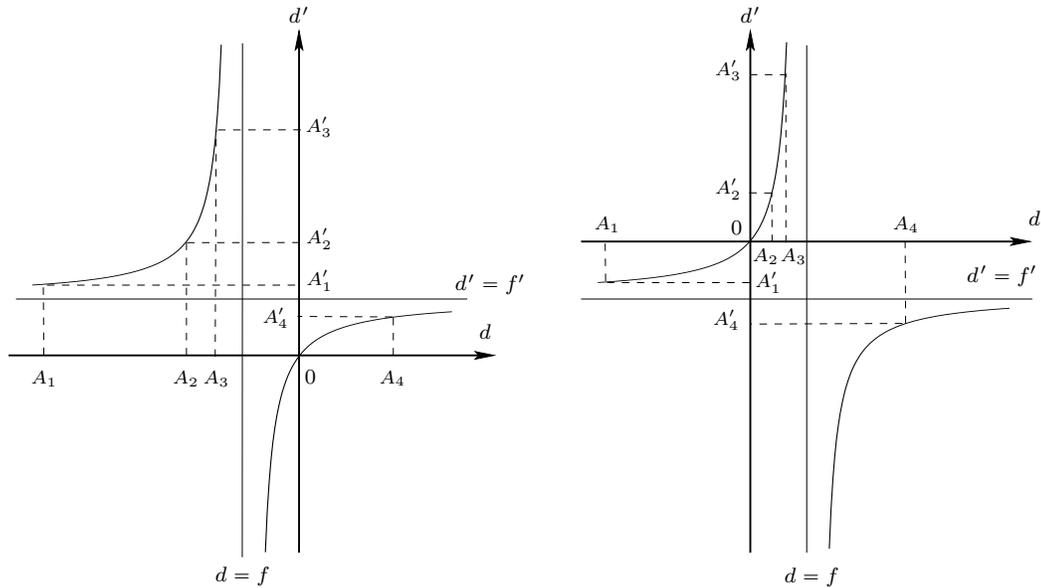
\caption{\small According to Eq.\ (\ref{eq232}) the mapping $h : d\,\longmapsto \, d\primespe$ is a homographic function. Left diagram: graph of $h$ for a convergent lens ($f'>0$); right diagram: graph of $h$ for a divergent lens ($f'<0$). In both cases  the mapping transforms the arrangement $(A_1A_2A_3A_4)$ on the $d$--axis  into  an arrangement of points $A'_j$, on the $d\primespe$--axis, which is a circular permutation of  $(A'_1A'_2A'_3A'_4)$, where $A'_j$ is the image of $A_j$. \label{fig9}}
\end{figure}
Figure \ref{fig9} shows the graph of $h$ for a convergent lens ($f'>0$, left diagram) and for a divergent lens ($f'<0$, right diagram): there is a horizontal asymptote whose equation is $d\primespe =f'$, and a vertical one, whose equation is $d=f$. The image of $]-\infty, f[$ is $]f',+\infty[$, and the image of $]f, \infty[$ is $]-\infty , f '[$. The function $h$ is strictly increasing on both intervals.

Let us consider a sequence of points $A_j$ ($j=1,2,3,4$), in the object space,  arranged on the lens axis according to $(A_1A_2A_3A_4)$, and let $d_j=\overline{HA_j}$.  For every $j$, let $d\primespe_j=\overline{H'A'_j}$, where $A'_j$ is the image of $A_j$ formed by ${\cal L}$. In Fig.\ \ref{fig9}, the $d$--axis represents the lens axis in the object space, with $d=0$ taken at point $H$; and the $d\primespe$--axis represents the lens axis in the image space, with $d'=0$ at point $H'$.

By inspection, we note that the points $A'_j$ form a circular permutation of $(A'_1A'_2A'_3A'_4)$. That is Theorem \ref{th3}.

If the system is afocal, since the lateral magnification is constant, the longitudinal magnification, say $m$,  is also constant, and is positive. Then for every pair of indices $(i,j)$, we have $\overline{A'_iA'_j} =m\, \overline{A_iA_j}$.
If we take $A_1$ as origin, the arrangement $(A_1A_2A_3A_4)$ is such that $\overline{A_1A_2}<\overline{A_1A_3}<\overline{A_1A_4}$,
and since $\overline{A'_1A'_j}=m \,\overline{A_1A_j}$, with $m>0$, we obtain $\overline{A'_1A'_2}<\overline{A'_1A'_3}<\overline{A'_1A'_4}$,
which means that the points $A'_i$ are arranged along the axis according to $(A'_1A'_2A'_3A'_4)$ (see Remark \ref{rem52}).

The proof is complete. \qed


{\Large \bigskip \centerline{\sc Conclusion}} \label{conc} \bigskip\smallskip

Although equivalent, all expressions of fractional-order Fourier transformations have not equal practical interests. Let us cite Namias \cite{Nam}: ``{\its The  operator\,\footnote{The operator $A$, as  defined by Namias, is $A={1\over 2}\bigl(x^2-\displaystyle{\D^2\over \D x^2}-1\bigr)$. It is related to the harmonic oscillator operator.} ${\cal F}_\alpha=\E^{\I\alpha A}$, although quite useful in theoretical considerations, does not lend itself to the simple and direct evaluation of the fractional transforms. Even in the usual Fourier transform case, use of the operator $\E^{\I\pi A/2}$ is most unpractical and one resorts to the integral representation. Thus, evaluation of fractional transforms could be facilitated by means of a corresponding integral representation.''}  
The integral form, as in Eq.\ (\ref{eq26}), clearly facilitates comparison with standard diffraction integrals (e.g., Fresnel integrals) more than the exponential form $\exp(\I\alpha {\cal H})$ does (${\cal H}$ denotes the harmonic oscillator\footnote{The two-dimensional harmonic oscillator operator is
${\cal H}=\displaystyle{1\over 2}(x_1^2+x_2^2)-\displaystyle{1\over 8\pi ^2}\Bigl(\displaystyle{\partial^2\over\partial x_1^2}+{\partial^2\over\partial x_2^2}\Bigr)$.}).

Fractional-order Fourier transformations are Weyl pseudo-differential operators. Their theory can be incorporated into the framework of pseudo-differential calculus. Should we follow Namias's remark and consider that this incorporation leads only to theoretical consequences? An intermediate position seems more reasonable.
Of course, some results on fractional transformations can be directly deduced from their definitions---for example, ${\cal F}_{\pi /2}={\cal F}$, from Eq.\ (\ref{eq26}), and ${\cal K}_0=\I {\cal F}$, from Eq.\ (\ref{eq28}). Others,  such as ${\cal F}_0={\cal I}$, or ${\cal F}_\pi ={\cal P}$, require more complex proofs, which pseudo-differential calculus may help to establish.  In this article, we showed how manipulating fractional-transformation symbols---in the meaning of pseudo-differential operators---led to the proof of a basic theorem (Theorem~\ref{th1}), with actual applications to optics (Theorem \ref{th4}).  The theory of pseudo-differential operators affords additional mathematical tools for dealing with fractional Fourier transformations and their applications.

\appendix


\section{Notation on distributions}\label{appenA}
A test function is a complex-valued function,  defined on  ${\mathbb R}^d$ ($d$ a strictly positivre integer), infinitely differentiable with compact support. The space ${\cal D}({\mathbb R}^d)$ is the vector space of test functions, endowed with an appropriate topology. The space of distributions is ${\cal D}'({\mathbb R}^d)$, the topological dual  vector-space of ${\cal D}({\mathbb R}^d)$ \cite{Sch}. A distribution, say $T$,  is thus a continuous linear form on ${\cal D}({\mathbb R}^d)$. If $\varphi$ is a test function, we write $T(\varphi )=\langle T,\varphi \rangle\in{\mathbb C}$. If $\vec x\in {\cal D}({\mathbb R}^d)$, we also write $\langle T_{\vec x},\varphi (\vec x) \rangle$ or $\langle T({\vec x}),\varphi (\vec x) \rangle$, to make the variable conspicuous.

The tensor product of distributions $S\in{\cal D'}({\mathbb R}^d)$ and $T \in{\cal D'}({\mathbb R}^{d\primespe})$ is $S\otimes T$, defined by
\begin{equation}
  \langle S\otimes T,\varphi (\vec x,\vec y)\rangle =\bigl\langle S_{\vec x}\,,\langle T_{\vec y},\varphi (\vec x,\vec y)\rangle\bigr\rangle = \bigl\langle T_{\vec y}\,,\langle S_{\vec x},\varphi (\vec x,\vec y)\rangle\bigr\rangle\,,\end{equation}
where $\varphi \in{\cal D}({\mathbb R}^{d+d'})$. In particular, if $f\in  {\cal D}({\mathbb R}^{d})$ and $g\in {\cal D}({\mathbb R}^{d\primespe})$, then
\begin{equation}
  \langle S\otimes T,f(\vec x)\,g(\vec y)\rangle=\langle S,f\rangle\,\langle T,g\rangle\,.
  \end{equation}

If $\vec a=(a_1,\dots, a_d)$, the Dirac distribution with support $\{\vec a\}$ is denoted $\delta_{\vec a}$ and is  defined by $\langle \delta_{\vec a},\varphi\rangle =\varphi (\vec a)$.
We also use the (improper) integral notation
\begin{equation}\langle \delta_{\vec a},\varphi\rangle =
  \int_{{\mathbb R}^d}\delta_{\vec a}\,\varphi (\vec x)\,\D \vec x =\varphi (\vec a)\,.\end{equation}

For $\vec a=(a_1,\dots,a_d)\in {\mathbb R}^d$, the Dirac distribution $\delta_{\vec a}$ is an element of ${\cal D'}({\mathbb R}^d)$ and it is also the tensor product of Dirac distributions in ${\cal D'}({\mathbb R})$, explicitly
$\delta_{\vec a}=\delta_{a_1}\otimes \cdots\otimes \delta_{a_d}$.
We will abusively write  $\delta_{\vec a}=\delta_{a_1}\,\delta_{a_2} \cdots \delta_{a_d}$.

We use $0$ as a universal symbol, that is, we write $0$ in place of $\vec 0=(0,\dots, 0)$. The Dirac distribution with support $\{\vec 0\}$ is denoted $\delta$ (in place of $\delta_{\vec 0}$), and we write $\langle \delta, \varphi \rangle=\varphi (0)$.

We also abusively write $\delta_{\vec a}$ as a function of a variable, say $\vec x$, i.e. we write $\delta_{\vec a}=\delta (\vec x-\vec a)$, and
\begin{equation}
  \langle \delta_{\vec a},\varphi \rangle =\langle \delta (\vec x-\vec a),\varphi (\vec x)\rangle =\int_{{\mathbb R}^d}\delta (\vec x-\vec a)\,\varphi (\vec x)\,\D\vec x =\varphi (\vec a)\,.
\end{equation}

If $\vec a=(a_1,\dots,a_d)\in {\mathbb R}^d$ and   $\vec b=(b_1,\dots,b_{d\primespe})\in {\mathbb R}^{d\primespe}$, then
\begin{equation}
  \delta_{\vec a}\otimes\delta_{\vec b}=\delta_{a_1}\otimes\cdots\otimes\delta_{a_d}\otimes\delta_{b_1}\otimes\cdots\otimes\delta_{b_{d'}}\,,
\end{equation}
and with variables $\vec x=(x_1,\dots,x_d)\in {\mathbb R}^d$ and   $\vec y=(y_1,\dots,y_{d\primespe})\in {\mathbb R}^{d\primespe}$, we write
\begin{equation}
  \delta (\vec x-\vec a)\otimes\delta (\vec y-\vec b)=\delta (x_1-a_1)\otimes\cdots\otimes\delta(x_d-a_d)\otimes\delta(y_1-b_1)\otimes\cdots\otimes\delta(y_{d\primespe}-b_{d\primespe})\,.
\end{equation}
In particular, we omit the tensor symbol $\otimes$ and write  (for $\vec a=0$ and $\vec b=0$)
\begin{equation}
  \delta (\vec x)\,\delta (\vec y)=\delta (x_1)\cdots\delta (x_d)\,\delta (y_1)\cdots \delta (y_{d\primespe})\,.\end{equation}

If $A$ is a real number and $\vec x\in{\mathbb R}^d$, a property of the Dirac distribution is
\begin{equation}
  \delta (A\vec x)={1\over |A|^d}\,\delta (\vec x)\,.\end{equation}

In the present article, vectors $\vec x$, $\vec y$, $\vec \xi$, $\dots$ generally are two-dimensional vectors: $\vec x=(x_1,x_2)$, $\vec y=(y_1,y_2)$, $\vec \xi =(\xi_1,\xi_2)$, etc. 
Thus $\delta (\vec x)=\delta (x_1)\otimes\delta (x_2) =\delta (x_1)\,\delta (x_2)$. The distribution $\delta (\vec x)\,\delta (\vec \xi)$ is the tensor product $\delta (\vec x)\otimes\delta (\vec \xi)=\delta (x_1)\otimes\delta (x_2)\otimes\delta (\xi_1)\otimes\delta (\xi_2)$; it is an element of ${\cal D}'({\mathbb R}^4)$ and it can be written $\delta (\vec X)$, where $\vec X=(x_1,x_2,\xi_1,\xi_2)\in{\mathbb R}^4$.

Eventually, we recall that tempered distributions are elements of the vector space ${\cal S}'({\mathbb R}^d)$, which is the topological dual of ${\cal S}({\mathbb R}^d)$, the vector space  of rapidly decreasing functions.



\section{Composition of field-transfer operators: proof of Equation (\ref{eq2.23})}\label{appenB}

Notation is that of Sect.\ \ref{sect441}. We write Eq.\ (\ref{eq2.16}) in the form
\begin{equation}
  U_B(\vec r')=\int_{{\mathbb R}^2}h_{BA}(\vec r',\vec r)\,U_A(\vec r)\,\D\vec r=\bigl\langle h_{BA}(\vec r',\vec r),U_A(\vec r)\bigr\rangle\,,\label{eq2.248}
  \end{equation}
and we use similar notation for $U_C$ and $U_B$ in Eqs.\ (\ref{eq2.18}) and (\ref{eq2.20}). The explicit expression of $h_{BA}$, given by Eq.\ (\ref{eq2.17}), leads us to write
\begin{eqnarray}
  U_B(\vec r')\rap &=&\rap {\I\over \lambda D}\int_{{\mathbb R}^2} \exp\left(-{\I\pi r'^2\over \lambda R_B}\right)
   \exp\left(-{\I\pi \|\vec r'-\vec r\|^2\over \lambda D}\right)
   \exp\left({\I\pi r^2\over \lambda R_A}\right)\,U_A(\vec r)\,\D\vec r\nonumber \\
   \rap &=&\rap {\I\over \lambda D} \exp\left(-{\I\pi r'^2\over \lambda R_B}\right)\int_{{\mathbb R}^2}
   \exp\left(-{\I\pi \|\vec r'-\vec r\|^2\over \lambda D}\right)
   \exp\left({\I\pi r^2\over \lambda R_A}\right)\,U_A(\vec r)\,\D\vec r\,,\label{eq2.249}
\end{eqnarray}
or equivalently, with distributions
\begin{eqnarray}
  U_B(\vec r')\rap &=&\rap {\I\over \lambda D}\Bigl\langle\exp\left(-{\I\pi r'^2\over \lambda R_B}\right)
   \exp\left(-{\I\pi \|\vec r'-\vec r\|^2\over \lambda D}\right),
   \exp\left({\I\pi r^2\over \lambda R_A}\right)\,U_A(\vec r)\Bigr\rangle\nonumber \\
   \rap &=&\rap {\I\over \lambda D} \exp\left(-{\I\pi r'^2\over \lambda R_B}\right)\Bigl\langle
   \exp\left(-{\I\pi \|\vec r'-\vec r\|^2\over \lambda D}\right),
   \exp\left({\I\pi r^2\over \lambda R_A}\right)\,U_A(\vec r)\Bigr\rangle \nonumber \\
   \rap &=&\rap {\I\over \lambda D} \exp\left(-{\I\pi r'^2\over \lambda R_B}\right)\Bigl\langle
   \exp\left(-{\I\pi \|\vec r'-\vec r\|^2\over \lambda D}\right)
   \exp\left({\I\pi r^2\over \lambda R_A}\right),U_A(\vec r)\Bigr\rangle\,.\label{eq2.250}
\end{eqnarray}
(In the last line we have used:  if $T$ is a distribution and $\varphi$ a test function, then $\langle fT,\varphi\rangle = \langle T,f \varphi\rangle $, where $f$ is a function of class $C^{\infty}$.)
   
For the proof of Eq. (\ref{eq2.23}) let us introduce the auxiliary functions
\begin{equation}
  V_A(\vec r)=\exp\left({\I\pi r^2\over \lambda R_A}\right)U_A(\vec r)\,,
\end{equation}
\begin{equation}
  V_B(\vec r ')=\exp\left({\I\pi r'^2\over \lambda R_B}\right)U_B(\vec r')\,,
  \end{equation}
\begin{equation}
   V_C(\vec r'')=\exp\left({\I\pi r''^2\over \lambda R_C}\right)U_C(\vec r'')\,.
\end{equation}

According to Eq.\ (\ref{eq2.16}) and (\ref{eq2.17}),
the field transfer from ${\cal A}$ to ${\cal B}$  can be expressed by a convolution product, in the form
 \begin{equation} V_B(\vec r')=g_{BA}*V_A (\vec r')\,,\label{eq251}
 \end{equation}
 where
 \begin{equation}
   g_{BA}(\vec r')={\I\over \lambda D}\exp\left(-{\I\pi r'^2\over \lambda D}\right)\,.\label{eq252}\end{equation}
By Fourier transformation, Eq.\ (\ref{eq251}) becomes
\begin{equation}
  \widehat{V}_B=\widehat g_{BA}\,\widehat V_A\,,\label{eq257}\end{equation}
which is an equivalent way of representing the field transfer ${\cal G}_{BA}$ from ${\cal A}$ to ${\cal B}$.

 Another way is obtained by the composition of ${\cal G}_{CA}$ and ${\cal G}_{BC}$; it is as follows. 
According to Eqs.\ (\ref{eq2.18}) and (\ref{eq2.19}), the field transfer from ${\cal A}$ to ${\cal C}$ takes the form 
\begin{equation} V_C(\vec r'')=g_{CA}*V_A (\vec r'')\,,\label{eq253}
\end{equation}
where
\begin{equation}
  g_{CA}(\vec r'')={\I\over \lambda D'}\exp\left(-{\I\pi r''^2\over \lambda D'}\right)\,,\label{eq254}\end{equation}
and, according to Eqs.\ (\ref{eq2.20}) and (\ref{eq2.21}), the field transfer from ${\cal C}$ to ${\cal B}$ takes the form
 \begin{equation} V_B(\vec r')=g_{BC}*V_C (\vec r')\,,\label{eq255}
\end{equation}
where
\begin{equation}
  g_{BC}(\vec r'')={\I\over \lambda D''}\exp\left(-{\I\pi r''^2\over \lambda D''}\right)\,.\label{eq256}\end{equation}
By Fourier transformation, Eqs.\ (\ref{eq253}) and (\ref{eq255}) become
\begin{equation}
  \widehat{V}_C=\widehat g_{CA}\,\widehat V_A\,,\end{equation}
and
\begin{equation}
  \widehat{V}_B=\widehat g_{BC}\,\widehat V_C\,,\end{equation}
and they combine to represent ${\cal G}_{BA}$ in the form
\begin{equation}
  \widehat{V}_B=\widehat g_{BC}\,\widehat V_C=\widehat g_{BC}\;\widehat g_{CA} \;\widehat V_A\,.\label{eq260}
\end{equation}
According to the Huygens--Fresnel principle, the two methods must give the same result, so that by comparing Eqs.\ (\ref{eq257}) and (\ref{eq260}), we obtain
\begin{equation}
  \widehat g_{BA}=\widehat g_{BC}\;\widehat g_{CA}\,.\label{eq261}\end{equation}

Let us check that Eq.\ (\ref{eq261}) holds for $g_{BA}$,  $g_{CA}$ and $g_{BC}$ given by Eqs.\ (\ref{eq252}), (\ref{eq254}) and (\ref{eq256}).
In explicit form, we use Eq.\ (\ref{eq3}) and from Eqs.\ (\ref{eq252}), (\ref{eq254}) and (\ref{eq256}), we obtain
\begin{equation}
  \widehat g_{BA}(\vec F)=\exp (\I\pi \lambda DF^2)\,,\hskip .3cm  \widehat g_{CA}(\vec F)=\exp (\I\pi \lambda D'F^2)\,,\hskip .3cm 
  \widehat g_{BC}(\vec F)=\exp (\I\pi \lambda D''F^2)\,,
\end{equation}
where $\vec F$ is a spatial frequency, the conjugate variable of spatial variables $\vec r$, $\vec r ''$ and $\vec r'$ (with $F=\|\vec F\|$). Since $D=D'+D''$, we obtain
\begin{equation}
  \widehat g_{BC}(\vec F)\;\widehat g_{CA}(\vec F) =
  \exp [\I\pi \lambda (D'+D'')F^2]\,=\exp (\I\pi \lambda DF^2)= \widehat g_{BA}(\vec F)\,.
\end{equation}

To complete the proof, we write Eq.\ (\ref{eq251}) in the form
\begin{equation}
  V_{B}(\vec r')=\langle g_{BA}(\vec r'-\vec r),V_A(\vec r)\rangle\,,
  \end{equation}
and Eqs.\ (\ref{eq253}) and (\ref{eq255}) as
\begin{equation}
  V_{C}(\vec r'')=\langle g_{CA}(\vec r''-\vec r),V_A(\vec r)\rangle\,,\hskip .5cm \mbox{and}\;\;\;
  V_{B}(\vec r')=\langle g_{BC}(\vec r'-\vec r''),V_C(\vec r'')\rangle\,,
  \end{equation}
so that  Eq.\ (\ref{eq260}) is equivalent to
\begin{equation}
  V_B(\vec r')=\bigl\langle  g_{BC}(\vec r'-\vec r''),\langle  g_{CA}(\vec r''-\vec r),V_A(\vec r) \rangle\bigr\rangle\,.
\end{equation}
We use the notation of Eq.\ (\ref{eq2.250}) and we derive
\begin{eqnarray}
  U_{B}(\vec r')\rap &=&\rap \exp\left(-{\I r'^2\over \lambda R_B}\right)\,V_B(\vec r')\nonumber \\
  \rap &=&\rap \exp\left(-{\I r'^2\over \lambda R_B}\right)
  \Bigl\langle  g_{BC}(\vec r'-\vec r''),\bigl\langle  g_{CA}(\vec r''-\vec r),\exp\left({\I r^2\over \lambda R_A}\right)U_A(\vec r) \bigr\rangle\Bigr\rangle\nonumber \\
  \rap &=&\rap
  \left\langle  \exp\left(-{\I r'^2\over \lambda R_B}\right) \exp\left({\I r''^2\over \lambda R_C}\right) g_{BC}(\vec r'-\vec r''),\right.\nonumber \\
  & & \hskip 2cm\left.
  \Bigl\langle \exp\left(-{\I r''^2\over \lambda R_C}\right)\exp\left({\I r^2\over \lambda R_A}\right)  g_{CA}(\vec r''-\vec r),U_A(\vec r)\Bigr\rangle
  \right\rangle\nonumber \\
  \rap &=&\rap
  \bigl\langle h_{BC}(\vec r',\vec r''),\langle h_{CA}(\vec r'',\vec r), U_A(\vec r)\rangle\bigr\rangle\,.
\end{eqnarray}
Since $U_B(\vec r')=\langle h_{BA}(\vec r',\vec r), U_{A}(\vec r)\rangle$, we obtain
\begin{equation}
  \langle h_{BA}(\vec r',\vec r), U_{A}(\vec r)\rangle= \bigl\langle h_{BC}(\vec r',\vec r''),\langle h_{CA}(\vec r'',\vec r), U_A(\vec r)\rangle\bigr\rangle\,,
\end{equation}
which is Eq.\ (\ref{eq2.23}), written with distributions. The proof is complete. \qed



\begin{thebibliography}{references}\label{refe2}

  \leftskip = -1cm

  {


\bibitem{Nam} V. Namias, ``The Fractional order Fourier transform and
its application to quantum mechanics,'' {\its J. Inst. Maths Applics} {\bf 25}
(1980) 241--265.
\bibitem{Wie} N. Wiener, ``Hermitian polynomials and Fourier analysis,'' J. Math. Phys. MIT {\bf 8} (1929) 70--73.
\bibitem{Con}  E. U. Condon, ``Immersion of the Fourier transform in a continuous
  group of functional transformations,'' {\its Proc. Nat. Acad. Sc. USA} {\bf 23}
  (1937) 158--164.

\bibitem{Kob} H. Kober, ``Wurzeln aus der Hankel-, Fourier- und aus anderen stetigen Transformationen,'' {\its Quart. J. Math. (Oxford)} {\bf 10} (1939) 45--59.

\bibitem{Pat}  A. L. Patterson, ``Function spaces between crystal space and Fourier-transform space,'' {\its Zeits. Kristal.} {\bf 112} (1959) 22--32.

\bibitem{Mcb} A. C. McBride,  F. H. Kerr, ``On Namias's fractional Fourier
  transforms,'' {\its IMA J. Appl. Math.} {\bf 39} (1987) 159--175.

\bibitem{Kha} R. S. Khare, ``Fractional Fourier analysis of defocused images,'' {\its Opt. Comm.} {\bf 12} (1974) 386--388.

\bibitem{Lud} L. F. Ludwig, ``General thin-lens action on spatial intensity distribution behaves as non-integer powers of Fourier transform,'' {\its Proceedings of the Spatial Light Modulators and Applications Conference}, South Lake Tatoe (1988) 173--176.

\bibitem{Man} D.  Mendlovic,  H. M.  Ozaktas, ``Fractional
Fourier transforms and their optical implementation I,''  {\its J. Opt. Soc. Am.
  A} {\bf 10} (1993) 1875--1881.

\bibitem{Oza1} H. M.  Ozaktas, D.\  Mendlovic, ``Fractional
Fourier transforms and their optical implementation II,''  {\its J. Opt. Soc. Am.
  A} {\bf 10} (1993) 2522--2531.

\bibitem{Loh} A. W. Lohmann, 
``Image rotation, Wigner rotation, and the fractional Fourier transform,''
  {\its J. Opt. Soc. Am. A} {\bf 10} (1993) 2181--2186.

\bibitem{PPFbog} P. Pellat-Finet, {\its Transformaci\'on de Fourier fraccional y propagaci\'on del campo electro\-magn\'etico,} Memos de investigaci\'on 121, Universidad de los Andes, Bogot\'a, 1993.

\bibitem{PPF1} P. Pellat-Finet, ``Fresnel diffraction and the fractional-order Fourier transform,'' {\its Opt. Lett.} {\bf 19} (1994) 1388--1390.

\bibitem{PPF2} P. Pellat-Finet, G. Bonnet, ``Fractional-order Fourier transform and Fourier optics,'' {\its Opt. Comm.} {\bf 111} (1994) 141--154.

\bibitem{PPF3} P. Pellat-Finet, ``Transfert du champ \'electromagn\'etique par diffraction et transformation de Fourier fractionnaire,'' {\its C. R. Acad. Sc. Paris} {\bf 320 IIb} (1995) 91--97.

\bibitem{Ali} T. Alieva, V. Lopez, V. Agullo-Lopez, L. B. Almeida, ``The fractional Fourier transform in optical propagation problems,'' {\its Journal of Modern Optics} {\bf 41} (1994) 1037--1044.

 \bibitem{Oza2} H. M.  Ozaktas, D.  Mendlovic, ``Fractional Fourier optics,''  {\its J. Opt. Soc. Am.
  A} {\bf 12} (1995) 743--751.

\bibitem{PPF4} P. Pellat-Finet, {\its Optique de Fourier. Th\'eorie m\'etaxiale
  et fractionnaire}, Springer, Paris, 2009.

  \bibitem{Alm} L. B. Almeida, ``The fractional Fourier transform and time-frequency repesentations'', {\its IEEE Transactions on Signal Processing} {\bf 42} (1994) 3084--3091.

  \bibitem{Oza3} H. M. Ozaktas, Z. Zalevsky, M.A. Kutay, {\its The Fractional Fourier Transform, with Applications in Optics and Signal Processing}, John Wiley \& Sons, Chichester, 2001.

\bibitem{PPF7} P. Pellat-Finet, \'E. Fogret, ``Effect of diffraction on Wigner distributions of optical fields and how to use it in optical resonator theory. II -- Unstable resonators,'' {\its arXiv} 2203.14546 (2022)  25 p.

\bibitem{PPF5} P. Pellat-Finet, ``Hyperbolic fractional-order Fourier transformations in scalar theory of diffraction,'' {\its arXiv}: 2509.21581v1 (2025) 35 p.

\bibitem{Mar} A. Mar\'echal, M. Fran\c con, {\its Diffraction, structure des images}, Masson, Paris, 1960.  

\bibitem{Hor3} L. H\"ormander, {\its The Analysis of Linear Partial Differential Operators III. Pseudo-Differential Operators}, Springer-Verlag, Berlin, 1994.
\bibitem{Ego} Yu. V. Egorov, A. I. Komech, M.  A. Shubin, {\its Elements of the Modern Theory of Partial Diffe\-rential Equations}, Springer-Verlag, Berlin, 1999. (Translated from Russian.)

\bibitem{Unt} A. Unterberger, {\its Quantification et analyse microlocale}. {\its Premi\`ere partie}, Cours de DEA, Universit\'e de Paris 6 (Universit\'e Pierre et Marie Curie), 1991.

\bibitem{Nou} J. Nourrigat, {\its Quantification et analyse microlocale}. {\its Deuxi\`eme partie : compl\'ements de calcul pseudo-diff\'erentiel}, Cours de DEA, Universit\'e de Paris 6 (Universit\'e Pierre et Marie Curie), 1991.

\bibitem{Har}  A. Harsoyo, {\its M\'ethodes fractionnaires en \'electromagn\'etisme et en optique,} Thesis, Universit\'e de Bretagne Sud, Lorient, 2003.
}

\bibitem{Pra} A. Prasad, P. Kumar, ``Pseudo-differential operator associated with the fractional Fourier transform,'' {\its Math. Comm.} {\bf 21} (2016) 115--126.

\bibitem{Das} S. Das, K. Mahato, A. I. Zayed, ``Characterization of pseudo-differential operators  associated with the coupled fractional Fourier transform,'' {\its Axioms} {\bf 13} (2024) 296--316.

  \bibitem{GB1}  G. Bonnet, ``Introduction \`a l'optique m\'etaxiale. Premi\`ere partie :
diffraction m\'etaxiale dans un espace homog\`ene : trilogie structurale, dioptre
sph\'erique,'' {\it Ann. T\'el\'ecomm.} {\bf 33} (1978) 143--165.

\bibitem{GB2} {G. Bonnet}, ``Introduction \`a l'optique m\'etaxiale. 
Deuxi\`eme partie :
syst\`e\-mes dioptriques centr\'es (non diaphragm\'es et non aberrants),''
{\it Ann. T\'el\'ecomm.}
{\bf 33}  (1978) 225--243.

\bibitem{PPF6} P. Pellat-Finet, \'E. Fogret, ``Effect of diffraction on Wigner distributions of optical fields and how to use it in optical resonator theory. I -- Stable resonators and Gaussian beams,'' {\its arXiv}: 2005.13430 (2020)  20 p.

\bibitem{PPF8} P. Pellat-Finet, \'E. Fogret, ``Effect of diffraction on Wigner distributions of optical fields and how to use it in optical resonator theory. III -- Ray tracing in resonators,'' {\its arXiv}: 2203.16905 (2022)  32 p.

\bibitem{Fog} \'E. Fogret, P. Pellat-Finet,  ``Agreement of fractional Fourier optics with the Huygens--Fresnel principle,'' {\its Opt. Comm.} {\bf 272} (2007) 281--288.

\bibitem{Sch} L. Schwartz, {\its Mathematics for the Physical Sciences}, Dover, Mineola, 2008. (English translation of {\its M\'ethodes math\'ematiques pour les sciences physiques}, 2nd edn, Hermann, Paris, 1965.)

\end{thebibliography}
\end{document}